\documentclass[a4paper,11pt]{article}
\usepackage{jheppub} % for details on the use of the package, please see the JINST-author-manual
\usepackage{lineno}
\usepackage[utf8]{inputenc}
\usepackage{amsmath}
\usepackage{amssymb}
\usepackage{amsfonts}
\usepackage{amsthm}
\usepackage{xcolor}
\usepackage{hyperref}
\usepackage{rotating}
\usepackage{array}
\usepackage{appendix}
\usepackage{bbold}
\usepackage{textcomp}
\usepackage{float}

\usepackage{tikz}
\usetikzlibrary{arrows,chains,matrix,positioning,scopes}

\theoremstyle{definition}

\def\sc{\textit{sc}}

\def\ads{\ell_{\text{{\tiny AdS}}}}

\title{Thermal One-point Functions\\
and Their Partial Wave Decomposition}
\author[a]{Ilija Buri\'c,}
\author[a]{Francesco Russo,}
\author[b]{Volker Schomerus}
\author[a]{and Alessandro Vichi}

\affiliation[a]{Department of Physics, University of Pisa and INFN, \\Largo Pontecorvo 3, I-56127 Pisa, Italy}
\affiliation[b]{Deutsches Elektronen Synchroton DESY, Notkestr. 85, 22607 Hamburg, Germany}

\emailAdd{ilija.buric@df.unipi.it}
\emailAdd{francesco.russo@phd.unipi.it}
\emailAdd{volker.schomerus@desy.de}
\emailAdd{alessandro.vichi@unipi.it}

\abstract{In this work we address partial wave decompositions of thermal one-point functions 
in conformal field theories on $S^1 \times S^{d-1}$. With the help of Casimir differential 
equations we develop efficient algorithms to compute the relevant conformal blocks for an 
external field of arbitrary spin and with any spin exchange along the thermal circle, at least in three dimensions. This 
is achieved by identifying solutions to the Casimir equations with a special class of 
spherical functions in the harmonic analysis of the conformal group. The resulting blocks 
are then applied to study the decomposition of one-point functions of the scalar $\phi^2$  
and the stress tensor $T$ for a three-dimensional free scalar field $\phi$. We are able to 
read off averaged OPE coefficients into exchanged fields of high weight and spin for a complete set 
of tensor structures. We also extract an asymptotic behaviour of conformal blocks and use it to analyse the density of heavy-heavy-light OPE coefficients for spinning operators, comparing it with semi-classical predictions, such as the dimensions of operators at large charge.}

\begin{document}

\maketitle

\section{Introduction}

Since the rekindling of interest in the conformal bootstrap program, \cite{Rattazzi:2008pe}, studies of conformal field theories continue to use increasingly diverse probes. Beyond the most well-understood case of four-point functions of local operators, observables of greatest interest include non-local defect operators, \cite{Liendo:2012hy,Billo:2016cpy}, and multi-point functions, \cite{Antunes:2021kmm,Poland:2023vpn,Antunes:2020yzv}. The former are of significant experimental relevance, being used to model boundaries or impurities in statistical mechanics systems, \cite{Cuomo:2021kfm,Cuomo:2022xgw}, and involve new CFT data going beyond the spectrum and OPE coefficients of local fields. By contrast, the latter can be used to place constraints on the same data present in four-point functions, however organised in a different way. Such re-organisation makes certain scaling dimensions and OPE coefficients more easily accessible, \cite{Poland:2023bny,Harris:2024nmr}. Still more constraints on CFTs are expected to follow from their consistency on geometries other than the flat space. This is the question addressed in the present work.
\smallskip

The simplest manifold which is not conformally equivalent to $\mathbb{R}^d$ is $S^1\times\mathbb{R}^{d-1}$. Correlation functions on this background may be considered as finite temperature correlators on $\mathbb{R}^d$, in the limit when the temperature becomes infinite. Quite generally, high-temperature behaviour of correlation functions encodes properties of CFT data at large scaling dimensions. This intuition has recently been formalised by the notion of thermal field theory, \cite{Benjamin:2023qsc,Benjamin:2024kdg}, that can be understood as a higher-dimensional replacement for modular invariance. In this sense, the high-temperature limit controls universal properties of CFTs that complement those in other regimes of large quantum numbers, such as spin \cite{Komargodski:2012ek,Fitzpatrick:2012yx} or charge \cite{Hellerman:2015nra}. Conjectures about behaviour of OPE coefficients in the high-temperature regime have been made in the context of the eigenstate thermalisation hypothesis (ETH), \cite{Lashkari:2016vgj}, and further explored in several examples, \cite{Karlsson:2021duj}.
\smallskip

Correlation functions on $S^1\times\mathbb{R}^{d-1}$ have been studied in the context of holography, \cite{El-Showk:2011yvt,Alday:2020eua,Dodelson:2022yvn,Dodelson:2023vrw,Karlsson:2022osn,Esper:2023jeq,Ceplak:2024bja,Rodriguez-Gomez:2021mkk,Parisini:2023nbd}, where they relate to physics of black holes, and the bootstrap, \cite{Iliesiu:2018fao,Iliesiu:2018zlz,Petkou:2018ynm,Manenti:2019wxs,Tejerinamasters,Marchetto:2023fcw,Marchetto:2023xap,David:2023uya,David:2024naf,Barrat:2024aoa}. Within the bootstrap, the most studied observable on $S^1\times \mathbb{R}^{d-1}$ is the two-point function of local fields. This is the simplest correlator not fixed by symmetry and it obeys the KMS constraint, \cite{El-Showk:2011yvt}, which assumes the role of the crossing equation. In this context, analytic bootstrap techniques, including the Lorentzian inversion formula, have been developed. Due to the lack of positivity, the KMS constraint has been analysed by truncating the resulting system of equations. This led to estimates of several thermal one-point functions, including the free energy density in the 3d Ising model, in good agreement with Monte-Carlo simulations, \cite{Iliesiu:2018zlz}.
\smallskip

It is, however, clear that many manifolds other than $S^1\times\mathbb{R}^{d-1}$ may be used to try and constrain the space of CFTs. In this respect, ‘genus two' manifolds of \cite{Benjamin:2023qsc}, as well as higher-dimensional tori and spaces of the form $\mathbb{R}\times\mathbb{T}^{d-1}$ have been explored, \cite{Shaghoulian:2015kta,Belin:2016yll}. 

In this work, we shall consider correlation functions of local operators on the geometry $S^1\times S^{d-1}$. Since these are essentially the same as thermal correlators on the space $\mathbb{R}\times S^{d-1}$ which is conformally-equivalent to $\mathbb{R}^d$, correlation functions on $S^1 \times S^{d-1}$ directly encode flat-space CFT data. However, as of now, very little is known about them. The only existing studies above two dimensions are \cite{Gobeil:2018fzy,Alkalaev:2024jxh}. Compared to $S^1\times\mathbb{R}^{d-1}$, the geometry $S^1\times S^{d-1}$ is much richer. Two distinguishing differences that will be most important for our purposes are: 1) On $S^1\times S^{d-1}$, one-point functions of local operators are not fixed by conformal symmetry, as they are on $S^1\times\mathbb{R}^{d-1}$. In the case of a scalar 
field, for example, these one-point functions depend on $d$ conformal invariants. 2) In addition to the temperature $\beta^{-1}$, the geometry $S^1\times S^{d-1}$ is characterised by rank$(SO(d))$ angular {\it chemical} potentials $y_a$. While it is possible to restrict all considerations to correlators at vanishing chemical 
potentials, i.e. set $y_a=1$, we shall argue that it is both natural and useful to explore the full dependence on these potentials. Only through the dependence on both $\beta$ and $y_a$ can one decompose thermal one-point functions into conformal blocks and read off appropriate OPE coefficients. Let us also note that $S^1 \times \mathbb{R}^{d-1}$ may be obtained as the high-temperature limit of $S^1\times S^{d-1}$. Indeed, as $\beta\to0$, the $S^1$ factor shrinks to zero - one can re-interpret 
this as the sphere $S^{d-1}$ becoming infinitely large and thus approaching $\mathbb{R}^{d-1}$. This intuition will be turned into a relation between correlators on the two backgrounds below.
\smallskip

We shall begin by discussing kinematics of general $n$-point functions on $S^1\times S^{d-1}$ and deriving the Ward identities that they satisfy. This is followed by a detailed analysis of one-point functions of operators of arbitrary spin in three dimensions, for which we shall compute conformal blocks. The latter are complicated functions of four variables and we write them in a low-temperature series expansion. The knowledge of thermal blocks opens the path to various investigations. These range from obtaining CFT data in examples where one-point functions are computable, such as in holographic theories, obtaining spin-refined asymptotics of averaged heavy-heavy-light (HHL) OPE coefficients (in the limit \mbox{$\Delta_{\text{heavy}}\to\infty$)}, to setting up crossing equations that involve correlators on different manifolds. Some of these applications are discussed in more detail in the concluding section. In the present work, we content ourselves with an analysis of block decompositions of thermal one-point functions in the free scalar theory. This already leads to interesting new results for averaged OPE coefficients. 
\smallskip

We proceed to give more details about results of the following sections.

\paragraph{Summary of results} Let us summarise the main new results of the present work. We shall start in Section \ref{S:CFT at finite temperature} by reviewing thermal $n$-point correlation functions in the presence of chemical potentials and deriving Ward identities that they satisfy. This is followed by the definition of thermal conformal blocks. Much like ordinary four-point conformal blocks, the thermal blocks satisfy Dolan-Osborn-like Casimir differential equations, \cite{Dolan:2003hv}. We will show in Section \ref{S:Harmonic analysis} that the Ward identities and Casimir equations allow to interpret thermal blocks as a certain type of spherical functions called vector characters, \cite{Etingof:1993gk}. In contrast to the vector characters usually studied in the mathematics literature, which are functions on a Lie group valued in some finite-dimensional vector space, conformal blocks are identified with spherical functions valued in an infinite-dimensional carrier space of a tensor product of field representations. Nevertheless, large parts of the theory of spherical functions apply in the infinite-dimensional setting and this will allow us to obtain explicit form of Casimir equations using Harish-Chandra's radial component map, \cite{HarishChandra}.
\smallskip

While the above statements apply to all thermal $n$-point functions, they will be written particularly explicitly for one-point functions in three dimensional CFTs, which are the main focus of this work. For such a one-point function, we shall establish, in Section \ref{S:Thermal blocks in 3d}, a conformal block decomposition of the form
\begin{equation}\label{Intro:block-decomposition}
    \mathcal{Z}(q,y)\langle\phi(x^\mu,z^\nu)\rangle_{q,y} = (-1)^\frac{\Delta_\phi}{2} r^{-\Delta_\phi} \left(z_2^2 + z_3^2\right)^{\frac{\ell_\phi}{2}} \sum_{\mathcal{O},a} \lambda_{\phi\phi\mathcal{O}}^a\, g_{\Delta_\mathcal{O},\ell_\mathcal{O}}^{\Delta_\phi,\ell_\phi,a}(q,y,p,z)\ .
\end{equation}
Here, $\phi$ is a field of dimension and spin $(\Delta_\phi,\ell_\phi)$, inserted at a point $x^\mu$ and whose indices have been contracted with a null polarisation vector $z^\nu$. The variable $q$ is related to the temperature by $q = e^{-\beta}$ and $y$ is another chemical potential, associated with rotations about the $x^1$-axis. The $\mathcal{Z}(q,y)$ is the partition function of the theory. On the right-hand side of equation~\eqref{Intro:block-decomposition} are conformal blocks $g_{\Delta_\mathcal{O},\ell_\mathcal{O}}^{\Delta_\phi,\ell_\phi,a}$ - they are labelled by quantum numbers of external and internal fields, $\phi$ and $\mathcal{O}$, as well as three-point tensor structures for the correlator $\langle\phi\mathcal{O}\mathcal{O}\rangle$. In addition to $(q,y)$, the blocks depend on two conformal invariants $(p,z)$. They will be given as a low-temperature series expansion
\begin{equation}\label{block-series-expansion}
    g^{\Delta_\phi,\ell_\phi,a}_{\Delta_\mathcal{O},\ell_{\mathcal{O}}}(q,y,p,z) = q^{\Delta_\mathcal{O}}\sum_{n_i,\varepsilon_j} A_{n_1 n_2 n_3 n_4}^{\varepsilon_1 \varepsilon_2} q^{n_1} y^{n_2} p^{n_3} z^{n_4} \left(1-p^2\right)^{\frac{\varepsilon_1}{2}} \left(1+z^2\right)^{\frac{\varepsilon_2}{2}}\ .
\end{equation}
Here, $n_i$ are integers whose precise range will be specified below. For now, we mention that $n_1\geq0$ and $n_{2,3,4}$ are bounded in terms of $n_1$ and spins $\ell_\phi$, $\ell_{\mathcal{O}}$ of external and internal fields. The $\varepsilon_j$ run over $\{0,1\}$. By imposing the Casimir equation on functions of the form~\eqref{block-series-expansion}, we find a number of linearly independent solutions equal to the number of $\langle\phi\mathcal{O}\mathcal{O}\rangle$ three-point tensor structures. Further, we provide a map that associates to any tensor structure the appropriate corresponding solution. The paper is accompanied by a Mathematica code which contains coefficients $A_{n_1\dots n_4}^{\varepsilon_1 \varepsilon_2}$, available at \href{https://gitlab.com/russofrancesco1995/thermal_blocks}{gitlab.com/russofrancesco1995/thermalblocks}. Our results are consistent with and considerably extend those of \cite{Gobeil:2018fzy}, where blocks for the case $\ell_\phi = \ell_{\mathcal{O}} = 0$ were found, again in the series expansion equation~\eqref{block-series-expansion}.
\smallskip

While in the present work we mostly focus on the low-temperature expansion \eqref{block-series-expansion}, some further exact results for thermal partial waves are obtained. One of these is the construction of the spherical weight-shifting operator $\mathfrak{q}_{\Delta_\phi,\ell_\phi} = \mathfrak{q}(q,y,p,z)$ which increases external spin of a block by one,
\begin{equation}\label{external-weight-shift}
    \mathfrak{q}_{\Delta_\phi,\ell_\phi} \cdot g_{\Delta_\mathcal{O},\ell_{\mathcal{O}}}^{\Delta_\phi,\ell_\phi,a} = g_{\Delta_\mathcal{O},\ell_{\mathcal{O}}}^{\Delta_\phi-1,\ell_\phi+1,b}\ .
\end{equation}
The existence of such an operator follows from the interpretation of thermal partial waves as spherical functions. The general construction of weight-shifting operators in the context of spherical functions was carried out in \cite{Buric:2022ucg,Buric:2023ykg}, and the computation of $\mathfrak{q}$ is an instance of it. However, unlike for four-point conformal blocks and partial waves for scattering amplitudes, the operator $\mathfrak{q}$ is not sufficient to construct arbitrary thermal blocks starting from the case of a scalar $\phi$ and a scalar $\mathcal{O}$. We note that $\mathfrak{q}$ does suffice to generate all thermal partial waves with exchanged scalars. Indeed, in these cases, there is a single tensor structure for  $\langle\phi\mathcal{O}\mathcal{O}\rangle$, irrespective of the spin of $\phi$. Therefore, indices $a$ and $b$ in equation~\eqref{external-weight-shift} become redundant and one has
\begin{equation}\label{scalar-exchange-blocks}
   g_{\Delta_{\mathcal{O}},0}^{\Delta_\phi-\ell_\phi,\ell_\phi} = \mathfrak{q}_{\Delta_\phi-\ell_\phi+1,\ell_\phi - 1} \cdot \cdot \dots \cdot \mathfrak{q}_{\Delta_\phi,0} \cdot g_{\Delta_\mathcal{O},\ell_{\mathcal{O}}}^{\Delta_\phi,0}\ .
\end{equation}
The explicit expression for $\mathfrak{q}_{\Delta_\phi,\ell_\phi}$ is given in equation~\eqref{weight-shifting-operator} and the Appendix \ref{A:Differential operators}. Furthermore, for vanishing chemical potential, $y=1$, it is expected that all scalar exchange blocks \eqref{scalar-exchange-blocks} admit a closed form expression in terms of standard special functions. We derive one result of this kind and write the scalar exchange block for external field of spin one using hypergeometric functions $_2F_1$ and $_3F_2$ in equation~\eqref{spin-1-exact}.
\smallskip

An important limit of thermal conformal blocks, particularly in view of asymptotic estimates of OPE coefficients, is that of large internal scaling dimensions, $\Delta_{\mathcal{O}}\to\infty$. We show that this limit has a simple interpretation in the harmonic analysis approach. Namely, after re-interpreting spherical functions as wavefunctions of a Calogero-Sutherland quantum mechanics problem, \cite{Olshanetsky:1983wh}, the limit becomes that of a free particle. We use this to write asymptotic expansions for thermal blocks as $\Delta_{\mathcal{O}}\to\infty$ in equation~\eqref{asymptotic-block}.
\smallskip

We test our new conformal blocks on the simplest example, that of the free bosonic theory, which already proves to be very interesting. We will begin by deriving the two-point function of the fundamental field at finite temperature and chemical potential,
\begin{align}\label{two-point-funtion}
    \langle\phi(x_1)\phi(x_2)\rangle_{q,y} &= \frac{1}{|x_{12}|}\\
    + &\sum_{l,m} \frac{(l-m)!}{(l+m)!} \frac{q^{l+1/2} y^m}{1 - q^{l+1/2} y^m} P_l^m(\cos\theta_1) P_l^m(\cos\theta_2) \left(r^l e^{-im\varphi} + r^{-l-1} e^{im\varphi}\right)\ .\nonumber
\end{align}
In this expression, $l$ runs from zero to infinity, $m=-l,-l+1,\dots,l$ and $P_l^m$ are associated Legendre polynomials. The $(r,\theta,\varphi)$ are spherical polar coordinates on $\mathbb{R}^3$ and we have partially fixed the frame by putting $r_1 = 1$ and $\varphi_1=0$. Coordinates of the second point are denoted $r_2 = r$ and $\varphi_2 = \varphi$. From the two-point function \eqref{two-point-funtion}, one can extract thermal one-point functions $\langle\mathcal{O}\rangle_{q,y}$ of operators that appear in the $\phi\times\phi$ OPE. We shall do this for the first two operators above the identity, which are $\phi^2$ and the stress tensor. For example, the one-point function of $\phi^2$ is
\begin{equation}\label{1pt-function-phi2}
    \langle\phi^2(x)\rangle_{q,y} = \frac{\sqrt{2}}{r} \sum_{l,m} \frac{(l-m)!}{(l+m)!} \frac{q^{l+1/2}y^m}{1 - q^{l+1/2}y^m} P_l^m(\cos\theta)^2\ .
\end{equation}
Upon decomposing $\langle\phi^2\rangle_{q,y}$ and $\langle T_{\mu\nu}\rangle_{q,y}$ in conformal blocks \eqref{block-series-expansion}, we read off OPE coefficients of the form $\lambda_{\phi^2\mathcal{O}\mathcal{O}}^a$ and $\lambda_{T\mathcal{O}\mathcal{O}}^a$. Some of such OPE coefficients are well known in the free theory and by re-deriving these we get a highly non-trivial check of the blocks \eqref{block-series-expansion}. In particular, decompositions also verify our treatment of shortening both for external and internal operators.

Free theory one-point functions at generic values of the temperature and chemical potential are given as series expansions such as equation~\eqref{1pt-function-phi2}. When the rotational potential is turned off, $y=1$, we are able to re-sum the series. For instance, the one-point function \eqref{1pt-function-phi2} becomes
\begin{equation}\label{1pt-function-zero-ch-pt}
    \langle\phi^2(x)\rangle_{q,1} = \frac{\sqrt{2}}{r} \frac{\log(1-q) + \psi_q(1/2)}{\log q}\,,
\end{equation}
where $\psi_q(z)$ is the $q$-digamma function.
\smallskip

We finish by performing a large-scale study, in the free theory, of averaged OPE coefficients $\lambda^a_{\phi^2\mathcal{O}\mathcal{O}}$ for fixed spin $\ell_{\mathcal{O}}$, as a function of the dimension $\Delta_{\mathcal{O}}$. Plots of OPE coefficients for spin $\ell_{\mathcal{O}}=0,2,4$, obtained using the thermal block decomposition, are given in Figures \ref{scalar-ope-plot} - \ref{spin4-ope-plot}. These can be used as raw data against which conjectures about the asymptotic behaviour of HHL OPE coefficients are to be made or checked. This is an important future direction and we dedicate Section \ref{S:Asymptotic OPE coefficients} to its further discussion. In particular we show that (in $3d$) a certain average of HHL OPE coefficients behaves as 
\begin{equation}
       \langle\langle \lambda_{HHL}\rangle\rangle \propto  \Delta_H^{\Delta_L/3}\,, 
\end{equation}
and we also compute the proportionality factor. This follows from the expected form of correlation functions at large temperature and the behavior of thermal blocks in the same limit. Applyting this reasoning to conserved $U(1)$ currents reproduces the semi-classical scaling of large charge operators $\Delta\sim Q^{3/2}$ obtained in \cite{Hellerman:2015nra, Monin:2016jmo}. Other perspectives are outlined in the concluding section.

The main text is accompanied by a number of technical appendices detailing computations. In Appendix \ref{app:holography} we give one more example of a thermal one-point function, namely the one-point function of the stress tensor in holographic theories. The latter is obtained as the Brown-York stress tensor of the Kerr-AdS spacetime.

\paragraph{Notation} Throughout the paper, we will be considering CFTs on several geometries. For convenience of the reader, we collect our notation in Table \ref{tab:my_label}

\begin{table}[h!]
    \centering
    \begin{tabular}{c|c|c|c}
       & $ \mathbb{R}^d $ & $ S^1\times S^{d-1}$ & $ S^1\times \mathbb{R}^{d-1}$ \\
        \hline
        coordinates & $x^\mu$ & $(\tau,\Omega)$ & $(\tau,\vec x)$\\
             \hline
   operators   & $\phi(x)$ &  $\Phi(\tau,\hat n)$ & $\Phi(\tau,\vec x)$ \\
        \hline
   correlators (no twist) & $\langle\ldots \rangle_{q}$  & $\langle \ldots \rangle_{S^1_\beta\times S^{d-1}}$ & $\langle \ldots \rangle_{S^1_\beta\times \mathbb{R}^{d-1}}$ \\
\hline
  correlators (with twist) & $\langle\ldots \rangle_{q,y_a}$  & $\langle \ldots \rangle_{S^1_\beta\times S^{d-1}_{\mu_a}}$ & 
    \end{tabular}
    \caption{Coordinates, fields and correlators on various geometries. The distinction with twist, no twist corresponds to presence or absence of chemical potentials. The empty cell corresponds to a case that is never used in this work.}
    \label{tab:my_label}
\end{table}

\section{Conformal Field Theory at Finite Temperature}
\label{S:CFT at finite temperature}

This section provides a smooth introduction to thermal correlators of conformal field theories on the geometry $S^1_\beta \times S^{d-1}$. In the first subsection, we shall introduce our main object of study, correlation functions of local operators at finite temperature and chemical potentials, see definition \eqref{def:Gn}. The second subsection is devoted to a comparison with the more familiar geometry $S^1_\beta \times \mathbb{R}^{d-1}$ that emerges in the limit in which the $(d-1)$-dimensional sphere has infinite radius. Returning to $S^1_\beta\times S^{d-1} $, we then discuss the Ward identities \eqref{Wardid} our correlation functions satisfy in the third subsection. As usual, these Ward identities allow us to reduce the dependence of the thermal correlators to functions of certain conformally invariant cross ratios. The resulting constraints on thermal one-point functions in $d=3$ are evaluated in the forth subsection, see in particular equation \eqref{eq:one-point-function}. Next, we define conformal blocks for thermal correlators and derive Casimir equations \eqref{tCFT-Casimir} these blocks obey. The section concludes with a brief discussion of conformal blocks for thermal one-point functions of spinning operators in three-dimensional CFTs. The blocks will be computed in later sections using techniques from harmonic analysis. Most of the material in the present section is well-known. In particular, our definitions agree with and generalise those of \cite{Gobeil:2018fzy}.

\subsection{Thermal correlation functions on $S^1_\beta \times S^{d-1}$}

We consider a conformal field theory on the flat Euclidean space $\mathbb{R}^d$. Thanks to the conformal 
equivalence of this space to the cylinder $\mathbb{R}\times S^{d-1}$,
\begin{equation}
    ds^2_{\mathbb{R}^d} = dr^2 + r^2 d\Omega_{d-1}^2 = e^{2\tau} \left(d\tau^2 + d\Omega_{d-1}^2 \right) 
    = e^{2\tau} ds^2_{\mathbb{R}\times S^{d-1}}, \qquad r = e^\tau\,,
\end{equation}
field operators and their correlation functions on the two backgrounds can be related to one another. 
Denoting quantum fields, possibly spinning, on the cylinder by $\Phi$, we have
\begin{equation}\label{Euclid-cylinder}
    \phi(x) = e^{-\Delta_\phi \tau} \Phi(\tau,\Omega) = e^{-\Delta_\phi \tau} \Phi(\tau,\hat n)\,,
\end{equation}
where $\Omega$ denotes a set of angular coordinates parametrising the sphere $S^{d-1}$ and $\hat n$ is 
the corresponding unit vector. By finite temperature correlators on the cylinder, we will understand the 
traces of the form
\begin{small}
\begin{equation}\label{thermal-correlators-cylinder}
    \langle\Phi^{(1)}(\tau_1,\Omega_1)\dots\Phi^{(n)}(\tau_n,\Omega_n)\rangle_{S^1_\beta\times S^{d-1}} 
    \equiv \frac{1}{\mathcal{Z(\beta)}}\text{tr}_\mathcal{H}\left(\Phi^{(1)}(\tau_1,\Omega_1)\dots
    \Phi^{(n)}(\tau_n,\Omega_n)e^{-\beta D}\right)\ ,
\end{equation}
\end{small}
where as usual we defined the partition function as 
\begin{equation}
   \mathcal{Z}(\beta)=\text{tr}_\mathcal{H}\left(e^{-\beta D}\right)\,.
\end{equation}
Here, $\mathcal{H}$ is the Hilbert space of the theory on the cylinder. In the path integral approach, the 
sum over states at time $\tau=0$ and $\tau=\beta$ is recast as summing over field configurations with 
periodic boundary conditions. Traces \eqref{thermal-correlators-cylinder} are then interpreted as correlation 
functions in the background geometry $S^1_\beta\times S^{d-1}$ with radii $\beta/2\pi$ and $R$, respectively. 
For a QFT on $S^1_\beta\times S^{d-1}$, correlations functions depend both on $\beta$, the length of the 
circle, and $R$, the radius of the sphere. In the case of a CFT, however, only their ratio $\beta/R$ matters. 
Without loss of generality we shall therefore set $R=1$. We can recover the infinite cylinder geometry of 
vanishing temperature by sending $\beta\rightarrow \infty$. Conversely, the limit $\beta\rightarrow 0$ (or 
equivalently $R\rightarrow \infty$) is that of large temperature. The resulting geometry is $S^1_\beta \times 
\mathbb{R}^{d-1}$, where one has blown up the sphere. 
\smallskip

Clearly, thanks to the relation \eqref{Euclid-cylinder}, we may equivalently consider the same type of 
correlators on $\mathbb{R}^d$. We will denote them as
\begin{equation}
    \langle\phi^{(1)}(x_1)\dots\phi^{(n)}(x_n)\rangle_q \equiv \prod_{i=1}^n e^{-\Delta_\phi \tau_i} 
    \langle\Phi^{(1)}(\tau_1,\Omega_1)\dots\Phi^{(n)}(\tau_n,\Omega_n)\rangle_{S^1_\beta\times S^{d-1}}\ .
\end{equation}
In the following, we shall be slightly more general and study correlation functions of the form
\begin{align}\label{thermal-correlators}
    \langle \phi_1(x_1)\dots \phi_n(x_n)\rangle_{q,y_a} = \frac{1}{\mathcal Z(q,y)}\text{tr}_\mathcal{H} 
    \left(\phi_1(x_1)\dots \phi_n(x_n) q^D y_2^{H_2}\dots y_r^{H_r}\right) \ ,
\end{align}
where the twisted partition function is
\begin{align}
  & {\mathcal Z(q,y_a)} = \text{tr}_\mathcal{H} \left( q^D y_2^{H_2}\dots y_r^{H_r}\right) \ .
\end{align}
Here, $q=e^{-\beta}$, while $H_2,\dots,H_r$ are Cartan generators of the rotation Lie algebra $\mathfrak{so}(d)$. 
We shall refer to $y_a=e^{\mu_a}$ as the {\it chemical potentials}. They can be interpreted as twisting the upper 
boundary condition on the cylinder before gluing the extremes. The parameters $\beta, \mu_a$ may be continued from 
the real axis into the complex domain. We only need $\beta > 0$ to ensure convergence. Points $x_i$ belong to 
$\mathbb{R}^d \cup \{\infty\}$ and are acted upon by conformal transformations in the familiar way - infinitesimal 
conformal transformations on the field $\phi_i$ are realised as first order differential operators that we denote 
by $X^{(i)}$. In case the field $\phi_i$ carries spin, these operators involve rotation matrices or, equivalently, 
they act as differential operators on the polarisation vectors, see e.g. 
equations~\eqref{principal-series-1}-\eqref{principal-series-4}. We shall often refer to this representation $X^{(i)} = 
\pi_i(X) $ of the conformal algebra as the field representation $\pi_i$. By a slight abuse of terminology, we call 
the correlators defined in equation~\eqref{thermal-correlators} \textit{thermal correlators}. These are the central 
objects we wish to study.
\smallskip

In this work we develop a general method to compute the conformal blocks for thermal correlation functions, focusing 
on one-point (1pt) functions. The presence of the partition function in the definition \eqref{thermal-correlators}, 
however, makes the conformal block decomposition ill-defined, since the function $\mathcal{Z}(q,y_a)$ is kinematically 
unconstrained. For this reason, it is more convenient consider block expansions of the \emph{unnormalised} correlation 
functions, which we denote as
\begin{align} \label{def:Gn} 
        G_n(x_i,q,y_a) &= \mathcal{Z}(q,y_a) \, \langle \phi_1(x_1)\dots \phi_n(x_n)\rangle_{q,y_a}\\
        &=\text{tr}_\mathcal{H} \left(\phi_1(x_1)\dots \phi_n(x_n) q^D y_2^{H_2}\dots y_r^{H_r}\right) \ .\nonumber
\end{align}
As we shall see below, these are the quantities that admit a nice conformal block decomposition. Before we 
enter our analysis of the unnormalised thermal correlators $G_n$ we want to pause for a moment and compare 
these quantities with a closely related set of thermal correlation functions that has received more 
attention in the past. 

\subsection{Comparison with thermal correlators on $S^1_\beta \times \mathbb{R}^{d-1}$}
\label{sec:kinematics}

As we mentioned above, the most common geometry studied in the context of thermal CFTs is $S^1_\beta \times \mathbb{R}^{d-1}$. We collect some of the salient features of correlators on this background. The operators on this geometry will be denoted as $\Phi_i(\tau, \vec x)$, where we have written points in $S^1_\beta\times\mathbb{R}^{d-1}$ as $(\tau,\vec{x})$ and the Euclidean time is periodic, $\tau\sim\tau+\beta$. The only local operators that have non-vanishing one-point functions are even-spin symmetric traceless tensor (STT) primaries. Moreover, these one-point functions are fixed up to a single coefficient, $b_\Phi$,
\begin{equation}\label{eq:1pt-large-beta}
   \langle\Phi^{\mu_1\dots\mu_{\ell_\phi}}(\tau,\vec{x})\rangle_{S^1_\beta \times \mathbb{R}^{d-1}} = \frac{b_\Phi}{\beta^{\Delta_\phi}}\left(e^{\mu_1}\dots e^{\mu_{\ell_\phi}} - \text{traces}\right)\ .
\end{equation}
 Here, $e$ is the unit vector in the $\tau$-direction. Indeed, by translation invariance, the one-point function does not depend on $\tau$ or $\vec{x}$, and the $SO(d-1)$-invariance in the $\vec{x}$-plane forces one to use the vector $e$ for its construction\footnote{The $SO(d-1)$ symmetry extends to $O(d-1)$ which includes the map $\tau\to-\tau$ composed with a space reflection.}. Descendants instead have a vanishing one-point function due to translation invariance.
\smallskip

The simplest correlator not fixed by symmetry is the two-point function. It admits the conformal block expansion that follows by performing the OPE. For two identical scalars $\Phi$, for example, one has 
\begin{small}
\begin{equation}\label{two-pt-function-Rdm1}
  \langle\Phi(\tau,\vec x)\Phi(0)\rangle_{S^1_\beta \times \mathbb{R}^{d-1}} = \sum_{\mathcal{O}} \frac{\ell_{\mathcal{O}}!}{2^{\ell_{\mathcal{O}}}(\frac{d-2}{2})_{\ell_{\mathcal{O}}}}\frac{\lambda_{\phi\phi\mathcal{O}}b_\mathcal{O}}{\beta^{\Delta_\mathcal{O}}} C_{\ell_\mathcal{O}}^{(\frac{d-2}{2})} \left(\frac{\tau}{|x|}\right) |x|^{\Delta_{\mathcal{O}}-2\Delta_\phi}\,, \quad |x| = \sqrt{\tau^2+|\vec{x}|^2}\ .
\end{equation}
\end{small}

The OPE, and consequently the blocks expansion, is valid for $|x|<\beta$. The KMS condition expresses periodicity in $\tau$ of the two-point function \eqref{two-pt-function-Rdm1}. This leads to constraints on one-point coefficients $b_{\mathcal{O}}$ appearing on the right-hand side of equation \eqref{two-pt-function-Rdm1} because individual blocks are not periodic. We refer the reader to \cite{El-Showk:2011yvt,Iliesiu:2018fao} for details.
\smallskip

%\paragraph{Kinematics on $S^1 \times S^{d-1}$} 
Let us contrast the above situation to the geometry $S^1 \times S^{d-1}$ we described in the previous subsection. When the chemical potentials $\mu_a$ are switched off, the correlators are invariant under the isometries of the background geometry $SO(2)\times SO(d)$. Hence, $n$-point correlation functions can only depend on differences of times $\tau_{ij}= \tau_i-\tau_j$, scalar products of vectors on the sphere $\hat n_i\cdot\hat n_j$ and eventually external polarisations $z_i^{\mu}$. For one-point functions of scalars this implies
\begin{equation}\label{1-pt-no-chemical}
   \langle \Phi(\tau,\hat n)\rangle_{S^1_\beta\times S^{d-1}} = \frac{b_\Phi f_\Phi(\beta)}{\beta^{\Delta_\phi}}\,,
\end{equation}
with $f_\Phi(\beta)$ a function not fixed by the symmetry\footnote{The function $f_\Phi$ is secretly a function of the ratio $\beta/R$, while the prefactor $\beta^{-\Delta_\phi}$ does not contain the radius $R$.}. In particular the above expression does not depend on $\hat n$. For general spin the above formula generalises to\footnote{See the equation \eqref{extended-Ward-invariance} for the precise reading of tensor indices in equations \eqref{eq:1pt-large-beta} and \eqref{eq:1pt-no-chemical-spin}.}
\begin{equation}\label{eq:1pt-no-chemical-spin}
   \langle\Phi^{\mu_1\dots\mu_{\ell_\phi}}(\tau,\hat n )\rangle_{S^1_\beta\times S^{d-1}} = \frac{b_\Phi f_\Phi(\beta)}{\beta^{\Delta_\phi}}\left(e^{\mu_1}\dots e^{\mu_{\ell_\phi}} - \text{traces}\right)\ .
\end{equation}
When the chemical potential are turned on, the kinematics is more complicated. Take one-point functions in $d=3$ as an example. In this case, there is only one chemical potential, which can be understood as inserting a rotation around the $x_1$ axis in the trace. This insertion identifies a preferred direction on the 2-sphere and now the azimuth angle $\theta$ becomes meaningful. The correlator is still covariant under rotations around $x_1$ but can now depend freely on the angle $\theta$. We discuss this case in detail in Section \ref{sec:One-point-blocks}, after having discussed the relevant Ward identities. 

As it is always the case in CFTs, whenever a correlator is not fixed by symmetry arguments, it admits a decomposition in conformal blocks. This is the case also for one-point functions on ${S^1_\beta \times S^{d-1}}$ and the purpose of this work is to compute them.

\paragraph{Limit to one-point functions on $S^1_\beta\times \mathbb{R}^{d-1}$} As we have mentioned above, it is possible to regard correlation functions on $S^1\times \mathbb{R}^{d-1}$ as high-temperature limits of correlators on $S^1\times S^{d-1}$,
\begin{equation}\label{flat-space-limit}
    \langle\Phi_1\dots\Phi_n\rangle_{S^1_\beta \times \mathbb{R}^{d-1}} =  \lim_{R\to\infty} \langle\Phi_1\dots\Phi_n\rangle_{S^1_\beta\times S_R^{d-1}}\ .
\end{equation}
Moreover, conformal invariance implies that the large radius limit $R\to\infty$ can be traded for that of high temperature $\beta\to0$. Applied to one-point functions, this implies that the function $f_\Phi$ in \eqref{1-pt-no-chemical} satisfies $f_\Phi(0)=1$.\footnote{However, let us point out that in reaching this conclusion, one assumes that no issues arise in the process taking the $R\to\infty$ limit. Below in equation \eqref{leading-divergence-1pt}, we will exhibit an explicit example of a one-point function for which the preceding analysis fails.} Finally, let us note that, unlike $S^1_\beta \times \mathbb{R}^{d-1}$, the geometry we consider in this work geometry is conformally equivalent to $\mathbb{R}^d$ and hence decompositions of correlation functions involve the usual flat space CFT data. Thus the preceding comments imply that one-point coefficients $b_\Phi$ are also fully determined by the flat space data, if in a complicated way. We shall come back and discuss this matter in more details in a later section.

\subsection{Ward identities for thermal correlation functions}

The thermal correlation functions \eqref{thermal-correlators} we introduced above satisfy a set of Ward identities. 
Note that the insertion of the $q$- and $y_a$-dependent factors in our thermal correlators breaks conformal symmetry 
down to the abelian subgroup $T$ that is generated by the $D$ along with the Cartan generators $H_a$ of the rotation 
group
\begin{equation}\label{eq:Torus}
    T = \{q^D y_2^{H_2}\dots y_r^{H_r} \ |\ q,y_a\in \mathbb{C}\} \subset SO(d+1,1)_{\mathbb{C}}\ .
\end{equation}
Hence, our thermal correlators are expected to satisfy Ward identities with respect to this restricted set of 
generators only. Let $X = D, H_a$, be one of these generators. Then we can write
\begin{align}\label{commuting-1}
    & \text{tr} \left(X \phi_1\dots\phi_n q^D y_2^{H_2}\dots y_r^{H_r}\right) = 
    \text{tr} \left([X,\phi_1]\dots\phi_n q^D y_2^{H_2}\dots y_r^{H_r}\right) + \dots\\[2mm]
    & + \text{tr} \left(\phi_1\dots[X,\phi_n] q^D y_2^{H_2}\dots y_r^{H_r}\right) + \text{tr} \left(\phi_1\dots\phi_n X q^D y_2^{H_2}\dots y_r^{H_r}\right)\ .\nonumber
\end{align}
For brevity, we have omitted the subscript $\mathcal{H}$ from traces over the entire Hilbert space of the theory. 
In the last term, $X$ commutes with everything to the right of it inside the trace. Therefore, it can be moved to 
the rightmost position. By cyclicity, this equals the term on the left-hand side. Therefore, we obtain
\begin{equation} \label{Wardid} 
    \left(X^{(1)} + \dots + X^{(n)}\right) G_n(x_i,q,y_a) = 0\,,
\end{equation}
where $X^{(i)}$ are the usual differential operators that describe the action of the conformal generator $X$ on 
the field $\phi_i$. Note that these operators do not act on $q$ and $y_a$, hence the partition function is 
transparent to them. The thermal correlator $G_n$ may be considered as a function on the abelian group $T$. 
These functions map points on the torus to `vectors' in a tensor product of the field representations $\pi_i$ 
of the conformal group. The Ward identities \eqref{Wardid} state that thermal correlators $G_n$ take values 
in the  ‘zero-weight' subspace, i.e. the space of invariants under the Cartan subalgebra, 
\begin{equation}\label{solutions-Ward-space}
    G_n : T \to (\pi_1 \otimes\dots\otimes \pi_n)^T\ .
\end{equation}
We shall recover the same space \eqref{solutions-Ward-space} as that of certain spherical functions 
in equation~\eqref{space-reduced-spherical-functions} below.

It is actually useful to think of the thermal correlators as partially gauge fixed twisted correlators of 
the form 
\begin{equation}\label{twisted-correlators}
   G_n(x_i,k) = \text{tr}_\mathcal{H} \left(\phi_1(x_1)\dots \phi_n(x_n) k \right)\,,
\end{equation}
where $k \in SO(d+1,1)$ is an arbitrary element of the conformal group. The set of these twisted correlators 
satisfies Ward identities with respect to all generators $X$ of the conformal group, not just the generators 
$X=D, H_a,$ we considered in the case of thermal correlators. The Ward identities read 
\begin{equation}
    \left(X^{(1)} + \dots + X^{(n)} + X^\textrm{ad}\right) G_n(x_i,k) = 0\,,
\end{equation}
Here, $X^{(i)}$ denote the action of the conformal generator $X$ on the field $\phi_i$ and $X^\textrm{ad}$ 
are first order differential operators that describe the adjoint action of $X$ on the conformal group, i.e. 
$X^\textrm{ad}$ are just sums of left- and right-invariant vector fields that act on $k$. Obviously, one can 
use these Ward identities to rotate arbitrary group elements $k$ into the torus $T$ \eqref{eq:Torus}, i.e. 
assume a gauge in which the twisted correlator is given by the thermal one. This choice of a gauge breaks 
conformal symmetry down to the abelian algebra that is generated by $X = D, H_a$. In this sense, the Ward 
identities \eqref{Wardid} are just remnants that are left from the conformal symmetry of twisted correlators. 
\medskip 

\subsection{One-point functions and cross ratios}
\label{sec:One-point-blocks}

For most of this work, we focus on one-point functions of a single spinning field, $\phi$, and 
furthermore put $d=3$. This means that the field $\phi$ can at most be a symmetric traceless tensor. As 
usual we can trade the tensor indices of $\phi$ for some dependence on a polarisation vector $z_\mu$. The 
entries $z_\mu$ are considered complex and assumed to satisfy the lightcone condition $z^\mu z_\mu=0$. 
With the help of such a polarisation, we can map a primary field $\phi$ of spin $\ell_\phi$ to a function 
in $z_\mu$ that is homogeneous of order $\ell_\phi$ under rescalings of $z_\mu$. In higher dimensions, 
spinning fields can also transform as more general mixed symmetry tensors. Their description requires 
introducing further polarisations. It is not difficult to study such extensions that are relevant for 
$d > 3$, but since there is nothing fundamentally new for higher $d$, we have decided to stick to $d=3$. 
Let us also note that in three dimensions, the conformal group is of rank two, %$r=2$,
and we can write $H_1 = i D$, $H_2 = M_{23}$. The chemical potential associated with $H_2$ will be denoted 
by $y$ and it corresponds to a rotation around the $x_1$ direction.
\smallskip

Let us determine the form of the correlation function $\langle\phi\rangle_{q,y}$ as allowed by Ward 
identities. According to equation \eqref{solutions-Ward-space}, we are to find the zero-weight subspace 
of the field representation $\pi_1 = \pi$. The latter is described by the following set of differential 
operators 
\begin{align}
   & P_\mu = -\partial_{x^\mu}\,,\label{principal-series-1}\\[2mm]
   & M_{\mu\nu} = x_\mu \partial_{x^\nu} - x_\nu \partial_{x^\mu} + z_\mu \partial_{z^\nu} - z_\nu \partial_{z^\mu}\,,\\[2mm]
   & D = -x^\mu\partial_{x^\mu} - \Delta_\phi,\\[2mm]
   & K_\mu = -x^2 \partial_\mu + 2 x_\mu (x^\nu\partial_\nu + \Delta_\phi) + 2x^\nu (z_\mu \partial_{z^\nu} - z_\nu \partial_{z^\mu})\label{principal-series-4}\ .
\end{align}
These operators act on functions $\phi(x^\mu,z^\nu)$ which are homogeneous polynomials of degree $\ell_\phi$ 
in the variables $z_\mu$. To impose irreducibility we also restrict to the lightcone $z^\mu z_\mu = 0$ in the 
familiar way. 

From the explicit formulas for the differential operators $D, M_{23}$ it is easy to deduce that the zero-weight 
subspace is spanned by functions of the form
\begin{equation}\label{H-invariants-in-principal}
    F = (ir)^{-\Delta_\phi} Z^{\ell_\phi}\, f \left(\frac{X}{r},-\frac{iW}{X Z},\frac{z_1}{Z}\right) \equiv 
    (ir)^{-\Delta_\phi} Z^{\ell_\phi}\, f_{q,y} \left(p,z,\frac{z_1}{Z}\right)\,,
\end{equation}
where $r^2=x_\mu x^\mu$ and we have introduced the following two cross ratios 
\begin{equation} 
 p=\frac{X}r, \qquad z = \frac{-iW}{X Z}\,, 
\end{equation} 
with
\begin{equation*}
    X = \sqrt{x_2^2 + x_3^2}, \qquad Z = \sqrt{z_2^2 + z_3^2}, \qquad W = x_2 z_2 + x_3 z_3\ .
\end{equation*}
The variables $q,y$ here are just spectators. Finally, we should restrict to the locus $z^\mu z_\mu=0$ that enforces the lightcone condition and sets $z_1/Z=$constant, thus making the function $f_{q,y}$ a function of the two cross ratios $p,z$. All in all, we arrive at the condition
\begin{align} \label{eq:one-point-function}
    \langle \phi(x^\mu, z^\nu)  \rangle_{q,y} = (i r)^{-\Delta_\phi} Z^{\ell_\phi} f_{q,y}(p,z)\ .
\end{align}
We can get some better understanding of the cross ratios we introduced by passing to polar coordinates
\begin{align}
    & x_1 = r \cos{\theta}, &&x_2 = r \sin{\theta}\cos\varphi, && x_3=r \sin{\theta}\sin\varphi\,, \nonumber\\
    & z_1 = \pm i Z, &&z_2 = Z \cos\psi, && z_3= Z \sin\psi \ .
\end{align}
When expressed in terms of these coordinates, the two cross ratios $p,z$ read 
\begin{align}
    p = \sin\theta,\qquad i z = \cos(\varphi-\psi)\ .
\end{align}
As anticipated in Section \ref{sec:kinematics}, when the chemical potential is turned on, 
the angle with the polar direction $x_1$ is relevant, while for spinning operators, the 
rotation invariance around $x_1$ forces the dependence on $\varphi-\psi$.

In the following sections, we shall show how to write the Casimir operator as a differential 
operator in variables $(q,y,p,z)$. If $\phi$ is a scalar field, the variable $z$ is not present. 
The remaining variable $p$ is related to $s$ used in \cite{Gobeil:2018fzy} by $s=p^2$.
\smallskip 

\paragraph{Remark} If we set the fugacity $y$ to $y=1$, i.e. we turn off the angular potential, the one-point function satisfies additional Ward identities that determine the function $f_{q,y}$ in equation \eqref{eq:one-point-function}. In general, Ward identities express invariance with respect to all those generators of the conformal group $SO(d+1,1)$ that commute with the dilaton generator $D$ as well as those Cartan generators $H_a$ that appear with non-trivial fugacity $y_a \neq 1$. If only the dilation generator is present, the invariance must hold for all generators of the subgroup $SO(1,1)\times SO(d)$, i.e.\ it consists of dilations and all rotations. So, for the one-point function in $d=3$, we have four instead of just two Ward identities. Hence, the one-point function is now fully determined up to an overall constant prefactor $c$ to take the form $c\, r^{-\Delta_\phi - \ell_\phi} \left(x_\mu z^\mu\right)^{\ell_\phi}$. Alternatively, we can express this consequence of the extended Ward identities by saying that the function $f$ that appears on the one-point function takes the form 
\begin{equation}\label{extended-Ward-invariance}
    f_q(p,z) = f_{q,y=1}(q,z)= \left(pz + \sqrt{1-p^2}\right)^{\ell_\phi}\ .
\end{equation}
In deriving this results we have inserted the formula $x_\mu z^\mu = i Z r \left(pz + \sqrt{1-p^2}\right)$ to express the $SO(d)$ invariant $x_\mu z^\mu$ in terms of the variables $Z,r$ and the cross ratios we introduced earlier,

\subsection{Casimir equations for conformal blocks}

Thermal conformal blocks are defined by inserting a projector to an $\mathfrak{so}(d+1,1)$ irreducible representation inside the trace
\begin{equation}\label{thermal-block-definition}
    G_n\big|_\mathcal{O} = \text{tr} \left(P_\mathcal{O}\phi_1(x_1)\dots \phi_n(x_n) q^D y_2^{H_2}
    \dots y_r^{H_r}\right) \equiv \text{tr}_{\mathcal{H}_\mathcal{O}} \left(\phi_1(x_1)\dots \phi_n(x_n) 
    q^D y_2^{H_2}\dots y_r^{H_r}\right)\ .
\end{equation}
In particular, zero-point blocks are simply the characters of $SO(d+1,1)$. As the projectors $P_\mathcal{O}$ commute with conformal transformations, the argument from above shows that blocks satisfy the same Ward identities as the correlation functions. 
\smallskip

Further, Dolan-Osborn-like, equations for thermal blocks are derived by insertions of generators of the conformal group $SO(d+1,1)$ that do not belong to the Cartan subalgebra. At this moment it pays off that we may think of the thermal correlators as partially gauge-fixed twisted correlators. In fact, since twisted correlators satisfy the full set of conformal Ward identities, they certainly also satisfy Casimir equations that can be derived in the standard way. What we are about to see below are the incarnation of these Casimir equations for the thermal correlators. 

To be more concrete, let us first consider insertions of a generator  $X = e_\alpha$ for a non-vanishing root vector $\alpha = (\alpha_1,\dots,\alpha_r)$ of $\mathfrak{so}(d+1,1)$, see Appendix \ref{A:Conventions for Lie algebras}. It is convenient to write $D=i H_1$,  $q=e^{i \mu_1}$ and $y_a = e^{\mu_a}$. The manipulation \eqref{commuting-1} from above remains valid. However, now the last term does not cancel between two sides, but instead
\begin{align}
    \text{tr} \left(e_\alpha \phi_1\dots\phi_n q^D y_2^{H_2}\dots y_r^{H_r}\right) & = \left(e_\alpha^{(1)}+\dots +e_\alpha^{(n)}\right) G_n(x_i,q,y_a)\\
    & + e^{-\alpha\cdot\mu} \text{tr} \left(\phi_1\dots\phi_n q^D y_2^{H_2}\dots y_r^{H_r} e_\alpha\right)\ ,\nonumber
\end{align}
where we have applied the Baker-Campbell-Hausdorff formula. Using cyclicity of the trace,
\begin{equation}
     \text{tr} \left(e_\alpha \phi_1\dots\phi_n q^D y_2^{H_2}\dots y_r^{H_r}\right) =\frac{1}{1 - e^{-\alpha\cdot\mu}} \left(e_\alpha^{(1)}+ \dots +e_\alpha^{(n)}\right) G_n(x_i,q,y_a)\ .
\end{equation}
Again, the same equation is satisfied if we replace the trace over $\mathcal{H}$ by that of a single irreducible representation $\mathcal{O}$ on the left and $G_n$ by $G_\mathcal{O}$ on the right. The equation is not immediately useful, as there is no second independent way to express the quantity on the left hand side. However, by inserting a Casimir operator, we do get a differential equation. To ease notation, let us write $e^{(1\dots n)}$ for the sum of generators over insertion points,
\begin{equation}
    e^{(1\dots n)}_\alpha = e^{(1)}_\alpha + \dots + e^{(n)}_\alpha\ .
\end{equation}
Then we have
\begin{align*}
    & \text{tr} \left(e_{-\alpha}e_\alpha \phi_1\dots\phi_n q^D y_2^{H_2}\dots y_r^{H_r}\right) = e^{(1\dots n)}_\alpha \text{tr} \left(e_{-\alpha} \phi_1\dots\phi_n q^D y_2^{H_2}\dots y_r^{H_r}\right)\\
    & + e^{-\alpha\cdot\mu} \text{tr} \left(e_\alpha e_{-\alpha}\phi_1\dots\phi_n q^D y_2^{H_2}\dots y_r^{H_r}\right) = e^{(1\dots n)}_\alpha \frac{1}{1-e^{\alpha\cdot\mu}} e^{(1\dots n)}_{-\alpha} G_n\\
    & + e^{-\alpha\cdot\mu} \text{tr} \left((h_\alpha + e_{-\alpha} e_\alpha) \phi_1\dots\phi_n q^D y_2^{H_2}\dots y_r^{H_r}\right)\ .
\end{align*}
We now use the fact that inserting a Cartan generator $H_i$ is equivalent to deriving with respect to $\mu_i$,
\begin{align*}
    & \left(1 - e^{-\alpha\cdot\mu}\right)\text{tr} \left(e_{-\alpha}e_\alpha \phi_1\dots\phi_n q^D y_2^{H_2}\dots y_r^{H_r}\right) = \frac{e_\alpha^{(1\dots n)} e_{-\alpha}^{(1\dots n)}}{1-e^{\alpha\cdot\mu}} G_n + e^{-\alpha\cdot\mu} \alpha\cdot\partial_\mu G_n\ .
\end{align*}
Therefore, we obtain the Casimir equation
\begin{align}
    C_2(\mathcal{O}) G_n\big|_\mathcal{O} & = \text{tr}_\mathcal{O} \left(\left(\sum_{i=1}^r H_i^2 + \sum_{\alpha\in R_+} \{e_{-\alpha},e_\alpha\}\right) \phi_1\dots\phi_n q^D y_2^{H_2}\dots y_r^{H_r}\right)\nonumber\\
    &  = \left( \sum_{i=1}^r\partial_{\mu_i}^2 + \sum_{\alpha\in R_+}\left(\frac{e^{\alpha\cdot\mu}+1}{e^{\alpha\cdot\mu}-1}\alpha\cdot\partial_\mu + \frac{\{e_\alpha^{(1\dots n)}, e_{-\alpha}^{(1\dots n)}\}}{(1-e^{\alpha\cdot\mu})(1-e^{-\alpha\cdot\mu})}\right)\right) G_n\big|_\mathcal{O}\label{tCFT-Casimir}\\
    & \qquad\qquad\quad = \left(\sum_{i=1}^r\partial_{\mu_i}^2 + \sum_{\alpha\in R_+}\left(\coth\frac{\alpha\cdot\mu}{2} \alpha\cdot\partial_\mu - \frac{\{e_\alpha^{(1\dots n)}, e_{-\alpha}^{(1\dots n)}\}}{4\sinh^2\frac{\alpha\cdot\mu}{2}}\right)\right) G_n\big|_\mathcal{O}\nonumber\ .
\end{align}
On the left-hand side, $C_2(\mathcal{O})$ is the Casimir eigenvalue in the representation $\mathcal{O}$, obtained by acting with the Casimir on the projector $P_\mathcal{O}$. On the right, we have a differential operator in variables $(x,\mu_i)$ in which the $x$-dependence only sits in the numerator of the last term. We will show below that the operator \eqref{tCFT-Casimir} precisely matches (twice) the Laplacian acting on a certain space of {\it spherical functions}, equation~\eqref{one-sided-Laplacian}. Together with the identification of spaces of solutions to Ward identities \eqref{solutions-Ward-space} and spherical functions \eqref{space-of-spherical-functions}, this will allow for a complete embedding of the theory of thermal blocks into harmonic analysis.

\subsection{Expansion of one-point functions and tensor structures}
\label{SSS:Blocks and tensor structures}

Since this work is mostly concerned with thermal one-point functions, we would like to conclude this section with a 
brief discussion of thermal one-point blocks. By expanding the definition of the trace, we may write
\begin{small}
\begin{align}
    &\text{tr}_{\mathcal{O}} \left(\phi^A(x) q^D y^{M_{23}}\right) = \sum_{m=-\ell_{\mathcal{O}}}^{\ell_{\mathcal{O}}} \langle m|\phi^A(x) q^D y^{M_{23}} |m\rangle + \text{desc} = q^{\Delta_{\mathcal{O}}} \sum_{m=-\ell_{\mathcal{O}}}^{\ell_{\mathcal{O}}} y^m \langle m|\phi^A(x) |m\rangle + \text{desc}\nonumber\\
    & = q^{\Delta_{\mathcal{O}}} \sum_{m,m'=-\ell_{\mathcal{O}}}^{\ell_{\mathcal{O}}} y^m \lim_{x'\to\infty} \left(\langle\mathcal{O}^m(x')\mathcal{O}^{m'}(0)\rangle \langle \mathcal{O}_{m'}(x')\phi^A(x)\mathcal{O}_m(0)\rangle \right) + \text{desc}\ . \label{limit-3pt-fn}
\end{align}
\end{small}

Here, $A=\mu_1\dots\mu_{\ell_\phi}$ is the spin index of the external field. In the sum, we have separated the contribution of states from the leading level in the multiplet of $\mathcal{O}$. These have conformal weight  $\Delta_{\mathcal{O}}$ and are labelled by the eigenvalue $m = -\ell_{\mathcal{O}},\dots,\ell_{\mathcal{O}}$ of $M_{23}$. The remainder of the sum is over descendants. 

Let us denote a set of independent $\langle\mathcal{O}\phi\mathcal{O}\rangle$ three-point structures by $\{\mathbb{T}_3^a\}$. Clearly, each of these tensor structures gives rise to a different conformal block. Therefore, the blocks are appropriately labelled as $g^{\Delta_\phi,\ell_\phi,a}_{\Delta_{\mathcal{O}},\ell_{\mathcal{O}}}$. As functions of $q$, they expand as $q^{\Delta_{\mathcal{O}}}$ multiplied by a powers series. The choice of the tensor structure is encoded in the leading term in the series. Putting
\begin{equation}
    \tau^a_{mm'} (x^\mu,z^\mu) = \lim_{x'\to\infty} \left( \langle\mathcal{O}^m(x')\mathcal{O}^{m'}(0)\rangle(\mathbb{T}_3^a)_{m'm}^{\mu_1\dots\mu_{\ell_\phi}}(x',x,0)\right)\,,
\end{equation}
we have
\begin{equation}\label{blocks-definition}
     G^{\Delta_\phi,\ell_\phi,a}_{\Delta_{\mathcal{O}},\ell_{\mathcal{O}}}(q,y,x^\mu,z^\nu) = (ir)^{\Delta_\phi} Z^{-\ell_\phi} \left( q^{\Delta_{\mathcal{O}}} \sum_{m,m'} y^m \tau^a_{mm'}(x^\mu,z^\mu) + \text{desc} \right)\ .
\end{equation}
Thermal conformal blocks $g^{\Delta_\phi,\ell_\phi,a}_{\Delta_{\mathcal{O}},\ell_{\mathcal{O}}}$ are defined by stripping off from $G^{\Delta_\phi,\ell_\phi,a}_{\Delta_{\mathcal{O}},\ell_{\mathcal{O}}}$ the prefactor multiplying the bracket on the right hand side. In summary, the block decomposition of the one-point function reads
\begin{equation}
    \mathcal{Z}(q,y)\langle\phi(x^\mu,z^\mu)\rangle_{q,y} = (-1)^\frac{\Delta_\phi}{2} r^{-\Delta_\phi} \left(z_2^2 + z_3^2\right)^{\frac{\ell_\phi}{2}} \sum_{\mathcal{O},a} \lambda_{\phi\phi\mathcal{O}}^a\, g_{\Delta_\mathcal{O},\ell_\mathcal{O}}^{\Delta_\phi,\ell_\phi,a}(q,y,p,z)\ .
\end{equation}
Notice that for an exchanged scalar field, the index $a$ becomes redundant and there is a single conformal block. Using the shadow formalism, one can write an integral representation for thermal partial waves. This is briefly discussed in Appendix \ref{A:Shadow formalism}. We shall compute the blocks \eqref{blocks-definition} in a later section. For now, we satisfy ourselves with a few remarks. For scalar external and scalar exchange, there is a single block. The Casimir equation in this case was derived and solved in \cite{Gobeil:2018fzy}. The solution has an expansion of the form
\begin{equation}\label{scalar-int-ext-solution}
    g^{\Delta_\phi}_{\Delta_\mathcal{O}}(q,y,s) = q^{\Delta_\mathcal{O}}\sum_{k\geq m \geq n\geq 0} A_{kmn} q^k u^m s^n\,, \qquad u = y + y^{-1} - 2\ .
\end{equation}
The inequality $k\geq m$ is easily understood - it follows from the fact that on the $n$-th descendant level of a scalar conformal multiplet, the maximum spin of a state is $n$. If the external field is a scalar, but the internal allowed to be spinning, the three-point function $\langle\mathcal{O}\mathcal{O}\phi\rangle$ has $(2\ell_{\mathcal{O}}+1)$ tensor structures. Imposing parity leaves $(\ell_{\mathcal{O}}+1)$ structures. In the general case of spinning external and exchange, the number of three-point structures is the number of solutions to the equation
\begin{equation}
    n_1 + n_2 + n_3 = 0, \qquad -\ell_\phi \leq n \leq \ell_\phi, \quad -\ell_{\mathcal{O}}\leq n_2,n_3 \leq \ell_{\mathcal{O}}\ .
\end{equation}
In Section \ref{S:Thermal blocks in 3d}, we shall reproduce the partial waves of \cite{Gobeil:2018fzy} and solve the more difficult cases just described. In the same section, we also derive blocks when either $\phi$ or $\mathcal{O}$ satisfy conservation conditions.

\section{Thermal Blocks from Harmonic Analysis}
\label{S:Harmonic analysis}

In this section we want to embed thermal conformal blocks, as introduced in the previous section, into their proper mathematical home, which is the theory of spherical functions. The transition from the physics perspective we assumed above into the framework of harmonic analysis is made in the first subsection. Through a gentle reformulation of the previous discussion we shall extract that the spherical functions relevant for thermal blocks are those associated with the so-called diagonal Gelfand pairs $(K \times K, K_{\text{diag}})$, with $K$ the conformal group. Once the transition is made, the remainder of the section is then formulated in terms of harmonic analysis. In the second subsection, we define the relevant spherical functions, show that they may be equivalently regarded as conjugation-covariant functions on $K$, and exhibit examples of such functions as traces of $K$-intertwiners. In the third subsection, we review how spherical functions are represented as vector-valued functions of rank$(K)$ variables and the action of covariant differential operators in this representation. The latter rests on Harish-Chandra's theory of radial parts. The fourth subsection is concerned with the action of the Laplacian on spherical functions, presented in equations~\eqref{radial-Laplacian} and \eqref{one-sided-Laplacian}. Thereby we will be able to establish the equivalence with the action \eqref{tCFT-Casimir} of the Casimir operator on thermal correlation functions. The subsection also describes the construction of a weight-shifting operator \eqref{form-of-ws-operators}. The reader interested only in the solution theory of the Casimir \eqref{tCFT-Casimir} (equivalently equation~\eqref{one-sided-Laplacian}) can skip directly to the next section.

\subsection{Conformal blocks and harmonic analysis}
\def\rmT{\textrm{T}}
 
In order to leverage the powerful mathematical tools of harmonic analysis for the construction of thermal conformal blocks it is useful to reflect a bit on how this embedding into harmonic analysis works. Blocks are conformal invariants in some space of functions that depend e.g. on insertion points $x_i$ of local fields, their polarisations $z_i$ in case of fields with spin, etc. Our first goal is to give a group-theoretic construction of these geometric objects and to describe certain bundles thereon that realise the field representations $\pi_i$. Then we will paste these geometric spaces together in order to build the associated conformally invariant cross ratios. The choice of external fields determines bundles over the spaces of cross ratios. Conformal blocks, finally, provide some basis of sections in these bundles. This basis depends on the choice of Casimir operators. By definition, conformal blocks are eigenfunctions of some (maximal) set of Casimir operators. In more mathematical terms this means that blocks are spherical functions and hence there is a very substantial amount of literature available, going back to \cite{HarishChandra,Gelfand-spherical,Godement1952TheoryOS}.

\paragraph{Configuration spaces and field representations.}
Let us first see how all relevant geometric parameters appear in group theory. The simplest example is that of scalar local fields, which only depend on the insertion point $x$ of the field in the $d$-dimensional Euclidean space. We can represent this configuration space as the quotient 
\begin{equation} \label{eq:Cfd}
\mathcal{X}_d := K /P_0   \quad \text{where}  \quad P_0 = \left(SO(d) \times SO(1,1) \right) \ltimes N_d\,,
\end{equation} 
is the stabiliser of the point $x=0 \in \mathbb{R}^d$ with respect to the usual action of the conformal group $K = SO(d+1,1)$ on $d$-dimensional Euclidean space. Here, $N_d$ denotes the $d$-dimensional subgroup generated by special conformal transformations. Since $P_0$ contains all rotations, the dilations and special conformal transformations, the quotient \eqref{eq:Cfd} is $d$-dimensional and its points are in one-to-one correspondence with translations. Hence, the quotient indeed realises the space of insertion points $x$ of a local field. 

On this space we can now easily realise the field representation $\pi = \pi_\Delta$ that comes with the scalar field $\phi$ of weight $\Delta = \Delta_\phi$. The carrier space $V_\pi$ of this representations is given by, \cite{Dobrev:1977qv},
\begin{equation} \label{eq:piscalar}
V_\pi \cong \Gamma(\Delta) := \{\, f: K \rightarrow \mathbb{C} \, | \, f (k p^{-1}) = \chi_{\Delta}(p) f(k)\ , 
\, \forall\ k\in K,\, p \in P_0\,  \} \ .
\end{equation}
Here, $\chi_\Delta$ denotes the unique one-dimensional representation of the parabolic subgroup $P_0$ that satisfies $\chi_\Delta(D) = \Delta$ for the generator $D$ of dilations, while being trivial on rotations as well as special conformal transformations. One can think of elements in this space as sections of a line bundle on the quotient space $K/P_0$. Since only the right action of the subgroup $P_0$ on the conformal group $K$ is used to characterise the functions $f$, the space $\Gamma(\Delta)$ carries an action $\pi$ of the conformal group $K$ that is obtained from left multiplication. It is not difficult to verify that the generators of conformal transformations indeed act through the usual first order differential operators on the $d$-dimensional Euclidean space $K/P_0$. 

Instead of working with functions of real variables, it is often convenient to complexify coordinates and restrict to holomorphic functions thereof. From now on we shall assume that all groups and their quotients are complexified and that we are dealing with holomorphic functions $f$ even if this is not shown explicitly in our notation.
\smallskip 

Let us now extend the discussion of configuration spaces to fields with spin. As in the previous section we shall focus on symmetric traceless tensors (STTs) of spin $\ell$. The construction is similar to the one in equation~\eqref{eq:Cfd}, except that we need to replace the (complexified) rotation group $R = SO(d)$ in the definition of $P_0$ by an appropriate subgroup, since now rotations no longer leave a field that is inserted at $x=0$ invariant. The space of light-like rays $z^\mu$ we introduced above in order to describe 
the polarisations of spinning fields in STT representations is given by the quotient 
\begin{equation} 
\mathcal{Z}_d := SO(d)/R_1 \quad \text{where} \quad  R_1 = (SO(d-2) \times SO(2))\ltimes 
\mathbb{C}^{d-2} \subset SO(d)\,,
\end{equation} 
is the stabiliser group of a point $z_\ast^\mu$ in the space of polarisations under the action of the rotation group. This is analogous to the construction of the configuration space $\mathcal{X}_d$ above, except that that the dimension is reduced, i.e.\ $d \rightarrow d-2$, and we are now in (complexified) Euclidean space. From the definition of $R_1$ we can easily infer that the dimension of $\mathcal{Z}_d$ is given by $\textrm{dim} \mathcal{Z}_d = d-2$. This indeed agrees with the dimension of the space of rays that obey $z^\mu z_\mu =0$.  

Given this group-theoretic construction of the space $\mathcal{Z}_d$ of polarisations, the full configuration space of a spinning fields with STT spin is therefore given by 
\begin{equation}  \label{eq:CFd}
\mathcal{X}_d^\textrm{STT} := K / P_1  \quad \text{where} \quad P_1 = \left(R_1 \times SO(1,1) \right) 
\ltimes N_d \ . 
\end{equation} 
As we discussed, the stabiliser subgroup $P_1$ of a spinning field is smaller than the parabolic subgroup 
$P_0$ that appeared in equation~\eqref{eq:Cfd}. Therefore, the quotient $\mathcal{X}_d^\textrm{STT}$ is 
larger than the configuration space $\mathcal{X}_d$ for scalar fields and the additional parameters are 
used to describe polarisations $z^\mu$. The dimension of the quotient $\mathcal{X}_d^\textrm{STT}$ we 
defined in equation~\eqref{eq:CFd} is given by $\text{dim} \mathcal{X}_d^\text{STT} = 2d-2$, in perfect 
agreement with the number of parameters that are used to specify a spinning field $\phi(x^\mu,z^\nu)$. 

Once again we can also use the coset description \eqref{eq:CFd} in order to construct the associated 
field representation $\pi$. The spinning field $\phi$ is characterised by its conformal weight $\Delta 
= \Delta_\phi$ and the STT spin $\ell = \ell_\phi$. These quantum numbers define a representation 
$\chi_{\Delta,\ell}$ of the generator $D$ of dilations and the Cartan element $M_{23}$ that generates the 
subgroup $SO(2)$ we used in defining $R_1 \subset SO(d)$. This representation can be extended trivially 
to a one-dimensional representation $\chi_{\Delta,\ell}$ of the subgroup $P_1$ in the denominator of 
the coset space \eqref{eq:CFd}. Generalising equation~\eqref{eq:piscalar} we can construct the carrier space 
for the field representation $\pi = \pi_{\Delta,\ell}$ as
\begin{equation} \label{eq:GDl}
V_\pi \cong \Gamma(\Delta,\ell) := \{\, f: K \rightarrow \mathbb{C} \, | \, f (k p^{-1}) = 
\chi_{\Delta,\ell}(p) f(k)\ , \, \forall\ k \in K,\, p \in P_1 \, \} \ .
\end{equation} 
As before, the construction only uses right multiplication with elements $p \in P_1 \subset K$. Hence, 
the space carries a representation of the conformal group $K$ that is obtained from left multiplication. 
It is not difficult to see that the infinitesimal conformal transformations are realised through the 
first order differential operators listed in equations~\eqref{principal-series-1}-\eqref{principal-series-4}. 
\smallskip 

There are many extensions of this construction. In particular, it is possible to describe spinning fields in more general spin representations, see e.g.\  \cite{Buric:2021ttm} for a detailed description. Let us finally also mention that the parameter space of a conformal defect of dimension $p$ is given by 
\begin{equation} \label{eq:CDdp}
\mathcal{X}^D_{d,p} := K   / (SO(p+1,1)  \times SO(d-p) )\ .
\end{equation} 
For a defect of co-dimension $d-p=1$, for example, the quotient has dimension $(d+1)$. The coordinates correspond 
to the mid-point coordinates and the radius of a $(d-1)$-dimensional sphere in the $d$-dimensional ambient space. 
We shall not need these extensions in the rest of this work and hence will not give any further detail here. 

\paragraph{Conformal invariants and cross ratios.} Within the group-theoretic setup described, we 
now want to build conformal invariants and in particular conformally invariant cross ratios for 
correlation functions of local and non-local operators. Before we address the thermal correlations 
functions let us briefly recall how this works at zero temperature. To be specific we consider a 
correlation function of $n$ spinning fields with weights $\Delta_i$ and spins $\ell_i$.\footnote{We 
allow for some of these fields to be scalar, i.e. for some of the fields to have spin $\ell_i =0$.} 
Then the space of conformal invariants is given by the following left quotient 
\begin{equation}\label{eq:CR0} 
\text{CR}^{(0)}_d(n) :=  K \backslash \mathcal{X}^{\textrm{STT},n}_d  = K \backslash K^{(0)}_n / P^{(0)}_{n} \,,
\end{equation} 
with
$$  K^{(0)}_{n} = K^n \quad , \quad P^{(0)}_n = P_0^{n_0} \times P_1^{n-n_0}\,,
\  $$
where $n_0$ denotes the number of scalar field insertions. We have placed a superscript $(0)$ in order 
to distinguish the spaces we introduced here from the ones that appear in the context of thermal 
correlators with $T \neq 0$, see below. The left quotient is defined with 
the help of the left action of $K$ on the factors of the product of configuration spaces in the numerator. 
This action can have fixed points and hence the associated space of cross ratios may be singular. Observe 
that the numerator of the double coset formulation on the right contains one copy of the conformal group 
$K$ for every field insertion, regardless of whether it carries spin or not. The denominator, on the other 
hand, is sensitive to the number of spinning fields that we insert. It is not difficult to see that 
the dimension of $\text{CR}^{(0)}_d(n)$ coincides with the number of independent cross ratios of the 
correlation function, provided we include those cross ratios that involve polarisations of spinning 
fields in the count. All this extends straightforwardly to correlation functions involving non-local 
defect insertions. In particular, each insertion of a non-local operator gives rise to an additional 
factor $K$ in the numerator of the coset. The denominator also contains additional factors, but these 
depend on the dimension $p$ of the defect. 
\smallskip 

So far we have just discussed the spaces on which the correlation function live. As we mentioned above, correlation functions are sections of holomorphic line bundles CR$^{(0)}_d(n)$. The line bundles depend on the choice of operator insertions. We have already described how to assign one-dimensional representations $\chi_\Delta$ and  $\chi_{\Delta, \ell}$ of the subgroups$P_0$ and $P_1$ to each scalar and spinning field, respectively. Now we can combine these representations to define a one-dimensional representation of the denominator group $P^{(0)}_n$ and put 
\begin{equation} \label{def:Gamma0}
\Gamma^{(0)}(\Delta_i,\ell_i) = \{ f: K_n^{(0)} \rightarrow \mathbb{C}\, | \, 
f(gp^{-1}) = \prod_i \chi_{\Delta_i,\ell_i}(p_i) f(g)  \ \forall \ p \in P^{(0)}_n \, \} \ .
\end{equation} 
The elements $f$ of this space can be considered as section of a holomorphic line bundle on the 
space \eqref{eq:CR0} of cross ratios. The space carries an action of the $d$-dimensional conformal 
group by left multiplication. An $n$-point correlation function is 
an element of $\Gamma^{(0)}$ that satisfies conformal Ward identities, i.e, that is invariant with 
respect to the action of the conformal group by left multiplication. We shall denote this space of 
conformal invariant sections by 
\begin{equation} \label{def:G0K}
\Gamma^{(0)}(\Delta_i,\ell_i)^K = \{ f \in \Gamma^{(0)}(\Delta_i,\ell_i) \, | \, f(kg) = f (g) \ 
\forall\ k \in K \, \} \ . 
\end{equation} 
All this discussion was about the familiar correlation functions in the vacuum of the conformal 
field theory. The space \eqref{def:G0K} contains all solutions to the conformal Ward identities 
that are associated with $n$ local field insertions of weight $\Delta_i$ and STT spin $\ell_i$. 
\smallskip 

Let us now turn to thermal correlators. It turns out that the most natural 
extension of the constructions we sketched above targets the twisted correlation functions
we introduced in equation~\eqref{twisted-correlators}. In these correlations we inserted a factor 
$k \in K$ before taking the trace. We now include this copy of $K$ into the configurations 
space before we pass to conformal invariants, i.e. define the space of \textit{thermal 
cross ratios} 
\begin{equation} \label{eq:CRT} 
\text{CR}^{\rmT}_d(n) := K \backslash \left(\mathcal{X}^{\textrm{STT},n}_d \times K\right) = K \backslash K^{\rmT}_n / P^{\rmT}_n \,,
\end{equation} 
where 
$$  K^{\rmT}_n = K^{n+2} \quad , \quad P^{\rmT}_n = P^{(0)}_n  \times K \ . $$
In the first definition of this space we have just included an additional factor $K$ into the 
configuration space and assumed that the action of the conformal group $K$ on this last factor 
is the adjoint one. It is possible to trade the adjoint action of $K$ on itself for the left 
action on the quotient $K \times K /K$. This is the way we think about the space of cross 
ratios on the right hand side of the previous construction. With this in mind, the numerator 
$K^\rmT_n$ now contains two additional copies of $K$ and the numerator group one additional 
factor $K$. There is a second way to think about the space of cross ratios in which we 
gauge fix all but the maximal torus $T \subset K$, 
\begin{equation} \label{eq:CRT-gauge-fixed}
\text{CR}^{\rmT}_d(n) \cong T \backslash \left(\mathcal{X}^{\textrm{STT},n}_d \times T\right) \ .
\end{equation} 
This space obviously supports the thermal correlators we introduced above. Indeed, the factor $T$ in the numerator is what we need when we insert $q^D y^H \in T $ and the quotient by $T$ captures the Ward identities that are satisfied by thermal correlators.

Using the same notations as in the construction \eqref{def:Gamma0} of the spaces $\Gamma^{(0)}$
we can now define the following thermal analogue 
\begin{equation} \label{def:GammaT}
\Gamma^{\rmT}(\Delta_i,\ell_i) = \{ f: K_n^{\rmT} \rightarrow \mathbb{C}\, | \, 
f(gp^{-1}k^{-1}) = \prod_i \chi_{\Delta_i,\ell_i}(p_i) f(g)  \ \forall \ p \in P^{(0)}_n, 
k \in K \, \}\ .
\end{equation} 
We stress that here we chose the trivial representation for the last factor $K$ in the denominator subgroup $P^\rmT_n$. Once again, this space may be considered as a space of sections of a line bundle over the space \eqref{eq:CRT} of thermal cross ratios. It carries an action of the conformal group $K$ by left multiplication so that we can define 
\begin{equation} \label{def:GTK}
\Gamma^{\rmT}(\Delta_i,\ell_i)^K = \{ f \in \Gamma^{\rmT}(\Delta_i,\ell_i) \, | \, f(kg) = f (g) \ 
\forall\ k \in K \, \} \ . 
\end{equation} 
This space consists of all solutions to the conformal Ward identities that can describe thermal 
correlators with $n$ local field insertions of weight $\Delta_i$ and spin $\ell_i$. 
\smallskip 
 
It is useful to discuss a few examples. Let us begin with the case in which we do not have any insertions of either local or non-local fields, i.e. $n=0$. In this case we have 
\begin{equation} \label{eq:CRT0} 
\text{CR}^\rmT_d(0) = K \backslash K^\rmT_0 =  K_\text{ad} \backslash K = T\,,
\end{equation} 
where $T$ is the torus that is generated by $D$ and the Cartan generators of the rotation group. This is of course 
consistent with what we discussed above and corresponds to the $r$ parameters $q$ and $y_a$ we introduced in  equation~\eqref{thermal-correlators}. Let us now look at our main example, namely the case of a thermal one-point function 
for a spinning field of STT spin $\ell$, for which the space of cross ratios becomes
$$ \text{CR}^\rmT_d(1) = K  \backslash \left(\mathcal{X}^\textrm{STT}_d \times K \right)
\cong \frac{\mathcal{X}_d^\textrm{STT}}{T}  \times T \ . $$
The dimension of this space is the same as the dimension of the configuration space $\mathcal{X}^\textrm{STT}_d$ which we computed just after equation~\eqref{eq:CFd}. For $d=3$, this space of thermal cross ratios is four-dimensional and it can be parametrised by the four variables $q,y,p,z$, see equation~\eqref{H-invariants-in-principal} for precise definitions. If we specialise to scalar fields, we loose one of these crass-ratios, namely the cross ratio $z$ and hence remain with 
three cross ratios. More generally, the dimension of the space \eqref{eq:CRT} of thermal cross ratios is the same as the dimensional of the configurations space of all its insertions. This statement remains true even if we insert non-local (line-, surface- etc.) operators. 
\smallskip 

The example of thermal one-point functions for a spinning field with weight $\Delta_\phi$ and spin $\ell_\phi$ is our main focus in this work. Before we conclude this subsection we want to briefly discuss the space \eqref{def:GTK} of invariants. For a spinning thermal one-point function, definition \eqref{def:GTK} can be written as
\begin{equation} \label{eq:defGammaDl} 
\Gamma^\textrm{T}(\Delta_\phi,\ell_\phi)^K = \{ f: K^3 \rightarrow 
\mathbb{C} \, | \, f(g_1 p^{-1},k_l g_2 k_r^{-1},k_l g_3 k_r^{-1}) = 
\chi_{(\Delta_\phi,\ell_\phi)} (p) f(k_l^{-1} g_1,g_2, g_3) \}\ .
\end{equation} 
Note that we have slightly rewritten the invariance condition with respect to the diagonal left 
action of $K$ so that it allows us to exchange the action on the second and third argument for 
an action on the first. By construction, this space contains all the solutions to the Ward 
identities for thermal one-point functions. 

In order to prepare for the analysis in the next subsection we would like to slightly rewrite 
the previous construction of conformal invariants. To this end we recall that the character 
$\chi_{\Delta_\phi,\ell_\phi}$ of the subgroup $P_1 \subset K$ actually induces an infinite 
dimensional field representation $\rho = \rho_{\Delta_\phi,\ell_\phi}$ of the $d$-dimensional 
conformal group $K$ on the space \eqref{eq:GDl} on which $K$ acts by left multiplication, 
\begin{equation} \label{eq:rho} 
\rho(k) h(g)  := h\left(k^{-1} g\right) \quad \textrm{ for all} \quad h \in \Gamma(\Delta_\phi,\ell_\phi)\ .  
\end{equation} 
Now we can think of the complex valued functions $f(g_1,g_2,g_3)$ that feature
in the definition \eqref{eq:defGammaDl} as $\Gamma(\Delta_\phi,\ell_\phi)$ valued functions 
$F$ in two arguments $(g=g_2,g'=g_3)$, i.e. 
\begin{equation} 
\Gamma^\textrm{T}(\Delta_\phi,\ell_\phi) \cong \{ F: K  \times 
K  \rightarrow \Gamma(\Delta_\phi,\ell_\phi)\, | \, 
F(k_l g k_r^{-1},k_l g' k_r^{-1}) = \rho(k_l) F(g,g') \ \forall \ k_{l,r} \in 
K \} \ . 
\end{equation} 
This means that we have re-expressed the space of conformally invariant spinning one-point 
functions as some `vector-valued' functions on the space $\textrm{CR}^\textrm{T}(0)$ that 
is parametrised by $q,y$. The relation between the original construction in eq.\ 
\eqref{eq:defGammaDl} in terms of complex valued functions on $\textrm{CR}^\textrm{T}(1)$ 
can be understood through a Taylor expansion of conformal invariants in the variables $p,z$. 
The infinite set of Taylor coefficients forms the `vector-valued' function $F$ of the cross
ratios $q,y$. 

\paragraph{Casimir equations and spherical functions.}
Thermal conformal blocks form a basis in the space \eqref{def:GTK} of conformal invariants. This basis consists of eigenfunctions of some maximal set of Casimir operators. To construct the associated Casimir equations we recall that the numerator group in the construction of the space \eqref{eq:CRT} of thermal cross ratios consists of $n+2$ copies of the conformal group. Let $J$ denote any set $J = \{ 0 < i_1 < i_2 < \cdots < i_s < n+2\}$ of $s$ ordered integers between $0$ and $n+2$. Here, we are free to choose $s = |J|$ such that $1\leq s \leq n+1$. Such a set defines an action of the conformal group $K$ on the numerator group $K^\textrm{T}_n$ where $k \in K$ acts diagonally on $s$ copies of $K$ and trivially on the other $(n+2-s)$ components. Given this action, we can also construct 
the Casimir differential operators $C^{(p)}_J$ of order $p$ on functions in $n+2$ variables $k_i \in K$. These Casimir operators commute with action of the denominator groups $K$ and $P_n^\textrm{T}$ and hence provide a well defined set of differential operators on the space \eqref{def:GTK} of conformal invariants. Moreover, Casimir operators $C^{(p_1)}_{J_1}$ 
and $C^{(p_2)}_{J_2}$ that are associated with sets $J_1 \subset J_2$ commute with each other and hence can be diagonalised simultaneously. While these Casimir operators do not suffice in general to fully characterise conformal blocks, they do provide a very important set of differential equations.\footnote{In this work, we do not characterise all quantum numbers that label conformal blocks by differential operators, but use a single differential operator, the quadratic Casimir, together with an appropriate ansatz for eigenfunctions.} The Casimir equation for thermal one-point functions of spinning fields we constructed in the previous section is a special example in which we use the action of the conformal group $K$ on the second factor $K$ of the numerator group $K^3$. Now that this equation is fully embedded into the context of harmonic analysis, we can study its eigenfunctions within with the powerful tools from the theory of spherical functions.

\subsection{Spherical functions for diagonal Gelfand pairs}

For the remainder of this section, let $K$ be a simple Lie group and $G$ the product of two copies of $K$, 
$G = K \times K$. We denote by $K_d$ the diagonal subgroup of $G$, consisting of elements of the form $(k,k)$
with $k \in K$. The Lie algebra of $K_d$ is the fixed point set of the automorphism of $\mathfrak{g}$ that exchanges the two copies of $\mathfrak{k}$, making $(G,K_d)$ a Gelfand pair, \cite{Kirillov}. We shall refer to Gelfand pairs of this form as {\it diagonal}. Let $\rho$ and $\sigma$ be two representations of $K$ with carrier spaces $W_l$ and $W_r$, respectively.\footnote{In order to connect to the discussion at the end of the previous subsection one may choose 
$\sigma$ to be the trivial representation on $W_r \cong \mathbb{C}$ and $\rho$ to be the representation we defined
in equation~\eqref{eq:rho}.} The space of {\it spherical functions}, \cite{Gelfand-spherical,HarishChandra,Godement1952TheoryOS}, associated with the quadruple $(G,K_d,\rho,\sigma)$ is that of vector-valued functions on $G$, which are left-right covariant with respect to $K_d$,
\begin{equation}\label{space-of-spherical-functions}
    \Gamma_{\rho,\sigma} = \{ f:G\to \text{Hom}(W_r,W_l)\ |\ f(k_l g k_r) = \rho(k_l) f(g) \sigma(k_r)\}\ .
\end{equation}
If we write the group elements appearing in equation~\eqref{space-of-spherical-functions} in two-component notation,
\begin{equation}
    k_l = (k_l,k_l), \quad g = (k_1,k_2), \quad k_r = (k_r,k_r)\,,
\end{equation}
the spherical condition reads
\begin{equation}\label{spherical-condition}
    f\left(k_l k_1 k_r, k_l k_2 k_r\right) = \rho(k_l)\, f(k_1,k_2)\, \sigma(k_r)\,,
\end{equation}
the form we shall mostly use below. For the diagonal Gelfand pair $(K\times K,K_d)$, it is possible to regard spherical functions as conjugation-covariant vector-valued functions on the group $K$. Indeed, given a spherical function $f\in\Gamma_{\rho,\sigma}$, consider the function 
\begin{equation}\label{F-and-f}
    F : K \to \text{Hom}(W_r,W_l), \qquad F(k) = f(k,e)\ .
\end{equation}
The function $f$ is uniquely specified by $F$, since it can be recovered from the latter as
\begin{equation}\label{recover-f-from-F}
    f(k_1,k_2) = \rho(k_2) f(k_2^{-1}k_1,e) = \rho(k_1) F(k_2^{-1}k_1)\ .
\end{equation}
Furthermore, an $F$ defined through equation~\eqref{F-and-f} is not arbitrary, but has to be conjugation-covariant
\begin{equation}\label{conjugation-covariant-functions}
    F(k^{-1} k' k) = f(k^{-1} k' k,e) = \rho(k^{-1}) f(k',e) \sigma(k) = \rho(k^{-1}) F(k') \sigma(k)\ .
\end{equation}
Conversely, if $F$ satisfies the covariance property \eqref{conjugation-covariant-functions} and $f$ is defined by the right hand side of equation~\eqref{recover-f-from-F}, similar manipulations show that $f$ satisfies the spherical condition \eqref{spherical-condition}. It is therefore equivalent to work either with spherical functions $f$ or conjugation-covariant functions $F$. We shall use this observation to treat conjugation-covariant functions obeying equation~\eqref{conjugation-covariant-functions}, that naturally arise as thermal correlation functions in conformal field theory, using recently developed methods of the theory of spherical functions, \cite{Buric:2022ucg,Buric:2023ykg}.
\smallskip

The simplest examples of spherical functions are characters of $K$. Indeed, when $\rho$ and  $\sigma$ are trivial, $\rho,\sigma = 1$, equation \eqref{conjugation-covariant-functions} states that $F$ is invariant under conjugation. Irreducible characters of $K$ provide a basis for such functions. More involved examples when not both $\rho$ and $\sigma$ are trivial, are provided by traces of intertwining operators. Let us consider the case of so-called associated spherical functions, $\rho = 1$ ({\it vector characters} in the terminology of \cite{Etingof:1994kd}). For them, the function $F$ satisfies
\begin{equation}
    F(k^{-1} k' k) = F(k') \sigma(k)\ .
\end{equation}
Let $\pi$ be a representation of $K$ with the carrier space $V$ and $\Phi: V \to V\otimes W_r^\ast$ an intertwiner. Then the function
\begin{equation}\label{trace-of-intertwiner}
    F(k) = \text{tr}|_V (\Phi k)\,,
\end{equation}
satisfies the covariance property \eqref{conjugation-covariant-functions}. Indeed, let $\{e_\mu\}$ and $\{e_\alpha\}$ be bases for $V$ and $W$, respectively. With indices, equation~\eqref{trace-of-intertwiner} is written as
\begin{equation}
    F_\alpha(k) = \Phi_\mu{}^\nu{}_\alpha \pi^\mu{}_\nu(k)\ .
\end{equation}
On the other hand, the intertwining property of $\Phi$ can be written as
\begin{equation}
    \Phi_\nu{}^\rho{}_\alpha \pi^\nu{}_\mu(k)= \pi^\rho{}_\nu(k)\ \sigma^\ast(k)_\alpha{}^\beta\ \Phi_\mu{}^\nu{}_\beta\ .
\end{equation}
Using these two equations, the transformation of $F$ under conjugation is found
\begin{align*}
    F_\alpha (k^{-1} k' k) & = \Phi_\mu{}^\nu{}_\alpha\ \pi^\mu{}_\nu(k^{-1} k' k) = \Phi_\mu{}^\nu{}_\alpha\ \pi^\mu{}_\rho(k^{-1}) \pi^\rho{}_\sigma(k') \pi^\sigma{}_\nu(k)\\
    & = \pi^\nu{}_\alpha(k^{-1})\ \sigma^\ast(k)_\alpha{}^\beta\ \Phi_\rho{}^\lambda{}_\beta\ \pi^\rho{}_\sigma(k')\ \pi^\sigma{}_\nu(k) = \delta^\sigma_\lambda\ \sigma^\ast(k)_\alpha{}^\beta\ \Phi_\rho{}^\lambda{}_\beta\ \pi^\rho{}_\sigma(k')\\
    & = \sigma^\ast(k)_\alpha{}^\beta\ \Phi_\rho{}^\sigma{}_\beta\ \pi^\rho{}_\sigma(k') = \sigma^\ast(k)_\alpha{}^\beta\ F_\beta(k')\ .
\end{align*}
Thus, $F$ is conjugation-covariant and the corresponding function $f$ is spherical. Let us observe the similarity between equation~\eqref{trace-of-intertwiner} and thermal correlation functions studied in the previous section. In view of it, it is natural to expect that thermal conformal blocks arise as appropriate harmonic spherical functions, which is indeed shown below.

\subsection{Radial parts of differential operators}

Due to their covariance properties, spherical functions may be though of as depending on a smaller number of variables than $\text{dim}(G)$. This is manifested in the Cartan decomposition of $G$,
\begin{equation}\label{KAK-decomposition}
    G = K_d A_p K_d\,,
\end{equation}
by which we mean that almost any\footnote{all elements up to a set of measure zero} element $g\in G$ can be factorised in the form $g = k_l a k_r$, where $A_p$ is a particular abelian subgroup of $G$ that shall be specified below. In order to explain equation~\eqref{KAK-decomposition} and its consequences, we introduce some notation.
\smallskip

Bases for the two copies of $\mathfrak{k}$ in $\mathfrak{g} = \mathfrak{k} \oplus \mathfrak{k}$ shall be denoted by $\{X^{(1)}_\mu\}$ and $\{X^{(2)}_\mu\}$. The Cartan decomposition of the Lie algebra $\mathfrak{g}$ is the direct sum decomposition
\begin{equation}
    \mathfrak{g} = \mathfrak{k}_d \oplus \mathfrak{p}\,,
\end{equation}
into $\mathfrak{k}_d$ and its orthogonal complement in $\mathfrak{g}$ with respect to the Killing form. Bases for the two summands are the sums and differences of generators of the two copies of $K$,
\begin{equation}
    \mathfrak{k}_d = \text{span}\{X^{(1)}_\mu + X^{(2)}_\mu\}, \qquad \mathfrak{p} = \text{span}\{X^{(1)}_\mu - X^{(2)}_\mu\}\ .
\end{equation}
Let $\mathfrak{h}$ be a Cartan subalgebra of $\mathfrak{k}$, with a basis $\{H_i\}$. Then, $A_p$ in equation~\eqref{KAK-decomposition} is defined as the abelian group whose Lie algebra is spanned by
\begin{equation}
    \mathfrak{a}_p = \text{span}\{H^{(1)}_i - H^{(2)}_i\}\ .
\end{equation}
Recall that for $\rho=\sigma=1$, the spherical functions may be thought of as linear combinations of characters of $K$. It is well-known that characters may be regarded as functions on the Cartan subgroup of $K$, i.e. functions of $r=\text{rank}(K)$ variables, consistent with the above.
\smallskip

When $\rho$ or $\sigma$ are non-trivial, important constraints on restrictions of spherical functions to $A_p$ come from the fact that the Cartan factorisation \eqref{KAK-decomposition} is far from unique. Indeed, this non-uniqueness is seen to be measured by the centraliser of $A_p$ in $K_d$, a group that we shall denote by $M$. In the case at hand, $M$ is nothing but the Cartan subgroup of $K_d$ whose Lie algebra is spanned by
\begin{equation}
    \mathfrak m = \text{span}\{H^{(1)}_i + H^{(2)}_i\}\ .
\end{equation}
As a simple consistency check, one can verify the relation
\begin{equation}
    \text{dim}(G) = \text{dim}(K) + \text{dim}(A_p) + \text{dim}(K) - \text{dim}(M)\ .
\end{equation}
The existence of the centraliser $M$ implies that restrictions of spherical functions to $A_p$ take values in the space of $M$-invariants in $\text{Hom}(W_r,W_l)$,
\begin{equation}
    f(a) = f(m a m^{-1}) = \rho(m)\, f(a)\, \sigma(m^{-1})\ .
\end{equation}
In other words, $\mathcal{F} = f|_{A_p}$ is a function
\begin{equation}\label{space-reduced-spherical-functions}
    \mathcal{F} : A_p \to (W_l\otimes W_r^\ast)^{[0]}\,,
\end{equation}
into the zero-weight subspace of $W_l\otimes W_r^\ast$. Notice that, if we put $K = SO(d+1,1)$ and $\rho\otimes\sigma^\ast = \pi_1 \otimes \dots \pi_n$, the space \eqref{space-reduced-spherical-functions} is precisely that of solutions to thermal $n$-point Ward identities, equation~\eqref{solutions-Ward-space}.
\smallskip

\paragraph{Radial decomposition}
We end this subsection by explaining how differential operators acting on spherical functions $f\in\Gamma_{\rho,\sigma}$ are reduced to differential operators with matrix coefficients that act on the restrictions $\mathcal{F}$. This is achieved using the radial decomposition of the universal enveloping algebra $U(\mathfrak{g})$, the infinitesimal analogue of the Cartan decomposition of the group, \cite{Warner2}. The radial decomposition reads
\begin{equation}\label{radial-decomposition-U(g)}
    U(\mathfrak{g}) \cong U(\mathfrak{a}_p) \otimes U(\mathfrak{k}) \otimes_{U(\mathfrak{m})} U(\mathfrak{k})\ .
\end{equation}
To realise \eqref{radial-decomposition-U(g)}, we we fix an element $h\in A_p$ and for any $X\in\mathfrak{g}$ write $X' = h^{-1} X h$. Let $u\in U(\mathfrak{k})$ - it is a polynomial in $X^{(1)}_\mu$ and $X^{(2)}_\mu$. By radially decomposing $u$, we mean writing it in term of a new set of variables
\begin{equation}
    X^+_\mu = X^{(1)}_\mu + X^{(2)}_\mu, \quad X'^+ = h^{-1} \left(X^{(1)}_\mu + X^{(2)}_\mu\right)h, \quad H^{(1)}_i - H^{(2)}_i\,,
\end{equation}
while imposing the ordering that in every monomial terms involving $X'^+_\mu$ appear in the leftmost positions and those involving $X^+_\mu$ in the rightmost. For details, we refer the reader to \cite{Warner2,10.1215/S0012-7094-82-04943-2,Stokman:2020bjj}. By assembling radial decompositions that we described for different elements $h\in A_p$, one arrives at Harish-Chandra's radial component map, \cite{HarishChandra},
\begin{equation}\label{radial-component-map}
    \Pi : U(\mathfrak{g}) \to \text{Fun}(A_p) \otimes U(\mathfrak{a}_p) \otimes U(\mathfrak{k}) \otimes_{U(\mathfrak{m})} U(\mathfrak{k}) \cong \mathcal{D}(A_p) \otimes U(\mathfrak{k}) \otimes_{U(\mathfrak{m})} U(\mathfrak{k})\ .
\end{equation}
Here, $\text{Fun}(A_p)$ is some space of complex valued functions on $A_p$, precise definition of which we will not need to be concerned about. In the last step, we have identified generators $\mathfrak{a}_p$ with partial derivatives with respect to coordinates on $A_p$, which allowed to regard $\text{Fun}(A_p)\otimes U(\mathfrak{a}_p)$ as the space of differential operators on $A_p$. Therefore, the space on the right-hand side of equation~\eqref{radial-component-map} is that of differential operators on $A_p$ with coefficients in $U(\mathfrak{k})\otimes_{U(\mathfrak{m})}\otimes U(\mathfrak{k})$.
\smallskip

We are prepared to spell out how differential operators acting on spherical functions are reduced to $A_p$. Let $u$ be some element of $U(\mathfrak{g})$, regarded as an invariant differential operator on $G$, $f\in\Gamma_{\rho,\sigma}$ a spherical function, and $\mathcal{F}$ its restriction to $A_p$. Then, a theorem of Harish-Chandra states that, \cite{HarishChandra,Warner2},
\begin{equation}\label{Harish-Chandra-theorem}
    (u f)|_{A_p} = \left(\left(\rho\otimes\sigma^\ast\right)\circ \Pi(u)\right) (\mathcal{F})\ .
\end{equation}
On the right-hand side, the notation means that coefficients of the differential operator $\Pi(u)$ have been evaluated in the representation $\rho\otimes\sigma^\ast$. The upshot of the construction is that the operator $\Pi(u)$ is universal and dependence on $\rho$ and $\sigma$ comes only at the very end. These remarks will become more evident in examples below.

\subsection{Casimir and weight-shifting operators}

In this subsection, we shall use the theory outlined above to compute the action of the quadratic Casimir on spherical functions, as well as the weight-shifting operator. The resulting operators will be written in terms of the root system of $\mathfrak{k}$. We shall denote the set of roots of $\mathfrak{k}$ by $R$ and its elements by $\alpha,\beta,\dots$. The corresponding root vectors are denoted by $e_\alpha$. Our conventions for Lie algebras and their root decompositions are collected in Appendix \ref{A:Conventions for Lie algebras}. The reader not interested in the derivations may jump straight to equations~\eqref{radial-Laplacian}, \eqref{one-sided-Laplacian} and \eqref{form-of-ws-operators}.
\smallskip

We wish to radially decompose the Laplacian on $G$. Let us denote the quadratic Casimirs on two copies by $C_2^{(1)}$ and $C_2^{(2)}$. The Laplacian is their sum
\begin{equation}\label{quadratic-Casimir}
    \Delta = C_2^{(1)} + C_2^{(2)} = \sum_{j=1}^2 \Big(\sum_{i=1}^r \left(H_i^{(j)}\right)^2 + \sum_{\alpha\in R_+}\{e^{(j)}_\alpha,e^{(j)}_{-\alpha}\} \Big) \ .
\end{equation}
To ease notation, for the remainder of this section we will write $X^{(1)}_\mu + X^{(2)}_\mu = X_\mu$. We set
\begin{equation}
    h = e^{t_i (H^{(1)}_i - H^{(2)}_i)} \equiv e^{t_i Y_i}\ .
\end{equation}
Then, by the Baker-Campbell-Hausdorff formula,
\begin{equation}
    H'^{(j)}_i = H^{(j)}_i, \qquad e'^{(1)}_\alpha = e^{-t_i \alpha_i} e^{(1)}_\alpha, \qquad e'^{(2)}_\alpha = e^{t_i \alpha_i} e^{(2)}_\alpha\ .
\end{equation}
Using these expressions, we radially decompose elements of the Cartan-Weyl bases for $\mathfrak{k}^{(1)}$ and $\mathfrak{k}^{(2)}$,
\begin{align}\label{linear-decomp-1}
    & H^{(1)}_i = \frac{H_i+Y_i}{2}, \qquad e_\alpha^{(1)} = \frac{e_\alpha - e^{-\alpha\cdot t} e'_\alpha}{1-e^{-2\alpha\cdot t}}, \qquad\ e_\alpha^{(2)} = \frac{e_\alpha - e^{\alpha\cdot t} e'_\alpha}{1-e^{2\alpha\cdot t}}\,,\\
    & H^{(2)}_i = \frac{H_i-Y_i}{2}, \qquad e_{-\alpha}^{(2)} = \frac{e_{-\alpha} - e^{-\alpha\cdot t} e'_{-\alpha}}{1-e^{-2\alpha\cdot t}}, \quad e_{-\alpha}^{(1)} = \frac{e_{-\alpha} - e^{\alpha\cdot t} e'_{-\alpha}}{1-e^{2\alpha\cdot t}}\ .\label{linear-decomp-2}
\end{align}
In these equations, it is assumed that $\alpha\in R_+$ is a positive root. As is clear from their form, one does not have to distinguish between positive and negative cases, but we have done so for later convenience. Expressions \eqref{linear-decomp-1}-\eqref{linear-decomp-2} are the starting point for radially decomposing any higher-degree element of $U(\mathfrak{g})$. In the new basis, the Casimir \eqref{quadratic-Casimir} becomes
\begin{equation*}
    C_2 = \frac12 \sum_{i=1}^r  \left(Y_i^2 + H_i^2\right)- \sum_{\alpha\in R_+}\frac{\{e'_\alpha,e'_{-\alpha}\} - \cosh(\alpha\cdot t) \left(\{e'_\alpha,e_{-\alpha}\} + \{e'_{-\alpha}, e_\alpha\}\right) +\{e_\alpha,e_{-\alpha}\} }{2\sinh^2(\alpha\cdot t)}\ . 
\end{equation*}
We still need to impose the ordering in which the primed root vectors appear in the leftmost positions. This is achieved with the help of bracket relations
\begin{equation}
    [e_\alpha^{(i)},e_{-\alpha}^{(i)}] = h_\alpha^{(i)} = \alpha\cdot H^{(i)}\ .
\end{equation}
Putting $h_\alpha = h_\alpha^{(1)} + h_\alpha^{(2)} = \alpha\cdot H$ and $y_\alpha = h_\alpha^{(1)} - h_\alpha^{(2)} = \alpha \cdot Y$, we find
\begin{equation}
    [e_\alpha,e'_{-\alpha}] = \cosh(\alpha\cdot t)\ h_\alpha + \sinh(\alpha\cdot t)\ y_\alpha, \quad [e_{-\alpha},e'_\alpha] = - \cosh(\alpha\cdot t)\ h_\alpha + \sinh(\alpha\cdot t)\ y_\alpha\ .
\end{equation}
Therefore, the Casimir becomes
\begin{align}\label{radial-Casimir}
    C_2 = \frac12 \sum_{i=1}^r  Y_i^2 & + \sum_{\alpha\in R_+} \coth(\alpha\cdot t) \ y_\alpha \\
    & - \sum_{\alpha\in R_+}\frac{\{e'_\alpha,e'_{-\alpha}\} - 2\cosh(\alpha\cdot t) (e'_\alpha e_{-\alpha} + e'_{-\alpha} e_\alpha) +\{e_\alpha,e_{-\alpha}\} }{2\sinh^2(\alpha\cdot t)} + \frac12 \sum_{i=1}^r H_i^2\ .\nonumber
\end{align}
Harish-Chandra's theorem \eqref{Harish-Chandra-theorem} implies that the action of the Laplacian on restrictions of spherical functions, $\mathcal{F} = f|_{A_p}$, reads
\begin{align}\label{radial-Laplacian}
    \Delta & = \frac12 \sum_{i=1}^r  \partial_{t_i}^2 + \sum_{\alpha\in R_+} \coth(\alpha\cdot t) \ \alpha_j\partial_{t_j} + \frac12 \sum_{i=1}^r \sigma^\ast\left(H_i^2\right)\,,\\
    & - \sum_{\alpha\in R_+}\frac{\rho\left(\{e_\alpha,e_{-\alpha}\}\right) + 2\cosh(\alpha\cdot t) (\rho(e_\alpha) \sigma^\ast(e_{-\alpha}) + \rho(e_{-\alpha})\sigma^\ast(e_\alpha)) +\sigma^\ast\left(\{e_\alpha,e_{-\alpha}\}\right) }{2\sinh^2(\alpha\cdot t)}\ .\nonumber
\end{align}
An interesting special case of equation~\eqref{radial-Laplacian} and one that will be used in applications below is the one-sided Laplacian, $\sigma = 1$,
\begin{equation}\label{one-sided-Laplacian}
    \Delta  = \frac12 \sum_{i=1}^r  \partial_{t_i}^2 + \sum_{\alpha\in R_+} \coth(\alpha\cdot t) \ \alpha_j\partial_{t_j} - \sum_{\alpha\in R_+}\frac{\rho\left(\{e_\alpha,e_{-\alpha}\}\right)}{2\sinh^2(\alpha\cdot t)}\ .
\end{equation}
As anticipated below equation \eqref{tCFT-Casimir}, the radial Laplacian \eqref{one-sided-Laplacian} coincides with ($1/2$ times) the Casimir operator acting on thermal correlation functions. Indeed, the match between the two is realised by putting $\mu_i = 2t_i$ and letting $\rho$ be the tensor product of conformal field representations, $\rho = \pi_1\otimes\dots\otimes\pi_n$. This shows that thermal conformal blocks may be identified with harmonic spherical functions.
\smallskip

\paragraph{Weight-shifting operators}
According to the general theory, \cite{Buric:2022ucg,Buric:2023ykg}, these operators arise upon decomposing $\mathfrak{g}$ under the adjoint action of $\mathfrak{k}_d$. Each irreducible component, apart from $\mathfrak{k}_d$ itself, gives rise to a pair of weight-shifting operators. They are constructed from left-invariant and right-invariant vector fields corresponding to the chosen Lie algebra elements. In the case at hand, the orthogonal complement $\mathfrak{p}$ of $\mathfrak{k}_d$ in $\mathfrak{g}$ transforms in the adjoint representation of $\mathfrak{k}_d$. Indeed, if we denote the structure constants of $\mathfrak{k}$ by $f_{\mu\nu}{}^\rho$, one gets
\begin{equation}\label{ws-module}
    [X_\mu,X^{(1)}_\nu - X^{(2)}_\nu] = f_{\mu\nu}{}^\rho \left(X^{(1)}_\rho - X^{(2)}_\rho\right)\ .  
\end{equation}
Since the adjoint representation is irreducible, we conclude that the Gelfand pair $(G,K_d)$ admits a single pair of weight-shifting operators. We will construct them in examples below by making use of radial decompositions of elements of $\mathfrak{p}$,
\begin{equation}\label{form-of-ws-operators}
    H^{(1)}_i - H^{(2)}_i = Y_i, \qquad e^{(1)}_\alpha - e^{(2)}_\alpha = \coth(\alpha\cdot t) e_\alpha - \frac{e'_\alpha}{\sinh(\alpha\cdot t)}\ .
\end{equation}

\section{Thermal Blocks in Three-dimensional CFT}
\label{S:Thermal blocks in 3d}

In this section, we compute thermal conformal blocks for three-dimensional CFTs. Therefore, we let $K = SO(4,1)$ be the three-dimensional conformal group. Our conventions for this group and its Lie algebra are collected in Appendix \ref{A:Conventions for the conformal algebra}. In the first subsection we write down the Laplacian and the weight-shifting operator following the discussion of Section \ref{S:Harmonic analysis}. This is followed by solutions for blocks, written as series expansions in radial-like variables introduced in Section \ref{S:CFT at finite temperature}. In the concluding, third, subsection we treat several further aspects of blocks, including conservation of external and exchanged fields, exact expressions at vanishing chemical potential, generation of scalar-exchange blocks via weight-shifting and asymptotic formulas for conformal blocks of large internal scaling dimensions.
\smallskip

In what follows we use the variables introduces in Section \ref{S:CFT at finite temperature}, see equation~\eqref{H-invariants-in-principal}
\begin{equation}\label{change-of-variables}
    (q,y,p,z) \equiv \left(e^{2i t_1}, e^{2 i t_2}, \sqrt{\frac{x_2^2+x_3^2}{x_1^2+x_2^2+x_3^2}}, -i\frac{x_2 z_2+x_3 z_3}{\sqrt{(x_2^2+x_3^2)(z_2^2+z_3^2)}}\right) \ .
\end{equation}
Here $x^\mu$ is the position of the field insertion, $z^\mu$ the corresponding polarisation vector and $t_{1,2}$ are the harmonic analysis variables used in the previous section. For reference, let us also state the relation to coordinates used in \cite{Gobeil:2018fzy}. The authors of \cite{Gobeil:2018fzy} use the $x^3$-axis for the definition of the chemical potential, as opposed to $x^1$ used in the present work. After accounting for this by relabelling axes, variables $u$ and $s$ of \cite{Gobeil:2018fzy} read
\begin{equation}
    u = y + y^{-1} - 2\,, \qquad s = p^2\ .
\end{equation}

\subsection{Laplacian and the weight-shifting operator}

Notation in this subsection follows that of Section \ref{S:Harmonic analysis}. We shall choose the Cartan generators of the conformal algebra $\mathfrak{g} = \mathfrak{so}(4,1)$ as
\begin{equation}
    H_1 = i D\,, \qquad H_2 = M_{23}\ .
\end{equation}
These are required to be unit-normalised, which determines our normalisation of the Killing form. The set of positive roots is
\begin{equation}
    R_+ = \{(i,0),(0,i),(i,i),(i,-i)\}\ .
\end{equation}
It is a simple matter to find the corresponding root vectors. They are written in Appendix \ref{A:Conventions for the conformal algebra}. We shall consider the action of the Laplacian on associated spherical functions with the trivial right-covariance behaviour, $\sigma=1$. On the other hand, $\rho$ will belong to the non-unitary principal series of $\mathfrak{so}(4,1)$, i.e. it is a CFT field representation. The representation is labelled by $(\Delta_\phi,\ell_\phi)$ and realised by differential operators \eqref{principal-series-1}-\eqref{principal-series-4}. The Laplacian is written down using the general theory of the previous section,
\begin{align}
    \Delta_{\Delta_\phi,\ell_\phi} & = -2q^2 \partial_q^2-2y^2\partial_y^2- 2q\partial_q-2y\partial_y-\frac{2y(y+1)}{y-1}\partial_y-\frac{2q(q+1)}{q-1}\partial_q \nonumber\\
    &+2\frac{q+y}{q-y}(y\partial_y-q\partial_q)-2\frac{qy+1}{qy-1}(y\partial_y+q\partial_q) + \frac{q}{(q-1)^2}\mathcal{D}^{(1)}_{\Delta_\phi,\ell_\phi}(p,z) \label{Laplace-Casimir-op}\\
    & + \frac{y}{(y-1)^2}\mathcal{D}^{(2)}_{\Delta_\phi,\ell_\phi}(p,z)+ \frac{qy}{(q-y)^2}\mathcal{D}^{(3)}_{\Delta_\phi,\ell_\phi}(p,z) + \frac{qy}{(qy-1)^2}\mathcal{D}^{(4)}_{\Delta_\phi,\ell_\phi}(p,z) \, . \nonumber
\end{align}
Explicit expressions for differential operator $\mathcal{D}^{(i)}_{\Delta_\phi,\ell_\phi}$ are given in Appendix \ref{A:Differential operators}. We have passed from variables $t_i$, discussed above, to their exponentials $q,y$ according to equation~\eqref{change-of-variables}. The latter will be used to write down solutions below.
\smallskip

Next to the Laplacian, the theory of radial decompositions allows us to construct a single weight-shifting operator corresponding to the adjoint representation of $K$. The Gelfand-Tsetlin labels of the adjoint representation of $\mathfrak{so}(5)$ are $(1,1)$. Therefore, one expects to have a family of operators $\mathfrak{q}_{\Delta_\phi,\ell_\phi}$, labelled by $\Delta_\phi$ and $\ell_\phi$, and satisfying the exchange relations with the Laplacians
\begin{equation}\label{exchange-relations}
     \mathfrak{q}_{\Delta_\phi,\ell_\phi} \Delta_{\Delta_\phi,\ell_\phi} = \Delta_{\Delta_\phi-1,\ell_\phi+1} \mathfrak{q}_{\Delta_\phi,\ell_\phi}\ .
\end{equation}
Indeed, we find such a family, of the form indicated in equation~\eqref{form-of-ws-operators}. They read
\begin{align}\label{weight-shifting-operator}
    \mathfrak{q}_{\Delta_\phi,\ell_\phi} & = 2iq \left(pz+\sqrt{1-p^2}\right) \partial_q -2ipy\sqrt{1+z^2}\partial_y +i\left(\frac{1+q}{1-q}\right) Q^{(1)}_{\Delta_\phi, \ell_\phi}(p,z) \\[2mm]
    & + i\left(\frac{y+1}{y-1}\right)Q^{(2)}_{\Delta_\phi, \ell_\phi}(p,z) + i\left(\frac{qy+1}{qy-1}\right) Q^{(3)}_{\Delta_\phi, \ell_\phi}(p,z) + i\left(\frac{q+y}{q+y}\right)Q^{(4)}_{\Delta_\phi, \ell_\phi}(p,z) \, ,\nonumber
\end{align}
where $Q^{(i)}_{\Delta_\phi,\ell_\phi}$ are first order differential operators in $p$ and $z$, collected in Appendix \ref{A:Differential operators}. Relations \eqref{exchange-relations} mean that the action of $\mathfrak{q}_{\Delta_\phi,\ell_\phi}$ increases the external spin of a conformal block by one.

\subsection{Thermal conformal blocks}

Here we summarise our results for thermal conformal blocks. We look for eigenfunctions of the Laplace-Casimir operator \eqref{Laplace-Casimir-op} in the form~\eqref{block-series-expansion},
\begin{equation}\label{ansatz}
    g^{\Delta_\phi,\ell_\phi,a}_{\Delta_\mathcal{O},\ell_{\mathcal{O}}}(q,y,p,z) =   q^{\Delta_\mathcal{O}}\sum_{n_i,\varepsilon_j} A_{n_1 n_2 n_3}^{\varepsilon_1 \varepsilon_2} q^{n_1} y^{n_2} p^{n_3} z^{n_4} \left(1-p^2\right)^{\frac{\varepsilon_1}{2}} \left(1+z^2\right)^{\frac{\varepsilon_2}{2}}\ .
\end{equation}
Here, $n_i$ are integers with $n_1\geq0$ and
\begin{equation}
    -n_1-\ell_{\mathcal{O}}\leq n_2 \leq n_1 + \ell_{\mathcal{O}}, \quad 0\leq n_3 \leq 2(n_1 + \ell_\phi + \ell_{\mathcal{O}}), \quad 0 \leq n_4 \leq \ell_\phi\ .
\end{equation}
Moreover, $\varepsilon_i\in\{0,1\}$. The range of $n_{1,2}$ can be understood by recalling the definition of a block as a sum over states in the conformal multiplet of $\mathcal{O}$ and inspecting the conformal dimensions and spins that these states have. On the other hand, the origin of $z$ from contracting indices of $\phi$ with a polarisation vector implies the bounds on $n_4$. The Laplace-Casimir equation is solved order by order in $q$. Notice that by the nature of the ansatz, at each order there are only finitely many coefficients $A^{\varepsilon_1\varepsilon_2}_{n_1\dots n_4}$. At order $q^{\Delta_{\mathcal{O}}}$, the number of linearly independent solutions turns out to be exactly the number of three-point structures for $\langle\phi\mathcal{O}\mathcal{O}\rangle$.\footnote{The number of solutions equals the number of three-point structures {\it before} imposing permutation invariance due to the operator $\mathcal{O}$ appearing twice. This over-counting is not an issue, since we in any case relate solutions to their corresponding three-point structures.} Each of these solutions is then extended to an arbitrary order $q^{\Delta_\mathcal{O}+n}$. Our results are given in the Mathematica notebook, see also Appendix \ref{app:More details on the computation} for some explanations.%
\smallskip

Variables $(q,y,p,z)$ are in many respects similar to radial coordinates for four-point conformal blocks, \cite{Hogervorst:2013sma}. This is manifested in the structure of the expansion \eqref{ansatz} - in particular in the range of $n_i$. From the point of view of spherical functions, the analogy is even clearer. Radial coordinates for four-point blocks, like the $(q,y)$ from above, are exponentials of Cartesian coordinates $t_i$ that parametrise the abelian factor $A_p$ of some Gelfand pair. The two cases are distinguished by the choice of the pair. The result is that the two Laplace-Casimir operators have different root systems, $BC_2$ for four-point blocks and $B_2$ in the present case.

\paragraph{Examples} 1) The block for scalar external and scalar exchange operator reads
\begin{equation}\label{scalar-external-scalar-exchange}
    g^{\Delta_\phi,0}_{\Delta_\mathcal{O},0} = q^{\Delta_{\mathcal{O}}} \left(1 + q\left(1 + \frac{1}{y} + y + \frac{(p^2\Delta_\phi-2)(y^2+1)-2y(1+\Delta_\phi(p^2-1))}{4y\Delta_{\mathcal{O}}}\right) + O(q^2)\right)\ .
\end{equation}
2) The block for external scalar and exchanged spin one field reads
\begin{equation}\label{scalar-external-spin1-exchange}
     g^{\Delta_\phi,0,a}_{\Delta_\mathcal{O},1} = q^{\Delta_{\mathcal{O}}} \left(\frac{c_1(1+y+y^2) - c_2(p^2(1-y)^2 +2y) + c_3\sqrt{1-p^2}(1-y^2)}{y} + O(q) \right)\,,
\end{equation}
where $c_{1,2,3}$ are arbitrary constants. Each of the three solutions may be extended to higher orders. E.g. setting $c_i=\delta_{i3}$, we have
\begin{equation}
     g^{2,0,a=3}_{3,1} = q^{\Delta_{\mathcal{O}}} \frac{\sqrt{1-p^2}(1-y^2)}{y}\left( 1 + q \frac{2y(9+5y) + 9p^2(1-y)^2 + 10}{16y}  + O(q^2)\right)\,,
\end{equation}
where we have put $\Delta_\phi=2$ and $\Delta_{\mathcal{O}}=3$ in order to fit the block in a single line. We can explicitly relate the above solutions to three-point functions following the discussion of Section \ref{SSS:Blocks and tensor structures}. For this, is it enough to specify the leading order coefficient in the $q$ expansion corresponding to each tensor structure. We have
\begin{equation}
    H_{12} \mapsto \frac{\pi^2 \left(1+y+y^2\right)}{9y}\,, \quad V_{1,23} V_{2,31} \mapsto -\frac{\pi^2\left(p^2 (1-y)^2 + 2y\right)}{18y}\ .
\end{equation}
Here, $H_{ij}$ and $V_{i,jk}$ are the usual building blocks for three-point functions defined in \cite{Costa:2011mg}. We see that the two structures constructed from them get mapped precisely, up to numerical multiples, to terms multiplying constants $c_1$ and $c_2$ in equation~\eqref{scalar-external-spin1-exchange}.
\smallskip

3) For external spin one and exchanged spin one, there are seven solutions. We present one of them. To first order in $q$
\begin{small}
\begin{align}
    g^{4,1,a}_{3,1} & = pz(y+\frac{1}{y}) +2\sqrt{1-p^2} + q\Big( \frac{pz\left(25p^2(1-y)^2(1-y+y^2)+2y(41-14y+41y^2)\right)}{16y^2}\nonumber\\
    & + \frac{2(2+9y+46y^2+9y^3+2y^4)-5p^2(1-y)^2(2-9y+2y^2)}{16y^2}\sqrt{1-p^2}\\    
    &\hskip4cm + \frac{p(y^2-1)(-2+5p^2(1-y)^2+11y-2y^2)}{8y^2}\sqrt{1+z^2} \Big) + O(q^2)\ .\nonumber
\end{align}
\end{small}

Again, we have have substituted concrete numbers for scaling dimensions to make the expression less bulky. As above, it is possible to relate different solutions to choices of the three-point tensor structure.
\medskip

The determination of the coefficients $A^{\varepsilon_1\varepsilon_2}_{n_1\dots n_4}$ in the expansion \eqref{ansatz} is the main result of this work. With the codes we provide it is possible to compute these coefficients to high orders for spinning external fields and any exchange along the thermal circle. Before we conclude this section and discuss a first application in the context of free field theory we want to discuss a few additional properties and refinements. 

\subsection{Comments, refinements and special cases}

In this final subsection we want to comment on some salient features of the thermal one-point blocks and discuss solutions a few special 
cases. Our first comment concerns the blocks that arise when the exchanged field $\mathcal{O}$ is a scalar. In this case, there is a single $\langle\phi\mathcal{O}\mathcal{O}\rangle$ tensor structure regardless of whether $\phi$ is spinning of not. Therefore, there is a unique 
solution to the Casimir equation of the form \eqref{block-series-expansion}. Consequently, to obtain these blocks, we need not solve the 
Casimir differential equation, but can instead act on blocks for which both $\phi$ and $\mathcal{O}$ are scalars with the weight-shifting 
operator \eqref{weight-shifting-operator}. That is, these blocks take the form 
\begin{equation}
   g_{\Delta_{\mathcal{O}},0}^{\Delta_\phi-\ell_\phi,\ell_\phi} = \mathfrak{q}_{\Delta_\phi-\ell_\phi+1,\ell_\phi - 1} \cdot \cdot \dots 
   \cdot \mathfrak{q}_{\Delta_\phi,0} \cdot g_{\Delta_\mathcal{O},\ell_{\mathcal{O}}}^{\Delta_\phi,0}\ .
\end{equation}
Recall that the scalar blocks that we act upon by our weight shifting operators on the right hand side were studied in \cite{Gobeil:2018fzy}
already. When we set $y=1$, the ansatz \eqref{ansatz} shows that the $p$-dependence of these solutions drops out, since we recover rotation 
invariance. The resulting single-variable function was found in \cite{Gobeil:2018fzy} already and it takes the form 
\begin{equation}\label{y=1-scalar-scalar}
    g^{\Delta_\phi,0}_{\Delta_{\mathcal{O}},0} = q^{\Delta_{\mathcal{O}}}(1-q)^{-2\Delta_{\mathcal{O}}}\ _3F_2\left(\Delta_{\mathcal{O}}-1,\Delta_{\mathcal{O}}-\frac{\Delta_\phi}{2},\Delta_{\mathcal{O}}-\frac{\tilde\Delta_\phi}{2};\Delta_{\mathcal{O}},2\Delta_{\mathcal{O}}-2;\frac{-4q}{(1-q)^2}\right)\ .
\end{equation}
We have denoted the scaling dimension of $\phi$-s shadow by $\tilde\Delta_\phi = 3-\Delta_\phi$. With the tools we developed above we can 
now obtain an analogous exact result for external spin one and scalar exchange,
\begin{align}\label{spin-1-exact}
    g^{\Delta_\phi-1,1}_{\Delta_{\mathcal{O}},0} & = i \left(p z + \sqrt{1-p^2} \right) \Big( (2 - \Delta_\phi) \frac{1+q}{1-q} g^{\Delta_\phi,0}_{\Delta_{\mathcal{O}},0} \\
    & + 2(\Delta_{\mathcal{O}} - 1) \frac{(1+q)q^{\Delta_\mathcal{O}}}{(1-q)^{2\Delta_{\mathcal{O}}+1}}\ _2F_1\left(\Delta_\mathcal{O}-\frac{\Delta_\phi}{2},\Delta_\mathcal{O}-\frac{\tilde\Delta_\phi}{2},2\Delta_{\mathcal{O}}-2;\frac{-4q}{(1-q)^2}\right) \Big)\ . \nonumber
\end{align}
This expression follows from the application of the weight-shifting operator to equation~\eqref{y=1-scalar-scalar}, which, in case under 
consideration, commutes with the $y\to1$ limit.\footnote{This is not true for $\ell_\phi \geq 2$, where the application of weight-shifting 
operators and the zero chemical potential-limit do not commute.}

\subsubsection{Conserved operators} 
\label{subsubsec:Conserved_blocks}
In the construction of blocks cases in which either external or exchanged operators are shortened by conservation laws require some extra care. If the external field $\phi$ is a conserved current, i.e. $\phi(x) = J(x)$, we can act on blocks by the Thomas-Todorov operator $D_\mu$, \cite{Dobrev:1975ru}, to re-instate the indices and then contract the result with a position vector $x^\mu$. The process leads to an operator in variables $(q,y,p,z)$ that can be applied directly to conformal blocks. For instance, for external spin one (which implies $\Delta_\phi=2$), the resulting operator reads
\begin{align}
    \mathcal{D}^{\text{ext}}_{2} = &\frac{z}{2} \Big( (1+z^2)^2\partial_z^3+p(1+z^2)(p\sqrt{1-p^2}+p^2z-z)\partial_p\partial_z^2 + 3p(1+z^2)(pz+\sqrt{1-p^2})\partial_z^2 \nonumber\\
    &+p(3+pz\sqrt{1-p^2} +2z^2-3p^2-2p^2z^2)\partial_p\partial_z+2p(p^2 z-z-2p\sqrt{1-p^2})\partial_p \nonumber\\
    &+ (3+3pz\sqrt{1-p^2}-9p^2-6p^2z^2)\partial_z + 6p(pz-2\sqrt{1-p^2})   \Big)\ .\label{external-conservation}
\end{align}
The operator for general $\ell_\phi$ is given in Appendix \ref{A:Differential operators}. By imposing that the conservation operator annihilates the block \eqref{block-series-expansion}, the number of linearly independent solutions reduces precisely to that of three-point structures $\langle J\mathcal{O}\mathcal{O}\rangle$.
\smallskip

Consider now the case when the exchanged operator is short. To explain how to proceed in such a setup, we focus on the situation in which $\mathcal{O}$ is the free field, so that the shortening condition reads $\partial^2\mathcal{O}=0$. By plugging $\Delta_{\mathcal{O}}=1/2$ in the Casimir equation and looking for solutions of the form \eqref{block-series-expansion}, we see that solution is no longer unique. To make it so, one would like to impose $P^2=0$ on the exchanged operator, since every state $|\psi\rangle$ in the representation of $\mathcal{O}$ satisfies $P^2|\psi\rangle = 0$. This can be achieved  by radially decomposing $P^2\equiv (P^{(1)})^2 + (P^{(2)})^2$ and demanding that the corresponding reduced operator, $\mathcal{D}_{\text{free}}$ annihilates the thermal block. The radial decomposition of $P^2$ is carried out in Appendix \ref{A:Radial decomposition algorithm}, leading to the reduced operator
\begin{small}
\begin{align}\label{internal-conservation-operator}
    \mathcal{D}_{\text{free}}^{\text{int}} & = \frac{2qz^2}{(q-y)(qy-1)} \Big( y(1+z^2)\partial_z^2 + yz\partial_z-\frac{p^2(p^2-1)(p^2q(y-1)^2-y(q^2-1))}{(q-1)^2}\partial_p^2 \nonumber\\
    &+ p\frac{2p^2(y+q(1+\Delta_\phi+y(q+y-4+y\Delta_\phi-2\Delta_\phi)))-y(q-1)^2-p^4q(y-1)^2(2\Delta_\phi+3)}{(q-1)^2}\partial_p \nonumber \\
    &+p^2\Delta_\phi\frac{ (1-\Delta_\phi)(y+yq^2)+q(1+\Delta_\phi-p^2(y-1)^2(\Delta_\phi+2)+y(y+y\Delta_\phi-4))}{(q-1)^2} \Big) \ .
\end{align}
\end{small}
There is a unique solution of the Casimir equation in the form \eqref{block-series-expansion} satisfying this additional condition.

\paragraph{Examples} 4) Above we considered blocks for external and exchanged spins equal to one and found seven linearly independent solutions. By imposing these are annihilated by equation~\eqref{external-conservation}, we get two linear constraints on the coefficients, leaving us with five solutions. An example of a conserved block reads
\begin{small}
\begin{align}
    & g^{2,1,a}_{3,1} = \frac{\left(1+y+y^2\right)\left(pz+\sqrt{1-p^2}\right)}{y} + q\Big(\frac{p(2y^4+3y^3-3y-2)}{16y^2}\sqrt{1+z^2}\\
    &+ \frac{28+70y+92y^2+70y^3+28y^4+3p^2(1-y)^2(2+y+2y^2)}{32y^2}(pz+\sqrt{1-p^2}) \Big) + O(q^2)\nonumber\ .
\end{align}
\end{small}

5) We have written the block for scalar external and scalar exchange up to order $O(q)$ in equation~\eqref{scalar-external-scalar-exchange}. The coefficient of $q^2$ in this solution has a pole at $\Delta_{\mathcal{O}} = 1/2$ and generic $\Delta_\phi$. This pole is not present for $\Delta_\phi=1$ and gives a well-defined (to all orders in $q$) function $g^{\Delta_\phi\to1,0}_{\Delta_{\mathcal{O}}\to1/2,0}$. However, this is not the correct block for an exchange of a free field $\mathcal{O}$. Imposing internal conservation using equation~\eqref{internal-conservation-operator}, one gets instead the linear combination
\begin{equation}\label{free-exchange-block}
    g^{1,0}_{1/2,0} = g^{\Delta_\phi\to1,0}_{\Delta_{\mathcal{O}}\to1/2,0} + \frac13 g^{\Delta_\phi\to1,0}_{\Delta_{\mathcal{O}}\to5/2,0}\ .
\end{equation}
Notice that internal quantum numbers of the second term on the right hand side are that of first the null descendant in the free field representation.

\subsubsection{Blocks at large scaling dimensions}
\label{SS:Blocks at large scaling dimensions}

For analysis of asymptotic OPE coefficients $\lambda^a_{\phi\mathcal{O}\mathcal{O}}$ in the limit when $\Delta_{\mathcal{O}}\to\infty$, we require the blocks in the same limit. These are determined by the structure of the Laplacian \eqref{one-sided-Laplacian}. For a general group $K$, by conjugating the Laplacian with the function
\begin{equation}\label{Weyl-denominator}
    \delta(t_i) = \prod_{\alpha\in\Phi_+} \sinh(\alpha\cdot t) = \sin t_1\, \sin t_2\, \sin(t_1+t_2)\, \sin(t_1-t_2)\,,
\end{equation}
first order derivatives $\partial_{t_i}$ are removed and we are left with a Schr\"odinger operator, \cite{Olshanetsky:1983wh},
\begin{equation}\label{Calogero-Sutherland-Hamiltonian}
    H = \delta \Delta \delta^{-1} = \frac12 \sum_{i=1}^r  \partial_i^2 - \sum_{\alpha\in R_+}\frac{\rho\left(\{e_\alpha,e_{-\alpha}\}\right)}{2\sinh^2(\alpha\cdot t)} + \text{const}\ .
\end{equation}
In the second step in equation~\eqref{Weyl-denominator}, we have written the function $\delta(t_i)$ for the three-dimensional conformal group. We refrain from spelling out the Hamiltonian \eqref{Calogero-Sutherland-Hamiltonian}, which follows directly from equation~\eqref{Laplace-Casimir-op}. Within the quantum-mechanical interpretation, the $\Delta_{\mathcal{O}}\to\infty$ limit is that of large energies. Therefore, we may drop the potential term to get the free wave equation
\begin{equation}
    \left( \partial_{t_1}^2 + \partial_{t_2}^2 \right) \delta^{-1}(t_i)g^{\Delta_\phi,\ell_\phi}_{\Delta_{\mathcal{O}},\ell_{\mathcal{O}}} = -4 \left(\Delta_{\mathcal{O}}(\Delta_{\mathcal{O}}-3) + \ell_{\mathcal{O}}(\ell_{\mathcal{O}}+1)\right) \delta^{-1}(t_i) g^{\Delta_\phi,\ell_\phi}_{\Delta_{\mathcal{O}},\ell_{\mathcal{O}}}\ .
\end{equation}
The appropriate wavefunctions read
\begin{equation}
    g^{\Delta_\phi,\ell_\phi}_{\Delta_{\mathcal{O}},\ell_{\mathcal{O}}} = \delta(t_i)\, e^{2 \sqrt{\Delta_{\mathcal{O}}(\Delta_\mathcal{O}-3)}\, t_1 + 2 \sqrt{\ell_{\mathcal{O}}(\ell_{\mathcal{O}}+1)}\, t_2}\, F(p,z)\,,
\end{equation}
where the function $F(p,z)$ can be determined e.g.~by comparing with the behaviour of solutions at small $q$. In the following we will only use the large-$\Delta_{\mathcal{O}}$ blocks for $t_2 = 0$. On this locus, $F(p,z)$ is fixed by Ward identities, see equation~\eqref{extended-Ward-invariance}. Using further $\Delta_{\mathcal{O}}(\Delta_\mathcal{O}-3)\approx\Delta_\mathcal{O}^2$, we get
\begin{equation}\label{asymptotic-block}
    g^{\Delta_\phi,\ell_\phi,a}_{\Delta_{\mathcal{O}},\ell_{\mathcal{O}}} = \frac{q^{\Delta_{\mathcal{O}}}}{(1-q)^3}\, \left(p z + \sqrt{1-p^2} \right)^{\ell_\phi} + O(\Delta_{\mathcal{O}}^{-1})\ .
\end{equation}
The asymptotic formula \eqref{asymptotic-block} is valid for $y=1$ and $\Delta_{\mathcal{O}}\gg \beta^{-1} \gg \ell_\phi, 1$. The overall normalisation of the right hand side is to be achieved by an appropriate choice of three-point tensor structures.\footnote{We have not imposed this choice in our code that generates conformal blocks.}

\section{Thermal Block Expansions in Free Field Theory}
\label{S:OPE coefficients in free field theory}

As the simplest application of conformal blocks found above, we will show how to obtain OPE coefficients in free bosonic theory by decomposing thermal one-point functions. In the first subsection, we briefly review the theory of a free conformally coupled scalar on the cylinder $\mathbb{R}\times S^2$, working in canonical quantisation. This is followed by a discussion of the partition function at finite chemical potential and its decomposition into conformal characters. In the third subsection, we compute the two-point function of the fundamental field, $\langle\phi(x_1)\phi(x_2)\rangle_{q,y}$,\footnote{In previous sections we called $\phi$ a generic spinning operator, not to be confused with the fundamental field $\phi$ in this section.} at finite temperature and chemical potential. This is then used to extract the one-point functions $\langle\phi^2\rangle_{q,y}$ and $\langle T_{\mu\nu}\rangle_{q,y}$ by means of the operator product expansion. By decomposing these one-point functions in thermal conformal blocks, we read off OPE coefficients of the form $\lambda^a_{\phi^2\mathcal{O}\mathcal{O}}$ and $\lambda^a_{T\mathcal{O}\mathcal{O}}$. In the concluding subsection, we give averaged OPE coefficients $\overline{\lambda_{\phi^2\mathcal{O}\mathcal{O}}^a}$ for fixed spin $\ell_{\mathcal{O}}=0,2,4$ as a function of $\Delta_{\mathcal{O}}$. 

\subsection{Free scalar in three dimensions}

We consider the theory of a scalar field $\Phi$ on the cylinder $\mathbb{R}\times S^2$, specified by the action
\begin{equation}\label{classical-action-free}
    S = \int\sin\theta dt d\theta d\varphi\  \left(\frac12\dot\Phi^2 + \Phi\Delta_{S^2}\Phi - \frac12 M^2\Phi^2\right)\ .
\end{equation}
Following canonical quantisation, the field is expanded in modes
\begin{equation}
    \Phi = \sum_{l,m} \frac{1}{\sqrt{2\omega_l}}\left(a_{l,m} e^{-i\omega_l t} Y_{l,m} + a_{l,m}^\dagger e^{i\omega_l t} Y_{l,m}^\ast\right)\,, \qquad \omega_l = \sqrt{l(l+1) + M^2}\,,
\end{equation}
with $Y_{l,m}$ the usual spherical harmonics. The modes are promoted to operators acting on the Fock space and satisfying the canonical commutation relations
\begin{equation}
    [a_{l,m},a^\dagger_{l',m'}] = \delta_{ll'}\delta_{mm'}, \quad [a,a] = [a^\dagger,a^\dagger] = 0\ .
\end{equation}
We shall be interested in the case of a conformally coupled field. The action \eqref{classical-action-free} is Weyl-invariant for $M^2=1/4$. Substituting this value into the expression for frequencies, we get
\begin{equation}\label{frequencies-free-field}
    \omega_l = l+1/2\ .
\end{equation}
The Hamiltonian is given by
\begin{equation}
    H = \int d\theta d\varphi \sin\theta \left(\frac12 \dot\Phi^2 - \Phi\Delta_{S^2}\Phi + \frac12 M^2 \Phi^2\right) \equiv \sum_{l,m} \omega_l a^\dagger_{l,m} a_{l,m}\ .
\end{equation}
In the second step we used the normal ordering to drop an infinite sum over modes. The spectrum of $H$ is given by sums $\sum\omega_{l_i}$ of frequencies of particles in the given state. Via the cylinder-to-plane map, the Hamiltonian is carried over to the dilation operator on $\mathbb{R}^3$ and energies to scaling dimensions of states in the CFT. We recognise equation~\eqref{frequencies-free-field} as the dimension of the CFT operator obtained from the fundamental field $\phi$ $(\Delta_\phi = 1/2)$ by an application of $l$ derivatives.

\subsection{The partition function}

At zero chemical potential, the partition function of the free CFT reads
\begin{equation}\label{free-partition-function-zero-ch-pt}
    \mathcal{Z}(q) = \prod_{l=0}^\infty \left(\frac{1}{1-q^{l+\frac12}}\right)^{2l+1} = (1 + q^{\omega_0} + q^{2\omega_0} + \dots) (1 + q^{\omega_1} + q^{2\omega_1} + \dots)^3 \dots\ .
\end{equation}
Let us briefly explain this formula. States in the free theory may be labelled by products of words of the form $\partial_{\mu_1}\dots\partial_{\mu_l}\phi$, which have energy $\omega_l$. For a given $l$, there are $(2l+1)$ independent words. A single term of the form $q^{n_0 \omega_0}\dots q^{n_k \omega_k}$ in the partition function \eqref{free-partition-function-zero-ch-pt} counts the contribution of a state of the form
\begin{equation}
    \phi^{n_0}\dots (\partial_{\mu_1}\dots\partial_{\mu_k}\phi)^{n_k}\ .
\end{equation}
Multiplying out all the brackets in equation~\eqref{free-partition-function-zero-ch-pt}, we get contributions from all possible words. At finite chemical potential, a similar counting argument leads to the expression for the partition function 
\begin{equation}\label{free-partition-function}
    \mathcal{Z}(q,y) = \prod_{l = 0}^\infty \prod_{m=-l}^l \frac{1}{1 - q^{l + \frac12} y^m}\ .
\end{equation}

The partition function decomposes in conformal characters. The latter are given by
\begin{equation}\label{characters}
    \chi_{\Delta,l}(q,y) = \frac{q^\Delta\left(1 - y^{2l+1}\right)}{y^l (1-y) (1 - q y) (1-\frac{q}{y}) (1-q)} \sim \frac{(2l+1)q^\Delta}{(1-q)^3}\ .
\end{equation}
In the second step, we have written the leading behaviour of characters as $y\to1$. We shall denote this function as $\chi_{\Delta,l}(q)$. The partition function decomposes as
\begin{equation}\label{character-decomposition}
    \mathcal{Z}(q,y) = 1 + \chi^{\text{free}}(q,y) + \sum_{l=2,4,\dots} \chi_l^{\text{short}}(q,y) + \mathcal{Z}_{\text{long}}(q,y)\ .
\end{equation}
The first term is the contribution of the identity, the second of the field representation of $\phi$ itself,
\begin{equation}
     \chi^{\text{free}}(q,y) = \chi_{1/2,0}(q,y) - \chi_{5/2,0}(q,y)\,,
\end{equation}
and the third of the tower of conserved currents
\begin{equation}
    \chi_l^{\text{short}}(q,y) = \chi_{l+1,l}(q,y) - \chi_{l+2,l-1}(q,y)\ .
\end{equation}
The last piece collects contributions from long multiplets. It can only be spectrally resolved by turning on the chemical potential $y$. To a few lowest orders,
\begin{equation}
    \mathcal{Z}_{\text{long}} = \chi_{1,0} + \chi_{\frac32,0} + \chi_{2,0} + \chi_{\frac52,0} + \chi_{3,0} + \chi_{\frac72,0} + \chi_{\frac72,2} + \dots\ .
\end{equation}
Of course, at higher orders, operators start appearing with multiplicities greater than one.

\subsection{Thermal two-point function}

In this subsection, we shall compute the two-point function of the fundamental field $\Phi$ at finite temperature and chemical potential. As in Section \ref{S:CFT at finite temperature} we shall consider the normalised expectation value on the twist cylinder which we restate here for convenience 
\begin{equation}
    \langle\mathcal{O}\rangle_{S^1_\beta\times S^2_\mu} =  \frac{1}{\mathcal{Z}} \text{tr}_{\mathcal{H}} \left(\mathcal{O} q^H y^{M_{23}} \right) \equiv \frac{1}{\mathcal{Z}} \text{tr}_{\mathcal{H}} \left(\mathcal{O} e^{-\beta H + \mu J}\right)\ .
\end{equation}
It is related to correlation functions on flat space in the usual way. By expanding the field in modes and using linearity of the thermal expectation value, we can write
\begin{align*}
    & \langle\Phi(\tau_1,\Omega_1)\Phi(\tau_2,\Omega_2)\rangle_{S^1_\beta\times S^2_\mu} = \sum_{l,l',m,m'} \frac{1}{\sqrt{(2l+1)(2l'+1)}} \Big(e^{i\omega_{l'}\tau_2 - i\omega_l \tau_1} Y_{l,m}(\Omega_1) \\ & Y^\ast_{l',m'}(\Omega_2) \langle a_{l,m} a^\dagger_{l',m'}\rangle_{S^1_\beta\times S^2_\mu} + e^{i\omega_l \tau_1 - i\omega_{l'} \tau_2} Y^\ast_{l,m}(\Omega_1)  Y_{l',m'}(\Omega_2) \langle a^\dagger_{l,m} a_{l',m'}\rangle_{S^1_\beta\times S^2_\mu} \Big)\ .
\end{align*}
We have used the fact that expectation values of the form $\langle a a\rangle_{\beta,\mu}$ and $\langle a^\dagger a^\dagger\rangle_{\beta,\mu}$ vanish, which follows from both $H$ and $J$ preserving the number of particles in a state. The remaining two expectation values read
\begin{equation}\label{elementrary-traces}
    \langle a_{l,m} a^\dagger_{l',m'}\rangle_{S^1_\beta\times S^2_\mu} = \frac{\delta_{ll'}\delta_{mm'}}{1-e^{-\beta\omega_l+\mu m}}, \qquad \langle a^\dagger_{l,m} a_{l',m'}\rangle_{S^1_\beta\times S^2_\mu} = \frac{e^{-\beta\omega_l+\mu m}}{1-e^{-\beta\omega_l+\mu m}}\delta_{ll'}\delta_{mm'}\ .
\end{equation}
These formulas can be derived by starting from the average of canonical commutation relations, and using the Baker-Campbell-Hausdorff formula
\begin{align*}
    & \delta_{ll'}\delta_{mm'} = \langle[a_{l,m},a^\dagger_{l',m'}]\rangle_{S^1_\beta\times S^2_\mu} = \mathcal{Z}^{-1} \text{tr} \left(a_{l,m} a^\dagger_{l',m'} e^{-\beta H + \mu J} - a^\dagger_{l',m'} a_{l,m} e^{-\beta H + \mu J}\right)\\
    & = \mathcal{Z}^{-1} \text{tr} \left(a_{l,m} a^\dagger_{l',m'} e^{-\beta H +\mu J} - a_{l,m} e^{-\beta H +\mu J} a^\dagger_{l',m'}\right) = (1-e^{-\beta \omega_{l'} + \mu m'}) \langle a_{l,m}a^\dagger_{l',m'}\rangle_{S^1_\beta\times S^2_\mu}\ .
\end{align*}
Equations \eqref{elementrary-traces} immediately follow. Plugging the result back in the two-point function and performing the sum over $l'$ and $m'$, we get
\begin{align*}
    & \langle\Phi(\tau_1,\Omega_1)\Phi(\tau_2,\Omega_2)\rangle_{S^1_\beta\times S^2_\mu} = \sum_{l,m} \frac{1}{2l+1}
    \frac{1}{1-e^{-\beta \omega_l + \mu m}}\\
    & \Big(e^{i\omega_l (\tau_2-\tau_1))} Y_{l,m}(\Omega_1) Y^\ast_{l,m}(\Omega_2) + e^{i\omega_l (\tau_1-\tau_2)} Y^\ast_{l,m}(\Omega_1)  Y_{l,m}(\Omega_2) e^{-\beta\omega_l+\mu m} \Big)\ .
\end{align*}
In applications below, we will be particularly interested in the OPE limit of this expression. We shall set $\tau_1=0$, $\varphi_1=0$ and denote $i \tau_2\equiv \tau$, $\varphi_2\equiv\varphi$. Then
\begin{align*}
    \langle\Phi\Phi\rangle_{S^1_\beta\times S^2_\mu} & = \frac{1}{4\pi}\sum_{l,m} \frac{(l-m)!}{(l+m)!} \frac{P_l^m(\cos\theta_1) P_l^m(\cos\theta_2)}{1-e^{-\beta \omega_l + \mu m}} \Big(e^{\omega_l\tau - im\varphi} + e^{-\omega_l\tau+im\varphi} e^{-\beta\omega_l+\mu m} \Big)\,,
\end{align*}
or, in flat space
\begin{align*}
   & 4\pi\langle\phi(x_1)\phi(x_2)\rangle_{q,y} = \sum_{l,m} \frac{(l-m)!}{(l+m)!} P_l^m(\cos\theta_1) P(\cos\theta_2) r^l e^{-im\varphi}\\
   & + \sum_{l,m} \frac{(l-m)!}{(l+m)!} \frac{q^{l+1/2} y^m}{1 - q^{l+1/2} y^m} P_l^m(\cos\theta_1) P_l^m(\cos\theta_2) \left(r^l e^{-im\varphi} + r^{-l-1} e^{im\varphi}\right)\ .
\end{align*}
In the following, we shall re-define our two-point function by absorbing into it the $4\pi$ factor. The first term may be evaluated exactly and coincides with the zero-temperature two-point function. This gives the result stated in the introduction, equation~\eqref{two-point-funtion},
\begin{align}
    \langle\phi(x_1)\phi(x_2)\rangle_{q,y} &= \frac{1}{|x_{12}|}\\
    + &\sum_{l,m} \frac{(l-m)!}{(l+m)!} \frac{q^{l+1/2} y^m}{1 - q^{l+1/2} y^m} P_l^m(\cos\theta_1) P_l^m(\cos\theta_2) \left(r^l e^{-im\varphi} + r^{-l-1} e^{im\varphi}\right)\ .\nonumber
\end{align}
The same formula may also be derived using the method of images.

\subsection{One-point functions and OPE coefficients}

We can extract one-point functions from the two-point function \eqref{two-point-funtion} using the operator product expansion. For scalar operators, the OPE takes the form
\begin{align}
    \phi(x_1) \times \phi(x_2) & \sim \sum_{\mathcal{O}} \frac{\lambda_{\phi\phi \mathcal{O}}}{|x_{12}|^{2\Delta_\phi - \Delta_\mathcal{O}+l}}  C_{\mathcal{O}}^{\mu_1 \dots \mu_l}(x_{12},\partial_{x_2})  \mathcal{O}_{\mu_1 \dots \mu_l}(x_2) \, , \\
    C_{\mathcal{O}}^{\mu_1 \dots \mu_l}(x_{12},\partial_{x_2}) &= x_{12}^{\mu_1} \dots x_{12}^{\mu_l} \left(a_1 + a_2 x_{12}^\mu \partial_{x_2^\mu} + \dots \right) \, ,
\end{align}
where the coefficients $a_i$ are fixed by conformal invariance. In the case of $\mathcal{O}$ also being a scalar, the differential operator $C_{\mathcal{O}}$ reads
\begin{equation}
    C_{\mathcal{O}}(x_{12},\partial_{x_2}) = 1 + \frac{1}{2} x_{12}^\mu \partial_{x_2^\mu} + \frac{\Delta_\mathcal{O}+2}{8(\Delta_\mathcal{O}+1)} x_{12}^\mu x_{12}^\nu \partial_{x_2^\mu} \partial_{x_2^\nu} + \frac{\Delta_\mathcal{O}}{8(2\Delta_\mathcal{O}-1)(\Delta_\mathcal{O}+1)} x_{12}^2 \partial_{x_2}^2 + \dots \, .
\end{equation}
Hence, for the free theory we may write
\begin{equation}\label{eq:OPE_2-pt_function}
    \langle \phi \phi \rangle_{q,y} = \frac{1}{|x_{12}|} + \lambda_{\phi \phi \phi^2} C_{\phi^2}(x_{12},\partial_{x_2}) \langle \phi^2 \rangle_{q,y} + \lambda_{\phi \phi T} C_T^{\mu \nu}(x_{12},\partial_{x_2}) \langle T_{\mu\nu} \rangle_{q,y} + \dots \, .
\end{equation}
A comparison of equations~\eqref{two-point-funtion} and \eqref{eq:OPE_2-pt_function} enables us to compute one-point functions of the low lying operators appearing in the $\phi\times\phi$ OPE. Both expansions start with $|x_{12}|^{-1}$, which is the contribution of the identity to the two-point function. In the remainder of this subsection, we shall compute the one-point functions $\langle\phi^2\rangle_{q,y}$ and $\langle T_{\mu\nu}\rangle_{q,y}$ and expand them in thermal conformal blocks. In the similar manner, one may continue to higher operators appearing in $\phi\times\phi$.

\subsubsection{$\langle\phi^2\rangle$ one-point function}

Since the operator $C_{\phi^2}(x,\partial)$ starts with one, to find $\lambda_{\phi \phi \phi^2} \langle \phi^2 \rangle_{q,y}$, we may set $\theta_1 = \theta_2 \equiv\theta$, $\varphi=0$ and $r=1$. This gives the one-point function
\begin{equation}\label{eq:one_point_phi2}
    \lambda_{\phi\phi\phi^2}\langle\phi^2(x)\rangle_{q,y} = \frac{2}{r} \sum_{l,m} \frac{(l-m)!}{(l+m)!} \frac{q^{l+1/2}y^m}{1 - q^{l+1/2}y^m} P_l^m(\cos\theta)^2\ .
\end{equation}
We have written the one-point function in general position by inserting the prefactor $r^{-1}$ coming from equation~\eqref{H-invariants-in-principal}. Owing to $\lambda_{\phi\phi\phi^2}=\sqrt{2}$, the last equation gives rise to equation~\eqref{1pt-function-phi2} written in the introduction. We can decompose the one-point function in conformal blocks by Taylor-expanding equation~\eqref{1pt-function-phi2} around $q=0$ and matching to blocks, which are given in the same expansion. Since the blocks behave as $q^{\Delta_{\mathcal{O}}}$ as $q\to0$, operators with increasing scaling dimensions start contributing at higher orders. The dependence of both correlators and blocks on the chemical potential $y$ allows to distinguish contributions of operators of different spin. Clearly, once more than a single operator with certain quantum numbers $(\Delta_{\mathcal{O}},\ell_{\mathcal{O}})$ appears in the spectrum, the conformal block decomposition of the one-point function can only determine the sum of corresponding OPE coefficients. For illustration, the expansion of equation~\eqref{1pt-function-phi2} to a few lowest orders reads (setting $r=1$)

\begin{small}
\begin{equation}
   \frac{\langle\phi^2(x)\rangle_{q,y}}{\sqrt{2}} = \sqrt{q} + q + \left(\frac{4y+p^2(y-1)^2}{2y}\right) q^{\frac32} + q^2 + \left( \frac{3p^4(y-1)^4+12p^2(y-1)^2 y+16y^2}{8y^2}\right) q^{\frac52} + \dots \ .
\end{equation}
\end{small}

On the other hand, the blocks of lowest operators contributing, divided for the partition function, are
\begin{small}
\begin{align*}
    & \mathcal{Z}^{-1} g^{\phi^2}_\phi(q,y,p) = \sqrt{q} - q + \left(1 + \frac12 p^2\Big(y+\frac{1}{y}-2\Big) \right) q^{\frac32} - \left(\frac{p^2(y-1)^2+2(y+1)^2}{2y}\right)q^2 + \dots\,,\\[2mm]
    & \mathcal{Z}^{-1} g^{\phi^2}_{\phi^2}(q,y,p) = q - q^{\frac32} + \left(\frac{p^2(y-1)^2+2(y+1)^2}{4y}\right) q^2 - \left(\frac{6+p^2(y-1)^2+8y+6y^2}{4y}\right) q^{\frac52} + \dots\ .
\end{align*}
\end{small}
Matching the expansions, we get
\begin{equation}
    \mathcal{Z} \langle\phi^2(x)\rangle_{q,y} = \sqrt{2} \left(g_\phi^{\phi^2} + 2 g_{\phi^2}^{\phi^2} + 3 g_{\phi^3}^{\phi^2} + \dots + 6 g_{\phi^6}^{\phi^2} + 4 g_T^{\phi^2} + 7 g_{\phi^7}^{\phi^2} + \dots \right)\ .
\end{equation}
We have slightly abused the notation in labelling blocks by external and intermediate fields rather than their quantum numbers in order to increase physical clarity. Clearly, we get the correct OPE coefficients for exchanged scalars\footnote{For exchanged operators, we use schematic notation $\phi^n$ to refer to the unit-normalised operator $:\phi^n:/\sqrt{n!}$. For the external operator $\phi^2$, the choice of normalisation does not matter, as it only contributes an overall factor that does not depend on $n$.}
\begin{equation}\label{phi^2-OPE-coefficients}
    \lambda_{\phi^2\phi^n\phi^n} = n\sqrt{2}\ .
\end{equation}
\paragraph{Remark} At $y=1$, the one-point function can be re-summed to give
\begin{equation}
    \langle\phi^2(x)\rangle_{q} = \frac{\sqrt{2}}{r} \frac{\log(1-q) + \psi_q(1/2)}{\log q}\,,
\end{equation}
where $\psi_q(z)$ is the $q$-digamma function. Expanding at small $\beta$ (setting $r=1$), we get
\begin{equation}\label{leading-divergence-1pt}
    \langle\phi^2(x)\rangle_{q} = -\frac{1}{\sqrt{2}} \left( \frac{\log\beta}{\beta} + \frac{\psi(1/2)}{\beta} + \frac{\beta}{288} + \dots\right)\ .
\end{equation}
Notice that the series does not behave as $b\beta^{-1}$ at high temperature, having a more singular term. This is consistent with the fact that $\phi^2$ does not have a well-defined one-point function on $S^1\times S^2$, see \cite{Iliesiu:2018fao} for a discussion.

\subsubsection{Stress tensor one-point function}

To read off the stress tensor one-point function, we subtract from $\langle\phi\phi\rangle_{q,y}$ the first two terms on the right-hand side of equation~\eqref{eq:OPE_2-pt_function}, expanding $C_{\phi^2}$ to second order in $x_{12}$. The result vanishes at order $O(x_{12})$, as expected. Stripping off $x_{12}^\mu x_{12}^\nu$, we read off $\langle T_{\mu\nu}\rangle_{q,y}$. To a few lowest orders,
\begin{small}
\begin{align}\label{eq:one_point_T}
   & \langle T_{\mu\nu}\rangle_{q,y} =  \\
   & \frac{e^{-\varphi M_{23}}}{r^3}\left(\frac18 \begin{pmatrix}
        2-3p^2 & 3p\sqrt{1-p^2} & 0\\
        3p\sqrt{1-p^2} & 3p^2-1 & 0\\
        0 & 0 & -1
    \end{pmatrix} (\sqrt{q} + q) + 
     \begin{pmatrix}
        T_{11}^{(3/2)} & T_{12}^{(3/2)} & T_{13}^{(3/2)}\\
        T_{12}^{(3/2)} & T_{22}^{(3/2)} &  T_{23}^{(3/2)}\\
        T_{13}^{(3/2)} & T_{23}^{(3/2)}  & T_{33}^{(3/2)}
        \end{pmatrix} q^{\frac32} + \dots \, \right) e^{\varphi M_{23}} , \nonumber
\end{align}
\end{small}

where $M_{23}$ is the $3\times 3$ matrix in the vector representation of $\mathfrak{so}(3)$ and
\begin{align}
    T_{11}^{(3/2)}(p,y) & = y^{-1}\left(4-15p^2(y-1)^2 +4(y+8)+6p^2(y^2-12y+1) \right)\,, \nonumber\\
    T_{12}^{(3/2)}(p,y) & = 3 y^{-1} p\sqrt{1-p^2}\left( 2+5p^2(y-1)^2+2y(y+8) \right)\,,\nonumber\\
    T_{13}^{(3/2)}(p,y) & = -12i p\sqrt{1-p^2} \left(y-y^{-1}\right)\,,\\
    T_{22}^{(3/2)}(p,y) & = y^{-1} \left(15p^4(y-1)^2-2\big(1+y(y+8) \big)+3p^2(y^2+18y+1)\right) \,,\nonumber\\
    T_{23}^{(3/2)}(p,y) & = 12 i p^2 \left(y-y^{-1}\right) \,,\nonumber \\
    T_{33}^{(3/2)}(p,y) & = -y^{-1}\left( 2+9p^2(y-1)^2+2y(y+8) \right)\ .\nonumber
\end{align}
As a nice check, one verifies that the thermodynamic relation
\begin{equation}\label{thermodynamic-relaion}
    \langle T^{00} \rangle_{S^1_\beta \times S^{d-1}} = \frac{1}{S_d} \partial_\beta \log \mathcal{Z}(\beta)\,,
\end{equation}
is satisfied. The one-point function (again, written in the frame $r=1$, $\varphi=0$) is expanded in conformal blocks,
\begin{equation}\label{stress-tensor-expansion}
     \mathcal{Z}\langle T_{\mu\nu}\rangle_{q,y} = \frac38 \left( g^T_\phi + 2 g^T_{\phi^2} + 3 g^T_{\phi^3} + 4 g^T_{\phi^4} + \dots \right)\ .
\end{equation}
Recall that the OPE coefficient $\lambda_{\mathcal{O}\mathcal{O}T}$ for a scalar operator $\mathcal{O}$ reads, \cite{Rattazzi:2010gj},
\begin{equation}
    \lambda_{\mathcal{O}\mathcal{O}T} = \frac{d \Delta_{\mathcal{O}}}{(d-1)S_d} = \frac{3\Delta_{\mathcal{O}}}{8\pi}\ .
\end{equation}
Therefore, equation~\eqref{stress-tensor-expansion} gives the correct OPE coefficients.

\subsection{Statistics of OPE coefficients}

The high-temperature partition function of a CFT is known to constrain large-$\Delta$ asymptotic behaviour of the density of states (either all states or primaries). In a similar way, the asymptotics of OPE coefficients $\lambda^a_{\phi\mathcal{O}\mathcal{O}}$ in the limit of $\Delta_\mathcal{O}\to\infty$ is constrained by the high-temperature behaviour of the thermal one-point function $\langle\phi\rangle_{q,y}$. The purpose of this subsection is to begin exploring the relation between thermal one-point functions and such HHL OPE coefficients by computing averaged OPE coefficients in the free theory. Thermal conformal blocks provide us with an efficient way of doing so.
\smallskip

\begin{figure}[H]
    \centering
    \includegraphics{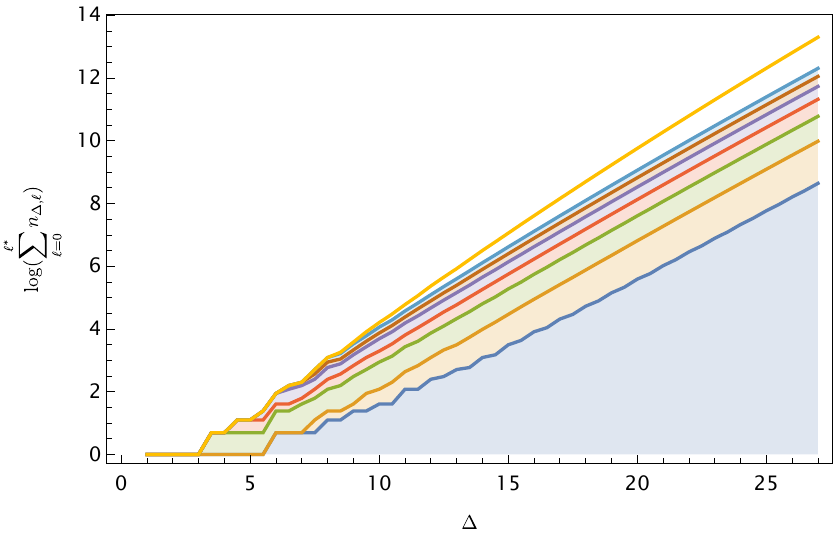}
    \caption{Multiplicities of operators in the free theory as a function of the dimension $\Delta$. The first line from above (yellow) is \eqref{eq:multiplicities_no_spin}, while the others are with smaller $\ell^\ast$. The first line from below (blue) represents multiplicities for only scalars, then next one (orange) both spin 0 and spin 1, and so on... . The lines get more and more dense, and we stop at $\ell^*=6$.}
    \label{spectrum}
\end{figure}

In order to discuss averaged OPE coefficients, we need degeneracies (or multiplicities) $n_{\Delta,\ell}$ of (non-conserved) primary operators of a given conformal weight and spin. These are obtained by character decomposition of the partition function, as explained above. E.g. we show in Figure \ref{spectrum} as the yellow curve the multiplicities summed over all spins as a function of $\Delta$. We will write
\begin{equation}\label{eq:multiplicities_no_spin}
    n_\Delta = \sum_\ell n_{\Delta,\ell} \ .
\end{equation}

One might consider comparing the exact data from Figure~\ref{spectrum} with the asymptotic formula presented in \cite{Benjamin:2023qsc} (see equation~\eqref{eq:rho-Delta-J}). However, given the specific range of validity of the latter, and the size of subleading corrections, a close match is not expected. In fact, the data significantly differs from the formula.
\smallskip

\begin{figure}[H]
    \centering
    \includegraphics[scale=1]{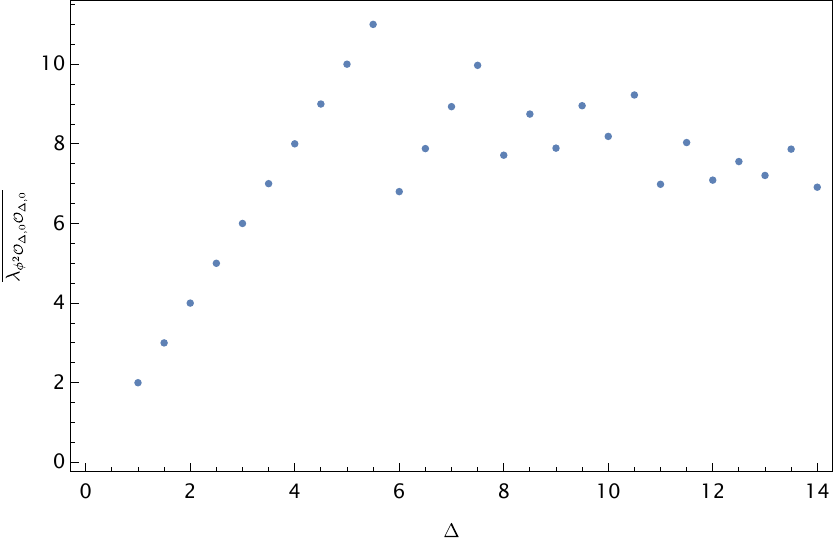}
    \caption{Average OPE coefficients in free theory \eqref{eq:average_OPE_coeff_Def} with $\ell_\mathcal{O}=0$. In this case we have no index $a$, since we have only one tensor structure.}
    \label{scalar-ope-plot}
\end{figure}

\begin{figure}[H]
    \centering
    \includegraphics[scale=1]{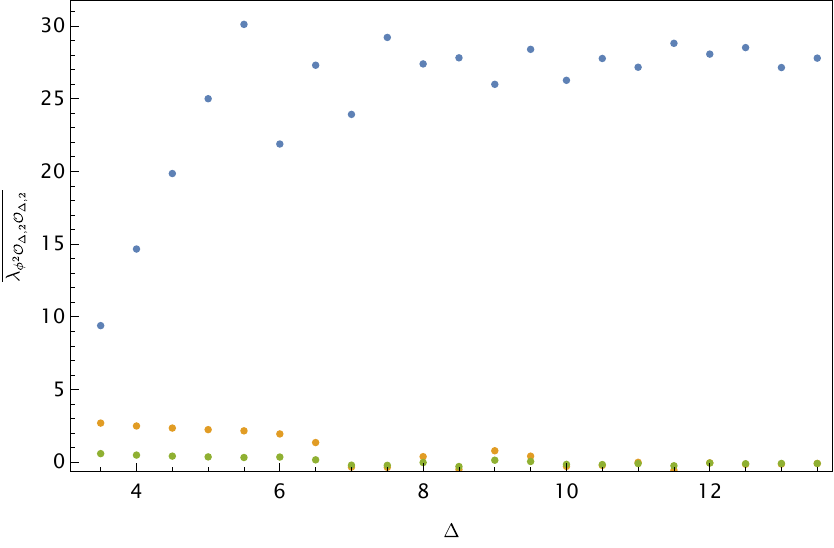}
    \caption{Average OPE coefficients in free theory \eqref{eq:average_OPE_coeff_Def} with $\ell_\mathcal{O}=2$. In this case $a=1,2,3$, and we label different tensor structures with different colours.}
    \label{spin2-ope-plot}
\end{figure}

\begin{figure}[H]
    \centering
    \includegraphics[scale=1]{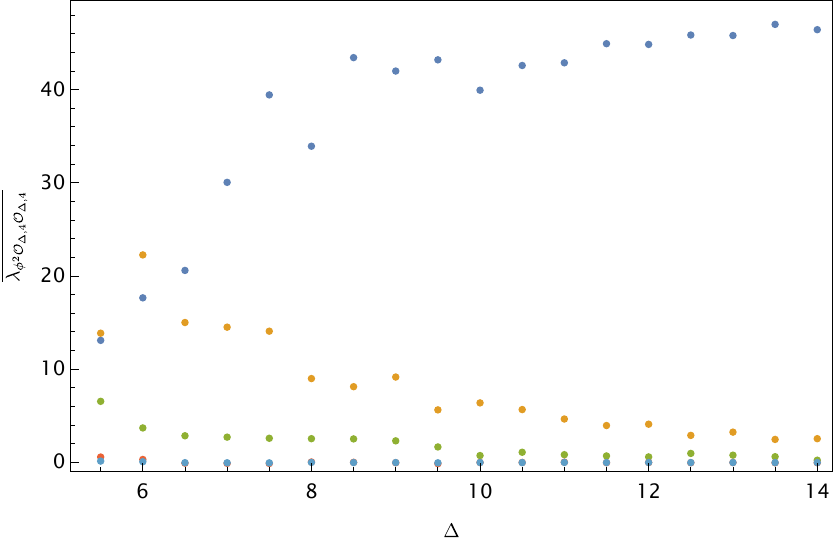}
    \caption{Average OPE coefficients in free theory \eqref{eq:average_OPE_coeff_Def} with $\ell_\mathcal{O}=4$. In this case $a$ runs from 1 to 5, and we label different tensor structures with different colours.}
    \label{spin4-ope-plot}
\end{figure}

Let us turn to OPE coefficients. We consider the one-point function $\langle\phi^2\rangle_{q,y}$. As explained in the last subsection, we read off OPE coefficients by expanding $\langle\phi^2\rangle_{q,y}$ in thermal conformal blocks, and we define averaged OPE coefficients as
\begin{equation}\label{eq:average_OPE_coeff_Def}
  \overline{\lambda^a_{\phi^2\mathcal{O}\mathcal{O}}} = \Big(\sum_i  \lambda^a_{\phi^2\mathcal{O}_i\mathcal{O}_i}\Big) / n_{\Delta_\mathcal{O},\ell_\mathcal{O}} \, ,
\end{equation}
where the sum runs over all (non-conserved) operators with the same quantum numbers $(\Delta_\mathcal{O},\ell_\mathcal{O})$, and $n_{\Delta_\mathcal{O},\ell_\mathcal{O}}$ is the multiplicity of primary operators. Plots in Figures \ref{scalar-ope-plot}, \ref{spin2-ope-plot} and \ref{spin4-ope-plot} show $\overline{\lambda^a_{\phi^2\mathcal{O}\mathcal{O}}}$ as the function of the dimension $\Delta_{\mathcal{O}}$. In each plot, the spin of $\mathcal{O}$ is fixed, to zero, two and four, respectively. In the second and third case, there are multiple tensor structures for the three-point function $\langle\mathcal{O}\mathcal{O}\phi^2\rangle$, which are distinguished by different colours.\footnote{Only parity even tensor structures appear in free theory, hence we have $(l_{\mathcal{O}}+1)$ tensor structures.} E.g. in the spin two case, there are three tensor structures
\begin{equation}
    V_{1,23}^2 V_{2,31}^2, \quad H_{12} V_{1,23} V_{2,31}, \quad H_{12}^2\ .
\end{equation}
Our blocks use the following linear combinations of these,
\begin{equation}
    \mathbb{T}_3^a = \frac{25}{36\pi^2} \begin{pmatrix}
        0  & -6 & -3\\
        0  &  6 &  4\\
        -1 & -4 & -2
    \end{pmatrix} \begin{pmatrix}
        V_{1,23}^2 V_{2,31}^2\\
        H_{12} V_{1,23} V_{2,31}\\
         H_{12}^2
        \end{pmatrix}\ .
\end{equation}
In Figure \ref{spin2-ope-plot}, values $a=1,2,3$ correspond to colours blue, grey and green, respectively. Similar remarks apply to the spin four plot. Our choice of tensor structures is dictated by our algorithm for computing the thermal blocks. For completeness, we note that the shown plots use 134, 254 and 409 lowest lying operators in the three spin sectors, respectively.

Let us discuss some features of the plots, starting with the scalar one, Figure \ref{scalar-ope-plot}. As we have mentioned in equation~\eqref{phi^2-OPE-coefficients}, operators $\phi^n$ have the OPE coefficient $\lambda_{\phi^2\phi^n\phi^n}$ equal to $n\sqrt{2}$. The first ten dots on the plot come from operators of this kind, where they are the only scalar operators of their respective dimensions. For higher scaling dimensions, however, other scalar operators can be constructed with the help of partial derivatives. These other operators have smaller OPE coefficients, leading to the average that grows slower than linearly in $\Delta_{\mathcal{O}}$. In certain circumstances arguments from the ETH, \cite{Lashkari:2016vgj}, and the bootstrap, \cite{Gobeil:2018fzy} predict that HHL OPE coefficients asymptotically grow as $\lambda_{HHL}\sim \Delta_H^{\Delta_L/3}$ (see also Section~\ref{S:Asymptotic OPE coefficients}). We see no such behaviour in Figure \ref{scalar-ope-plot}. This may not be surprising as neither of the arguments referred to quite applies to the case of Figure \ref{scalar-ope-plot}. Indeed, the ETH assumes that the theory under consideration is non-integrable, which is certainly not the case for the free theory. On the other hand, due to the fact that $\phi^2$ does not have a one-point function $b_{\phi^2}$ on $S^1\times \mathbb{R}^2$, it violates assumptions of the bootstrap analysis of \cite{Gobeil:2018fzy} and the asymptic analysis of next Section.
\smallskip

Turning to the spinning exchange plots in Figures \ref{spin2-ope-plot} and \ref{spin4-ope-plot}, we observe a clear asymptotic suppression of OPE coefficients in all directions in the tensor structure space except for one. The same phenomenon occurs for exchanges of all higher spin operators as well. Such behaviour certainly begs for an explanation and will be addressed in future work.
\smallskip

We are currently optimising the algorithm for computation of the blocks, which should allow to significantly extend each of the presented plots. With these, together with similar plots for the external stress tensor, we will be better equipped to discuss the asymptotics of HHL OPE coefficients.

\section{Asymptotic OPE Coefficients} 
\label{S:Asymptotic OPE coefficients}

Under certain assumptions, it has been argued in \cite{Gobeil:2018fzy} that averaged $\lambda_{\phi\mathcal{O}\mathcal{O}}$ OPE coefficients, at least when $\phi$ is a scalar, have universal asymptotics in the limit $\Delta_{\mathcal{O}}\to\infty$. The argument rests on the fact that the high temperature one-point functions on $S^1 \times S^2$ are fixed by their form on $S^1 \times \mathbb{R}^2$. On the other hand, the one-point blocks in the limit $\Delta_{\mathcal{O}}\to\infty$ reduce to characters, and are therefore well understood. In order to use the character approximation in the block expansion of the one-point function, one further needs to assume the ‘low spin dominance', i.e. that there is a function $J(\Delta)$ such that in the $\beta\to0$ limit only the exchange of operators with spin $\ell<J(\Delta)$ contributes to the one-point function. Furthermore, the ‘low spin' means that $J(\Delta)/\Delta\to0$ as $\Delta\to\infty$. 

Given these assumptions, one can proceed by considering the following ‘integrated density' of OPE coefficients,
\begin{equation}\label{integrated-density-OPE-def}
    \Lambda_\phi(\Delta_\mathcal{O}) = \sum_{\ell_\mathcal{O}=0}^{J(\Delta_{\mathcal{O}})} \sum_{a=1}^{N_3} \lambda^a_{\phi \mathcal{O} \mathcal O} \ .
\end{equation}
The sum is over all operators in the theory of fixed scaling dimension $\Delta_{\mathcal{O}}$ and spin less than the cutoff $J(\Delta_{\mathcal{O}})$, and over all tensor structures, whose number we denote by $N_3 = N_3(\phi,\mathcal{O},\mathcal{O})$. Notice that \eqref{integrated-density-OPE-def} requires us to specify the basis for three-point structures. This is done by requiring that the corresponding thermal blocks have the asymptotics \eqref{asymptotic-block}, i.e.~not allowing for multiples by a constant.\footnote{In our solutions in the attached code, we do not impose this requirement.} The object on the right-hand side is non-zero only for discrete set of values $\Delta_{\mathcal{O}}$ belonging to the physical spectrum. In the following, we shall treat it as a continuous density.

Under the assumptions discussed above, one shows that $\Lambda_\phi(\Delta)$ takes the asymptotic form
\begin{equation}\label{integrated-asymptotic-density-OPE}
    \Lambda_\phi(\Delta) \simeq \frac{2^{3-\Delta_\phi} \pi^{2/3-\Delta_\phi/3} f^{7/6 - \Delta_\phi/3} }{\sqrt{3}}\, b_\phi\ \Delta^{\frac13\Delta_\phi - \frac53}\ e^{3\pi^{1/3} f^{1/3} \Delta^{2/3}}\,,
\end{equation}
up to sub-leading corrections at large $\Delta$,\footnote{In some formulas we replace $\Delta_\mathcal{O}$ with $\Delta$ in order not to clutter the notation.} or at least that the form \eqref{integrated-asymptotic-density-OPE} is consistent both with the conformal block expansion and the limit to $S^1_\beta \times \mathbb{R}^2$. Here, $f$ is the free energy density of the CFT, see e.g. \cite{Benjamin:2023qsc}. For instance, in the free scalar theory, $f=\zeta(3)/(2\pi)$. We give a precise statement of assumptions and the definition of $\Lambda_\phi(\Delta)$ in Appendix \ref{A:Asymptotics of OPE coefficients}. While the factor multiplying $\Delta^{2/3}$ in the exponential is expected, as it also appears in the asymptotic density of (primary) states, the exponent $\Delta_\phi/3$ of $\Delta$ has been discussed as supporting the ETH, \cite{Gobeil:2018fzy}.
\smallskip

To elucidate these comments, consider the asymptotic behaviour of the density of primary states, $\hat\rho(\Delta_\mathcal{O})$. The latter may be obtained from the character decomposition of the partition function and reads
\begin{equation}\label{asymptotic-density-of-primaries}
 \hat \rho(\Delta) \simeq  \frac{8 \pi^{2/3} f^{7/6} \Delta^{-5/3}}{\sqrt{3}} e^{3 \pi^{1/3} f^{1/3}\Delta^{2/3}} \ .
\end{equation}
Compared to the asymptotic density of all states, \eqref{asymptotic-density-of-primaries} contains an additional factor of $\Delta^{-1}$, that arises from replacing exponentials $e^{-\beta\Delta}$ by characters in the decomposition of the partition function. Characters are approximately equal to $\beta^{-3}e^{-\beta\Delta}$ at large temperatures and through the saddle point approximation the factor $\beta^{-3}$ gives rise to $\Delta^{-1}$, see Appendix \ref{A:Asymptotics of OPE coefficients}. Now, taking the ratio of integrated densities \eqref{integrated-asymptotic-density-OPE} and \eqref{asymptotic-density-of-primaries}, we obtain
\begin{equation}\label{eq:average}
    \langle\langle \lambda_{\phi \mathcal O \mathcal O }\rangle\rangle\equiv \frac{\Lambda_\phi(\Delta_{\mathcal{O}})}{\hat \rho(\Delta_\mathcal{O})} = C_\phi \, \Delta_{\mathcal{O}}^{\Delta_\phi/3}\,, \qquad C_\phi = \frac{b_\phi}{2^{\Delta_\phi} \pi^{\Delta_\phi/3}  f^{\Delta_\phi/3}}\ .
\end{equation}
As it stands, the argument presented equally applies to scalar and spinning operators $\phi$, as both the asymptotics of one-point functions and blocks at vanishing chemical potential are known for the latter. Indeed, the latter was obtained in Section \ref{SS:Blocks at large scaling dimensions}. More work is required to establish/test the validity of other assumptions mentioned above, especially the low spin dominance. Similarly as the dependence of blocks on the chemical potential allowed for decompositions of one-point functions at low temperature, we expect that more precise estimates of blocks in the regimes of large quantum numbers $\Delta_{\mathcal{O}},\ell_{\mathcal{O}},a$ can lead to refined statements about asymptotic OPE coefficients. Indeed, the density $\hat\rho(\Delta)$ can be decomposed into different spin sectors,
\begin{equation}
    \hat \rho(\Delta_\mathcal{O}) = \int_0^{\Delta_\mathcal{O}-1} d\ell(2\ell_\mathcal{O}+1) \rho(\Delta_\mathcal{O},\ell_\mathcal{O})\,,
\end{equation}
where $\rho(\Delta_\mathcal{O},\ell_\mathcal{O})$ is the full density of primary states. The latter also admits a universal asymptotic form, computed for instance in \cite{Benjamin:2023qsc}, see \eqref{eq:rho-Delta-J}. Ideally we would like to obtain a similar asymptotic formula for averaged OPE coefficients of the form $\lambda^a_{\phi \mathcal O \mathcal O}(\Delta,\ell) / \rho_{\mathcal O}(\Delta,\ell)$, with an explicit dependence on both $\Delta$ and $\ell$, as well as the tensor structure label $a$. This requires asymptotic expressions for conformal blocks in various regimes, not only for $\Delta\gg \beta^{-1}\gg \ell, 1$, that was considered in Section~\ref{SS:Blocks at large scaling dimensions}, which instead allows only to control the integrated density \label{eq:integrated-density}. 
\smallskip

Nevertheless, the expression \eqref{eq:average} is quite suggestive, as it gives a simple scaling for the integrated asymptotic values of OPE coefficients. As a check  one can consider the operator $\phi$ to coincide with the stress tensor $T^{00}$, with dimension $\Delta_\phi = 3$. Our result is then consistent with the expectation that $T^{00}$ measures the dimension of a state, $T^{00}$ being the dilatation operator.

A second interesting application concerns conserved spin-1 currents, associated to a global $U(1)$ symmetry. In this case, form the Ward identity, we expect $\lambda_{J\mathcal{OO}} \propto Q_{\mathcal{O}}$, where $Q_{\mathcal{O}}$ is the $U(1)$ charge of the operator $\mathcal{O}$. Equation \eqref{eq:average} in this case gives\footnote{Strictly speaking equation \eqref{eq:average} does not apply to $U(1)$ currents because the trace over the Hilbert space contains states with opposite charge whose contributions cancel in the sum. We expect however that inserting a suitable twist one should be able to formalise the argument without affecting the overall result. }
\begin{equation}
   Q_\mathcal{O} \sim  \langle\langle \lambda_{J \mathcal O \mathcal O }\rangle\rangle \sim \Delta_\mathcal{O}^{2/3}\ .
\end{equation}
One can interpret the above expression as the average charge of primary states at large dimension. This is precisely the same scaling obtained by the semi-classical analysis of large charge operators \cite{Hellerman:2015nra,Monin:2016jmo}, where the dimension of the first operator of charge $Q$ is predicted to scale as $\Delta\simeq c_{3/2}Q^{3/2}+c_{1/2} Q^{1/2}+O(Q^0)$. The preceding analysis suggests that non only the leading operator obeys this scaling, but also the integrated density does.

\section{Summary and Perspectives}

In this work, we have initiated a systematic study of one-point conformal correlation functions at finite temperature and in presence of chemical potentials. After deriving Ward identities for any higher-point thermal correlator in the presence of these potentials and the quadratic Casimir equation for the corresponding conformal blocks, we focused on one-point functions in three dimensions. In this case, we solved the Casimir equation, thereby obtaining explicit expressions for thermal blocks, in a low-temperature series expansion. Our results apply to situations where both the external and exchanged field are spinning, either non-conserved or conserved. In the case where the external and intermediate fields are assumed to be scalar, the only one that has been previously analysed, we recover the know results from the literature. Our blocks are available at \href{https://gitlab.com/russofrancesco1995/thermal_blocks}{gitlab.com/russofrancesco1995/thermalblocks} .
\smallskip

The analysis of blocks in this work starts from their identification with harmonic spherical functions for the Gelfand pair $(SO(d+1,1)^2,SO(d+1,1)_{\text{diag}})$, an observation similar to those made for the usual four-point blocks, \cite{Schomerus:2016epl}, or partial waves for scattering amplitudes, \cite{Buric:2023ykg}. In the present context, this perspective allowed us to derive a weight-shifting operator that generates scalar exchange blocks with arbitrary spin of the external field, by acting on those that have a scalar external operator.
\smallskip

As we have pointed out, the presence of chemical potentials is necessary in order to extract OPE coefficients from thermal one-point functions. At zero chemical potentials, only certain linear combinations of the coefficients can be obtained. We gave first examples of thermal block decompositions by expanding the correlators $\langle\phi^2\rangle_{q,y}$ and $\langle T_{\mu\nu}\rangle_{q,y}$ in the free field theory and recovering the correct OPE coefficients. To this end, we first found the two point function $\langle\phi\phi\rangle_{q,y}$ and used the OPE to extract the relevant one-point functions. Further, we have obtained averaged OPE coefficients $\overline{\lambda_{\phi^2\mathcal{O}\mathcal{O}}^a}$ at fixed spin $\ell_{\mathcal{O}}$, as a function of the dimension $\Delta_{\mathcal{O}}$, allowing $\Delta_{\mathcal{O}}$ to be considerably large. These results, which follow much more efficiently from the thermal block decomposition than a first principle combinatorial calculation, can serve to test predictions/make conjectures about asymptotic behaviour of HHL OPE coefficients.
\medskip

The present work opens several interesting directions to pursue. One of these was sketched in Section \ref{S:Asymptotic OPE coefficients}. While it is expected that universal formulas for asymptotic (averaged) HHL OPE coefficients $\overline{\lambda^a_{\phi\mathcal{OO}}}$ exist and may be approached along the lines of that section, more work is needed in order to resolve their dependence on all quantum numbers. In the case of the tensor structure label, this was exemplified by our plots in Section \ref{S:OPE coefficients in free field theory}. To properly address these questions, a comprehensive further investigation of thermal conformal blocks is required. The first step that is needed is to establish asymptotic formulas for blocks similar to \eqref{asymptotic-block} in other regimes of large quantum numbers. A study along these lines, culminating with the inversion formula, is certainly a worthwhile future direction. We expect the spherical function interpretation of partial waves to be useful in this regard.
\smallskip

A downside of one-point functions considered in this work is that they do not, to our knowledge, satisfy any interesting crossing equation. There are, however, several related systems that do. One example is that of thermal two-point functions. As shown in Sections \ref{S:CFT at finite temperature} and \ref{S:Harmonic analysis}, these may again be related to spherical functions, and some of our results carry over to them. The full analysis of two-point blocks in presence of chemical potentials certainly appears as a worthwhile extension of the present work. Another context where thermal one-point blocks make an appearance is the decomposition of ‘genus-two' partition function of \cite{Benjamin:2023qsc} in the ‘dumbbell channel'\footnote{We thank David Simmons-Duffin and Yixin Xu for discussions on this topic.}. We intend to study the associated system of crossing equations in the future.
\smallskip

More in spirit of Section \ref{S:OPE coefficients in free field theory}, one may use our blocks to extract OPE coefficients by expanding thermal one-point functions, in situations where the latter are available. Examples of this kind include thermal Witten diagrams. A particularly interesting case in this direction is the holographic stress tensor of the thermal AdS space. This space dominates the gravitational path integral at low temperatures and the resulting stress tensor should be expandable in our blocks, working order by order in $q$. On the other hand, at high temperatures, after the Hawking-Page transition, the dominant gravitational solution is the AdS-Kerr black hole, \cite{Carter:1968ks}. As we review in Appendix \ref{app:holography}, the corresponding one-point function is exponentially suppressed and cannot be expanded in a Taylor series in $q$. However, by developing asymptotic arguments sketched above, the latter one-point function should give rise to asymptotics of HHL OPE coefficients in holographic theories.
\smallskip

Another interesting possibility would be to numerically decompose thermal one-point functions in the fuzzy sphere model of the 3d Ising CFT, \cite{Zhu:2022gjc,Hu:2023xak}. While the OPE coefficients can be computed directly in this model, obtaining them via block expansion suffers from a different kind of finite size effects. It would be interesting to see how the coefficients obtained by the two methods compare to each other.
\smallskip
 
Finally, let us mention that, although it was not the main focus of this work, our results of Sections \ref{S:Harmonic analysis} and \ref{S:Thermal blocks in 3d} (and their extension to the compact case $G = SO(5)\times SO(5)$) present a step forward in the theory of spinning spherical functions. An interesting open question is whether the weight-shifting operator we found may be supplemented by sufficiently many covariant differential operators that would provide the solution theory without the need for series expansions, akin to \cite{Buric:2022ucg,Buric:2023ykg}. In this direction, it might be fruitful to try and combine our techniques with those used for analysis of matrix spherical functions, \cite{Koelink_2012,koelink2017matrix}. We believe that a fully algebraic solution theory will require significant new ideas.

In a related vein, for multi-point correlation functions at zero temperature is was possible to derive a complete set of commuting differential operators from the relation with limits of Gaudin integrable models \cite{Buric:2020dyz,Buric:2021ywo}. It would be interesting to extend this approach to thermal correlators. For non-zero temperature, one expects 
the Gaudin model to be replaced by a Hitchin integrable system on an elliptic surface with punctures. In order to recover the Casimir operators we considered in this work, one has to consider a degeneration limit of the underlying elliptic 
surface. This limit is expected to simplify the solution theory. At least for some special cases one might also be able to leverage relations with elliptic Calogero-Moser models, \cite{Etingof:1993gk,Etingof:1994kd, Langmann2014}.

\paragraph{Acknowledgements:} We would like to thank Julien Barrat, Francesco Bertucci, Giulia Fardelli, Yin-Chen He, Eric Koelink, Jaeha Lee, Alessio Miscioscia, Sridip Pal, Elli Pomoni, David Simmons-Duffin, Petar Tadi\'c and Yixin Xu for discussions. This project received funding from the German Research Foundation DFG under Germany’s Excellence Strategy -- EXC  2121 Quantum Universe -- 390833306. I.B. is funded by a research grant under the project H2020 ERC STG 2017 G.A. 758903 "CFT-MAP".

\appendix

\section{Conventions for Lie Algebras}
\label{A:Conventions for Lie algebras}

In this appendix we collect our conventions for Lie algebras that are used throughout the text. Let $\mathfrak{g}_c$ be a complex simple Lie algebra, $\kappa$ the Killing form on $\mathfrak{g}_c$, normalised in some arbitrary way, and $\mathfrak{h}_c$ a Cartan subalgebra of $\mathfrak{g}_c$. We shall denote the rank of $\mathfrak{g}_c$ by $r$ and pick an orthonormal basis for $\mathfrak{h}_c$ with respect to $\kappa$, $\{H_1,\dots,H_r\}$. By the standard theory, $\mathfrak{g}_c$ decomposes as
\begin{equation}\label{root-decomposition}
    \mathfrak{g}_c = \mathfrak{h}_c \oplus \sum_{\alpha\in R} \mathfrak{g}_c^\alpha\ .
\end{equation}
Here, $\alpha$ are roots of $\mathfrak{g}_c$ and the set of all roots is denoted by $R$. We regard roots as linear maps $\mathfrak{h}_c\xrightarrow{}\mathbb{C}$ such that $\text{ad}_H = \alpha(H)$ on the space $\mathfrak{g}_c^\alpha$. The inner product between roots induced by the Killing form is denoted by angle brackets, e.g. $(\alpha,\beta)$. Root spaces are one-dimensional, $\mathfrak{g}_c^\alpha = \text{span}\{E_\alpha\}$. We denote the normalised root vectors by $e_\alpha$,
\begin{equation}
    e_\alpha = \frac{E_\alpha}{\sqrt{\kappa(E_\alpha,E_{-\alpha})}}\,, \qquad \kappa(e_\alpha,e_{-\alpha}) = 1\ .
\end{equation}
Then, also the following relations hold
\begin{equation}
    [e_\alpha,e_{-\alpha}] = h_\alpha\,, \quad [h_\alpha,e_{\pm\alpha}] = \pm(\alpha,\alpha) e_{\pm\alpha}\,, \quad \kappa(h_\alpha,H) = \alpha(H)\,, \quad \forall H\in\mathfrak{h}_c\ .
\end{equation}
By the usual theory, the bracket $[e_\alpha,e_\beta]$ can be non-vanishing only if $\alpha+\beta = 0$ or $\alpha + \beta$ is a root. In the latter case, the bracket is proportional to the root vector $e_{\alpha+\beta}$ and we write
\begin{equation}
    [e_\alpha,e_\beta] = n_{\alpha\beta} e_{\alpha+\beta}\ .
\end{equation}
We shall define $n_{\alpha\beta}=0$ if $\alpha+\beta\notin\{0\}\cup R$.

\section{Conventions for the Conformal Algebra}
\label{A:Conventions for the conformal algebra}

Here we spell out our conventions for the Euclidean conformal group $SO(d+1,1)$ and its Lie algebra. The non-vanishing Lie brackets in $\mathfrak{so}(d+1,1)$ read
\begin{align}
    & [M_{\mu\nu},P_\rho] = \delta_{\nu\rho} P_\mu - \delta_{\mu\rho} P_\nu,\quad [M_{\mu\nu},K_\rho] = \delta_{\nu\rho} K_\mu - \delta_{\mu\rho} K_\nu\,,\\
    & [M_{\mu\nu},M_{\rho\sigma}] = \delta_{\nu\rho} M_{\mu\sigma} - \delta_{\mu\rho} M_{\nu\sigma} + \delta_{\nu\sigma} M_{\rho\mu} - \delta_{\mu\sigma} M_{\rho\nu}\,,\\
    & [D,P_\mu] = P_\mu,\quad [D,K_\mu]=-K_\mu,\quad [K_\mu,P_\nu] = 2(M_{\mu\nu} - \delta_{\mu\nu}D)\,,
\end{align}
where $\mu,\nu = 1,\dots,d$. In the Lorentz-like notation, we write the generators as $\{L_{\alpha\beta}\}$, $\alpha,\beta = 0,1,\dots,d+1$. These obey the relations
\begin{equation}
    [L_{\alpha\beta},L_{\gamma\delta}] = \eta_{\beta\gamma} L_{\alpha\delta} - \eta_{\alpha\gamma} L_{\beta\delta} + \eta_{\beta\delta} L_{\gamma\alpha} - \eta_{\alpha\delta} L_{\gamma\beta}\,,
\end{equation}
where $\eta$ is the mostly-positive Minkowski metric. The relation between conformal and Lorentz generators reads
\begin{equation}
    L_{01} = D, \quad L_{0\mu} = \frac12(P_{\mu-1} + K_{\mu-1}), \quad L_{1\mu} = \frac12(P_{\mu-1} - K_{\mu-1}), \quad L_{\mu\nu} = M_{{\mu-1,\nu-1}}\ .
\end{equation}
The quadratic Casimir and its value in the representation $(\Delta,l)$ of $SO(d+1,1)$ are given by
\begin{equation}
    C_2 = -\frac12 L^{\alpha\beta} L_{\alpha\beta} = D^2 +\frac12 \{P_\mu,K^\mu\} -\frac12 M^{\mu\nu} M_{\mu\nu}, \quad C_2(\Delta,l) = \Delta(\Delta-d) + l(l+d-2)\ .
\end{equation}
In the $(d+2)$-dimensional vector representation, the Lorentz generators are
\begin{equation}
    L_{\alpha\beta} = \eta_{\alpha\gamma} E_{\gamma\beta} - \eta_{\beta\gamma} E_{\gamma\alpha}\,,
\end{equation}
where $ (E_{\alpha\beta})_{ij} = \delta_{\alpha i} \delta_{\beta j}$. 
\smallskip

With our conventions for root systems, the positive and negative root vectors for three-dimensional conformal group read
\begin{equation}\label{eq:Map_roots_to_Generators}
    e_{(i,0)} = -\frac{i P_1}{\sqrt{2}}\,, \quad e_{(0,i)} = \frac{M_{12} + i M_{13}}{\sqrt{2}}\,, \quad e_{(i,\pm i)} = \frac{P_2 \pm i P_3}{2}\,,
\end{equation}
and
\begin{equation}
    e_{(-i,0)} = -\frac{i K_1}{\sqrt{2}}\,, \quad e_{(0,-i)} = \frac{M_{12} - i M_{13}}{\sqrt{2}}\,, \quad e_{(-i,\pm i)} = \frac{-K_2 \pm i K_3}{2}\ .
\end{equation}

\section{Radial Decomposition Algorithm} 
\label{A:Radial decomposition algorithm}

In this appendix, we give an algorithm for how to radially decompose an arbitrary element $u\in U(\mathfrak{g})$, in the context of diagonal Gelfand pairs studied in Section \ref{S:Harmonic analysis}. We assume that $u$ is given to us as a polynomial in variables $\{Y_i,H_i,e_\alpha,e'_\alpha\}$. If this is not the case, all the linear factors appearing in $u$ are first to be expressed in the basis $\{Y_i,H_i,e_\alpha,e'_\alpha\}$, with $i=1,\dots,r$ and $\alpha\in R$. Furthermore, we shall describe only the decomposition required to compute the action on one-sided spherical functions, $\sigma = 1$. This means that any monomial in $u$ that contains a factor of $H_i$ is set to zero. Finally, $u$ is radially decomposed by repeatedly applying the rules
\begin{equation}
    Y_i e'_\alpha \to e'_\alpha Y_i + [Y_i,e'_\alpha], \quad e_\alpha Y_i \to Y_i e_\alpha - [Y_i,e_\alpha], \quad  e_\alpha e'_\beta \to e'_\beta e_\alpha + [e_\alpha,e'_\beta]\,,
\end{equation}
where the brackets are
\begin{align}
    & [Y_i,e_\alpha] = \alpha_i \left(\coth(\alpha\cdot t) e_\alpha - \frac{e'_\alpha}{\sinh(\alpha\cdot t)}\right)\,,\\
    & [Y_i,e'_\alpha] = -\alpha_i \left(\coth(\alpha\cdot t) e'_\alpha - \frac{e_\alpha}{\sinh(\alpha\cdot t)}\right)\,,\\
    & [e_\alpha,e'_{-\alpha}] = \alpha_i \left(\cosh(\alpha\cdot t) H_i + \sinh(\alpha\cdot t)Y_i\right),\\
    & [e_\alpha,e'_\beta] = n_{\alpha\beta} \left(\frac{\sinh(\alpha\cdot t)}{\sinh((\alpha+\beta)\cdot t)} e_{\alpha+\beta} + \frac{\sinh(\beta\cdot t)}{\sinh((\alpha+\beta)\cdot t)} e'_{\alpha+\beta}\right)\,, \quad \alpha+\beta\neq0\ .
\end{align}
Clearly, the process ends after a finite number of steps to produce a polynomial in radial form, i.e. one such that in each monomial factors $e'_\alpha$ appear in leftmost positions and $e_\beta$ in rightmost positions, possibly being separated by factors $Y_i$.

\paragraph{Radial decomposition of $(P^{(1)})^2+(P^{(2)})^2$} As explained in Section \ref{S:Thermal blocks in 3d} order to find thermal blocks for exchanged free field, we were required to radially decompose $u = (P^{(1)})^2+(P^{(2)})^2$. To achieve this, we first write the translation generators in terms of root vectors using equation~\eqref{eq:Map_roots_to_Generators}, 
\begin{equation}
    P_1^{(j)}= \sqrt{2} \ i \ e_{(i,0)}^{(j)} \,, \qquad P_2^{(j)} = e_{(i,i)}^{(j)}+e_{(i,-i)}^{(j)} \, , \qquad P_3^{(j)}= i \left(e_{(i,-i)}^{(j)}-e_{(i,i)}^{(j)}\right)\,,
\end{equation}
where $j=1,2$. Thus, upon substituting equation~\eqref{linear-decomp-1}, the building blocks of $u$ are written in the basis $\{Y_i,H_i,e_\alpha,e'_\alpha\}$. Executing the above algorithm, we obtain
\begin{equation}
    \left(P^{(1)}\right)^2 + \left(P^{(2)}\right)^2  \mapsto \frac{e'^2_{(i,0)}}{\sin^2 t_1} - \frac{2e'_{(i,i)} e'_{(i,-i)}}{\sin(t_1+t_2)\sin(t_1-t_2)}\ .
\end{equation}

\section{Differential Operators}
\label{A:Differential operators}

In this appendix, we give explicit expressions for the Laplacian \eqref{Laplace-Casimir-op} and the weight-shifting operator \eqref{weight-shifting-operator}. We also give the generalisation of the external conservation operator \eqref{external-conservation} to arbitrary spin $\ell_\phi.$

\paragraph{Weight-shifting operator} The terms $Q^{(i)}_{\Delta_\phi,\ell_\phi}$ appearing in the weight-shifting operator are
\begin{small}
\begin{align*}
    Q^{(1)}_{\Delta_\phi,\ell_\phi} & = p(1+z^2)\partial_z -(p^2-1)(pz+\sqrt{1-p^2}) (p\partial_p+\Delta_\phi)-\ell_\phi pz \, , \\
    Q^{(2)}_{\Delta_\phi,\ell_\phi} & =\sqrt{1+z^2} \bigg( \big(z-2p(pz+\sqrt{1-p^2})\big)\partial_z+p(p^2-1)\partial_p + \ell_\phi p^2  \bigg)\, ,\\
    Q^{(3)}_{\Delta_\phi,\ell_\phi} & = \frac{(2p\sqrt{1-p^2}-z)\sqrt{1+z^2}+p^2(2z\sqrt{1+z^2}-1-z^2)}{2p} \partial_z \\ &+ \frac{1}{2}\big( p(\sqrt{1-p^2}+pz)-\sqrt{1-z^2} \big) \big( (p^2-1)\partial_p+p \Delta_\phi \big) - \frac{\ell_\phi \sqrt{1-p^2}}{2} \, ,\\
    Q^{(4)}_{\Delta_\phi,\ell_\phi} & = \frac{(-2p\sqrt{1-p^2}+z)\sqrt{1+z^2}-p^2(2z\sqrt{1+z^2}+1+z^2)}{2p} \partial_z \\ &+ \frac{1}{2}\big( p(\sqrt{1-p^2}+pz)+\sqrt{1-z^2} \big) \big( (p^2-1)\partial_p+p \Delta_\phi \big) - \frac{\ell_\phi \sqrt{1-p^2}}{2}\, .
\end{align*}
\end{small}

\paragraph{Laplacian} The operators $\mathcal{D}^{(i)}_{\Delta_\phi,\ell_\phi}$ appearing in the Laplacian are
\begin{small}
\begin{align*}
    \mathcal{D}^{(1)}_{\Delta_\phi,\ell_\phi} & = 2p^2(1-p^2)\partial_p^2 + 4p\sqrt{1-p^2}(1+z^2)( p\partial_p \partial_z +\Delta_\phi) \\
    &-2p(2\ell_\phi pz\sqrt{1-p^2}-2\Delta_\phi +p^2(1+2\Delta_\phi))\partial_p-2\Delta_\phi(1+2\ell_\phi z\sqrt{1-p^2}+(1-p^2)\Delta_\phi) \, ,\\
    \mathcal{D}^{(2)}_{\Delta_\phi,\ell_\phi} & = 2(p^2-1)\partial_p^2-4(1+z^2)\sqrt{1-p^2}\partial_p \partial_z+2\frac{(1+z^2)(1-2p^2+2p^2 z\sqrt{1-p^2})}{p^2}\partial_z^2+\\
    & 2\frac{2p(p+\ell_\phi z\sqrt{1-p^2})-1}{p}\partial_p +2\frac{z-2p\big(pz+\sqrt{1-p^2}(\ell_\phi+z^2(\ell_\phi-1))\big)}{p^2}\partial_z+2\ell_\phi(\ell_\phi+1)\,,\\
    \mathcal{D}^{(3)}_{\Delta_\phi,\ell_\phi} & = \tilde{\mathcal{D}}^{(a)}+ \tilde{\mathcal{D}}^{(b)} \, ,\\
    \mathcal{D}^{(4)}_{\Delta_\phi,\ell_\phi} & = \tilde{\mathcal{D}}^{(a)}- \tilde{\mathcal{D}}^{(b)} \, , \\
    \tilde{\mathcal{D}}^{(a)} &= (p^2+1)\partial_p^2-(1-p^2)^{3/2}(1+z^2)\partial_p\partial_z+\frac{(1+z^2)\big( 2p(p-z\sqrt{1-p^2})-1 \big)}{p^2}\partial_z^2\\
    &+\frac{(p^2-1)\big( p(p+2\ell_\phi z\sqrt{1-p^2}+2p \Delta_\phi)-1 \big)}{p}\partial_p +\\
    & \frac{2z p^2-z-2p\sqrt{1-p^2}(z^2+(p^2\Delta_\phi -\ell_\phi)(1+z^2))}{p^2}\partial_z \\
    &+ \Delta_\phi (2\ell_\phi pz \sqrt{1-p^2}+p^2 \Delta_\phi-2) \, ,\\
    \tilde{\mathcal{D}}^{(b)} &= 2\sqrt{1+z^2}(z\sqrt{1+p^2}-p)\partial_p \partial_z + 2 \frac{\sqrt{1-p^2}}{p}(1+z^2)^{3/2}\partial_z^2 + 2\ell_\phi(1-p^2)^{3/2}\sqrt{1+z^2}\partial_p + \\
    & \frac{2\sqrt{1+z^2}\bigg( p-(\ell_\phi-1)z\sqrt{1+p^2}-p^2\Delta_\phi+p^2 z\Delta_\phi \sqrt{1+p^2} \bigg)}{p}\partial_z - 2\ell_\phi p\sqrt{1-p^2}\sqrt{1+z^2} \Delta_\phi \, .
\end{align*}
\end{small}

\paragraph{Conservation of external} The operator to impose conservation for the external field of general spin reads
\begin{small}
\begin{align}\label{eq:external-conservation_General-Spin}
    \mathcal{D}^{\text{ext}}_{\ell_\phi} &= \frac{z}{2} \Big( (1+z^2)^2\partial_z^3+p(1+z^2)(p\sqrt{1-p^2}+p^2z-z)\partial_p \partial_z^2+\nonumber\\
    &+\big( (1+\ell_\phi)(p\sqrt{1-p^2}+p^2 z)-2z(\ell_\phi-2) \big)\partial_z^2 \nonumber \\
    & +p\big(-1+p^2-2\ell_\phi(p^2-1)(1+z^2)+z(p\sqrt{1-p^2}+2p^2 z-2z)\big) \partial_p \partial_z \\
    &+\ell_\phi p \big( (\ell_\phi-1)(p^2-1)z-\ell_\phi p \sqrt{1-p^2} \big)\partial_p  \ell_\phi(1+\ell_\phi)p\big( pz(\ell_\phi-1)-\ell_\phi \sqrt{1-p^2} \big) \nonumber\\
    & \big( 1+p^2+pz\sqrt{1+p^2}(\ell_\phi+1)+2(1+p^2)z^2-\ell_\phi^2(2p^2-1)(1+z^2)-\ell_\phi(1+p^2+3z^2)  \big) \partial_z \Big) \, .\nonumber
\end{align}
\end{small}

\section{Shadow Formalism}
\label{A:Shadow formalism}

The shadow integral representation of one-point thermal partial waves in three dimensions reads
\begin{align}
    \text{tr}_{\mathcal{O}} \left(\phi^A(x) q^D y^{M_{23}}\right) & = \int d^3 x'\ \langle \mathcal{\tilde O}_m(x') \phi^A(x) q^D y^{M_{23}}\mathcal{O}^m(x')\rangle\\
    & = q^{\Delta_\mathcal{O}} \left(y^{M_{23}}\right)^m{}_{m'} \int d^3 x'\ \langle \mathcal{\tilde O}_m(x') \phi^A(x)\mathcal{O}^{m'}\left(q^D y^{M_{23}}\cdot x'\right)\rangle\nonumber\ .
\end{align}
Here $\mathcal{\tilde O}$ is the shadow of the operator $\mathcal{O}$. Partial waves satisfy the same Casimir equation as conformal blocks, but obey different boundary conditions. The shadow formalism was used in \cite{Gobeil:2018fzy} to obtain exact results for scalar-external, scalar-exchange blocks at vanishing chemical potential. This result was recently extended to higher dimensions, \cite{Alkalaev:2024jxh}.

\section{More Details on the Computation}
\label{app:More details on the computation}
In this appendix, we give more details on how to compute in practice the blocks for different cases and decompose the one-point functions. An implementation in {\it Mathematica} is available \href{https://gitlab.com/russofrancesco1995/thermal\_blocks}{gitlab.com/russofrancesco1995/thermal\_blocks}. We will refer to some of these functions and explain how to easily generate the blocks.

\subsection{Blocks}
In this section we refer to functions in {\it Blocks.wl}. Here, we have the main functions to generate the blocks, i.e. the Casimir differential equation and its action on the ansatz to get the blocks at fixed power of $q$. The notation here is as in the main text, $\phi$ being the external operator with quantum numbers $(\Delta_\phi,\ell_\phi)$ and $\mathcal{O}$ the internal one with $(\Delta_\mathcal{O},\ell_\mathcal{O})$.

\paragraph{Definitions}
The one-sided Laplacian \eqref{one-sided-Laplacian}, conjugated by the factor $(ir)^{\Delta_\phi}$, is written as the function \texttt{lapQtyp[$\Delta_\phi,\ell_\phi$]}. In a similar way we define the weight-shifting operator \texttt{qQtyp[$\Delta_\phi,\ell_\phi$]} in equation~\eqref{weight-shifting-operator} and the conservation operator \texttt{dvQtyp[$\Delta_\phi,\ell_\phi$]} in equations~\eqref{external-conservation} and \eqref{eq:external-conservation_General-Spin}. The latter represents the action of the derivative $\partial_\mu \langle \phi^{\mu,\dots} \rangle_{\beta,y}$, reduced onto our space of functions.

\paragraph{General solution}
Blocks are eigenfunctions of the Casimir equation defined above. We compute them by acting on the ansatz \eqref{ansatz} and imposing that it is a solution of our Laplacian at some order $q$ fixed. The function to generate a block at order $q^k$ is \texttt{GeneralSolution[$\Delta_\phi,\ell_\phi,\Delta_\mathcal{O},\ell_\mathcal{O}$][$k$]}.

A function to generate the block for an external conserved operator at order $q^k$ is implemented as \texttt{GeneralSolutionCons[$\ell_\phi,\Delta_\mathcal{O},\ell_\mathcal{O}$][$k$]}. In practice, the function takes the blocks previously computed and imposes that the action of \eqref{eq:external-conservation_General-Spin} is zero. This reduces the number of tensor structures and gives back the correct block.

\paragraph{Remark} The correct way to compute a block is by leaving the parameters $\Delta_\phi,\Delta_\mathcal{O}$ in the function general and to fix them only after the block is obtained. If one needs to compute blocks faster, one can in principle put specific values for the dimensions, but this may not give the whole block. In fact, the action of the Laplacian on the ansatz may be trivial on some polynomials, which leads to some extra free parameters in the block. This phenomenon is observed to be pretty rare,\footnote{It happens mostly in free theory, where dimensions are integers or half integers. This is not expect to happen for general dimensions.} so substituting for dimensions at the outset is a viable way to compute blocks more efficiently.

Another way to speed up the computation for external spinning blocks is to use in combination the function to generate blocks with the weigh-shifting operator. This has not been implemented in a function, but tested in various situation to be more efficient.

\subsection{Decomposing one-point functions}
We now deal with details on the decomposition of the one-point functions in {\it Free\_theory.nb}. 

\paragraph{Spectrum} The first step is to compute the spectrum, and we do it by decomposing the partition function \eqref{free-partition-function} into characters \eqref{character-decomposition}. In practice, we expand the partition function at some fixed power $q^k$, and order by order we match with the same expansion of characters. This procedure gives us the spectrum of the free theory up to operators with $\Delta \leq k$, where $k$ is the power of $q$. We store the spectrum up to $\Delta=27$, which means around 2.7 million primaries, see Figure \ref{spectrum}.

\paragraph{One-point function of $\phi^2$} We turn to the first one-point function in the OPE of $\phi \times \phi$, i.e. $\langle \phi^2 \rangle_{q,y}$, equation~\eqref{eq:one_point_phi2}. The strategy is the same as for the partition function. We expand the non-normalised one-point function\footnote{We take $\mathcal{Z}\langle \phi^2 \rangle_{q,y}$, since the blocks are not normalised by $\mathcal{Z}$.} to some power of $q$, and we match with the blocks. Since we have the spectrum from the previous computation, we know exactly the blocks that we need to compute in order to decompose. The only subtlety is in the computation of the blocks for conserved internal operator, see Section \ref{subsubsec:Conserved_blocks}. To address this, we first get rid of the pole appearing at the unitarity bound, and then we add an extra block with quantum numbers $(\Delta+1,\ell -1)$. Apart for the block of the stress tensor, we do not need to add this extra block, since it is already present in the spectrum. We cannot distinguish operators with the same quantum numbers in the decomposition, which means that we cannot disentangle the contribution of the subtraction of the conserved block from other operators. This implies that some of the averaged OPE coefficients that we compute are not precise due to the contribution of the short blocks. This is happening only for odd spins, since conserved operators have even spin.

Once we have the blocks, we expand in $q$ and we solve a linear system to compute the averaged OPE coefficients, see Figures \ref{scalar-ope-plot} and \ref{spin2-ope-plot}. We store the other values in {\it OPE\_data\_q14.nb}.

\paragraph{One-point function of $T_{\mu \nu}$} We can follow the same strategy to decompose the one-point function of the stress tensor $\langle T_{\mu \nu} \rangle_{q,y}$, equation~\eqref{eq:one_point_T}. The first issue is that it takes more time to produce the blocks, which means that we push the decomposition at a lower power of $q$ with respect to $\langle \phi^2 \rangle_{q,y}$. The second difference is that we act with the Thomas-Todorov operator \cite{Dobrev:1975ru} on the blocks before decomposing. We attach the values for the OPE coefficients up to power $q^4$.

\subsection{Tensor structures}
In {\it Study of tensor structures.nb}, we implement a map to go from our definition of OPE coefficients to the one related to three-point function at zero temperature, see Section \ref{SSS:Blocks and tensor structures}, in the embedding formalism \cite{Costa:2011mg}. Starting with a tensor structure in the embedding space, we can use the command \texttt{ruleToReal} to project it into real space and then call \texttt{mapToFrame[$\ell_{\mathcal{O}}$]} to obtain the corresponding tensor structure in our basis. This has been done for a series of examples both with scalar and spinning operators. In practice, the function \texttt{mapToFrame} decomposes our tensor structure into spherical harmonics, so that we know the contribution of $y^{M_{23}}$.

\section{Asymptotics}
\label{A:Asymptotics of OPE coefficients}

\subsection{Asymptotics density of states}
\label{A:Asymptotics-density}

We begin by recalling the standard argument that allows to extract the asymptotic density of primary states in a CFT. At large temperatures, the free energy is determined by dimensional analysis and the partition function is given by
\begin{equation}
    Z_{\text{asym}}(\beta) = e^{\frac{4\pi f}{\beta^2}}\ .
\end{equation}
The above expression has been generalised to account for the presence of a non-vanishing chemical potential, see \cite{Shaghoulian:2015lcn,Benjamin:2023qsc},
\begin{equation}
     Z_{\text{asym}}(\beta,\mu) = e^{\frac{4\pi f}{\beta^2 - \mu^2}}\ .
\end{equation}
By taking a double inverse Laplace transform in the variables $\beta_{L,R}= \beta \pm \mu$ it is possible to extract the asymptotic density of primary states with dimension $\Delta$ and spin $\ell$, \cite{Benjamin:2023qsc},
\begin{equation}\label{eq:rho-Delta-J}
     \rho(\Delta,\ell) \simeq \frac{8 \pi ^{2/3} f^{5/3} (2 \ell +1) \Delta }{\sqrt{3} (\Delta+\ell+1/2)^{7/3} (\Delta-\ell-1/2)^{7/3}} e^{3 \pi^{1/3} f^{1/3} (\Delta+\ell+1/2)^{1/3}(\Delta-\ell-1/2)^{1/3}}\ .
\end{equation}
In what follows we will need a different notion of density of states, that is summed over all possible spins of the field. To extract its behaviour, we start again from the decomposition of the partition function at zero chemical potential in characters, which for $\mu=0$ read (see \eqref{characters})
\begin{equation}
    \chi_{\Delta,\ell}(q,1) =\frac{(2\ell+1)q^{\Delta}}{(1-q)^3}\ .
\end{equation}
We can then define
\begin{equation}
     Z_{\text{asym}}(\beta) = \int d\Delta \sum_\ell \rho(\Delta,\ell) \chi_{\Delta,\ell}(q,1) \equiv \int d\Delta \hat\rho(\Delta) \frac{q^{\Delta}}{(1-q)^3}\ .
\end{equation}
At this point one can proceed as before and consider the inverse Laplace transform in the variable $\beta$ to obtain an asymptotic expression for the integrated density of states
\begin{equation}\label{eq:rho-hat}
    \hat \rho (\Delta) \simeq  \frac{8 \pi^{2/3} f^{7/6} \Delta^{-5/3}}{\sqrt{3}} e^{3\pi^{1/3} f^{1/3}\Delta^{2/3}}\ .
\end{equation}
It is worth mentioning that the quantity \eqref{eq:rho-hat} differs from what one would obtain by summing \mbox{$(2\ell+1)\rho(\Delta,\ell)$} given in \eqref{eq:rho-Delta-J} over all values of $\ell\in[0,\Delta-1]$. Indeed, this is expected, since the density \eqref{eq:rho-Delta-J} is  only  applicable in the regime where both $\Delta+\ell$ and $\Delta-\ell$ are large. It is conceivable that the density of states may follow a different expression when $\ell\sim \Delta$, potentially leading to the scaling \eqref{eq:rho-hat}. 

\subsection{Asymptotics of OPE coefficients}

The analysis of asymptotics of OPE coefficients $\lambda^a_{\phi\mathcal{O}\mathcal{O}}$ follows the broadly the same logic as above. To appreciate their relation to high temperature limits of one-point functions, consider the conformal block decomposition
\begin{equation}\label{decomposition-1pt-fn}
    \mathcal{Z}(q,y) \langle\phi\rangle_{q,y} = \sum_{\mathcal{O},a} \lambda^a_{\phi\mathcal{O}\mathcal{O}} (ir)^{-\Delta_\phi} Z^{\ell_\phi} \left(pz+\sqrt{1-p^2}\right)^{\ell_\phi} \frac{q^{\Delta_{\mathcal{O}}}}{(1-q)^3} h^{\Delta_\phi,\ell_\phi,a}_{\Delta_{\mathcal{O}},\ell_{\mathcal{O}}}(q,y,p,z)\ .
\end{equation}
For convenience, we have written the conformal block as a prefactor multiplied by a function $h$. For the moment, this simply defines the latter. From now on, whenever $a$ appears as a repeated index, a sum over it will be understood. We split the sum \eqref{decomposition-1pt-fn} into two pieces,
\begin{equation}\label{converse-Maloney-equality}
     (ir)^{\Delta_\phi}Z^{-\ell_\phi}\left(pz + \sqrt{1-p^2}\right)^{-\ell_\phi} \mathcal{Z}(q,y) \langle\phi\rangle_{q,y} = \sum_{\ell_{\mathcal{O}}<J(\Delta_{\mathcal{O}})} \lambda^a_{\phi\mathcal{O}\mathcal{O}} \frac{q^{\Delta_{\mathcal{O}}}}{(1-q)^3} h^{\Delta_\phi,\ell_\phi,a}_{\Delta_{\mathcal{O}},\ell_{\mathcal{O}}} + R(J)\ .
\end{equation}
The splitting is determined by a, as of now unspecified, real function $J(\Delta)$. By the unitarity bound, one may assume without loss of generality that $J(\Delta)\in[0,\Delta]$. We shall further assume that $J(\Delta)/\Delta\to0$ as $\Delta\to\infty$. At zero chemical potential, $y=1$, the first term in equation~\eqref{converse-Maloney-equality} may be approximated by
\begin{small}
\begin{align}
 & \sum_a \sum_{\ell_{\mathcal{O}}<J(\Delta_{\mathcal{O}})} \lambda^a_{\phi\mathcal{O}\mathcal{O}} \frac{q^{\Delta_{\mathcal{O}}}}{(1-q)^3} \left(1 + O(\Delta_{\mathcal{O}}^{-1})\right) \equiv  \sum_{\ell_{\mathcal{O}}<J(\Delta_{\mathcal{O}})} N_3 \bar\lambda_{\phi\mathcal{O}\mathcal{O}} \frac{q^{\Delta_{\mathcal{O}}}}{(1-q)^3} \left(1 + O(\Delta_{\mathcal{O}}^{-1})\right)\nonumber\\
 & \equiv \int_0^{\infty} d\Delta \int_0^{J(\Delta)} d\ell\ N_3(\ell) \Lambda_\phi(\Delta,\ell) \frac{q^{\Delta}}{(1-q)^3} \left(1+O(\Delta^{-1})\right) \equiv \int_0^\infty d\Delta \Lambda_\phi(\Delta) \frac{q^\Delta}{(1-q)^3}\left(1 + O(\Delta^{-1}) \right)\ . \label{integral-saddle}
\end{align}
\end{small}

In the first step, we have defined the averaged OPE coefficient over the $\langle\phi\mathcal{OO}\rangle$ tensor structures, the number of which is denoted $N_3$,
\begin{equation}
    \bar\lambda_{\phi\mathcal{O}\mathcal{O}} = \frac{1}{N_3} \sum_a \lambda^a_{\phi\mathcal{O}\mathcal{O}}\ .
\end{equation}
In the second step, this quantity is replaced by a continuous density $\Lambda_\phi(\Delta,\ell)$. Finally, the integral over spin up to $J(\Delta)$ defines the density $\Lambda_\phi(\Delta)$
\begin{equation}\label{integrated-density-OPE}
    \Lambda_\phi(\Delta) = \int_0^{J(\Delta)} d\ell\ N_3(\ell)\, \Lambda_\phi(\Delta,\ell)\ .
\end{equation}
It is the density $\Lambda_\phi(\Delta)$ that can be constrained by considering the high temperature limit of the one-point function. At $y=1$, the latter is given by
\begin{equation}\label{high-T-lim-one-pt}
    (ir)^{\Delta_\phi}Z^{-\ell_\phi} \left(pz + \sqrt{1-p^2}\right)^{-\ell_\phi} \mathcal{Z}(\beta) \langle\phi\rangle_\beta \sim \frac{b_\phi e^{\frac{4\pi f}{\beta^2}}}{\beta^{\Delta_\phi}}\,,
\end{equation}
as follows from the one-point function on $S^1 \times \mathbb{R}^2$.  Let us consider an ansatz for $\Lambda_\phi(\Delta)$ of the form
\begin{equation}
    \Lambda_\phi(\Delta) = a_0 \Delta^{a_1\Delta_\phi + a_2} e^{a_3 \Delta^{2/3}}\,,
\end{equation}
where $a_0,\dots,a_3$ are constants independent of $\Delta_\phi$. Then the integral \eqref{integral-saddle} is dominated by the saddle
\begin{equation}
    \Delta_{\text{saddle}} = \left(\frac{2 a_3}{3\beta}\right)^3\ .
\end{equation}
For small $\beta$, the saddle is large, so it is justified to neglect corrections subleading in $\Delta$. If one can drop the remainder term $R(J)$, i.e.
\begin{equation}\label{low-spin-dominance-assumption}
   \lim_{\beta\to0} \frac{R(J)}{e^{\frac{4f\pi}{\beta^2}}\beta^{-\Delta_\phi}} = 0\,,
\end{equation}
then by demanding that the saddle-point approximation of equation~\eqref{integral-saddle} matches equation~\eqref{high-T-lim-one-pt}, we arrive at a unique solution for the coefficients $a_i$,
\begin{equation}
    a_0 = \frac{2^{3-\Delta_\phi} \pi^{2/3-\Delta_\phi/3} f^{7/6 - \Delta_\phi/3} }{\sqrt{3}} b_\phi\,, \quad a_1 = \frac13\,, \quad a_2 = -\frac53\,, \quad a_3 = 3\pi^{1/3} f^{1/3}\ .
\end{equation}
Through the definition \eqref{integrated-density-OPE} of the integrated density $\Lambda_\phi(\Delta)$, these values translate into information about asymptotic OPE coefficients.

\section{Holographic One-point Functions}
\label{app:holography}

Given an asymptotically AdS spacetime $(\mathcal{M},g)$, a method for computing the expectation value of the stress tensor in the tentative boundary CFT was given in \cite{Myers:1999psa,deHaro:2000vlm}. In the first step, one expresses the metric in the Fefferman-Graham coordinates
\begin{equation}
    g_{\mu\nu} = \frac{\ell_{\text{AdS}}^2}{z^2} \left(dz^2 + g_{bc}(z,x) dx^b dx^c\right)\ .
\end{equation}
Here, $\ell_{\text{AdS}}$ is the AdS radius and Latin indices run over the boundary coordinates $1,\dots,d$. The tensor $g_{bc}$ expands in $z$ as
\begin{equation}
    g_{bc}(z,x) dx^b dx^c = h_0 + z^2 h_2 + \dots + z^d h_d + o(z^d)\,, 
\end{equation}
where, up to the order $(d-1)$, only the even powers of $z$ appear. The leading term simply denotes the boundary metric. The one-point function of the CFT stress tensor reads
\begin{equation}\label{CFT-stress-tensor-holography}
    \langle T_{bc} \rangle = \frac{d\, \ell_{\text{AdS}}^{d-1}}{16\pi G_N} (h_d)_{bc} + X_{bc}[h_0,h_d]\ .
\end{equation}
The discussion of the second term in \eqref{CFT-stress-tensor-holography} can be found in \cite{deHaro:2000vlm} - the only fact that we shall need is that it vanishes for odd $d$. The above statements acquire a more definite meaning in the context of black holes. For the rest of this discussion, we focus on $d=3$ and assume that $(\mathcal{M},g)$ is the Kerr-AdS black hole solution. This solution depends on two parameters, usually taken to be the mass and spin. For us, it is more convenient to us to use the temperature $T$ and the angular momentum $\Omega_h$, as measured by an observer at the asymptotic infinity. Then
\begin{equation}
    \text{tr}\left(T_{bc}(x) e^{-T^{-1} D} e^{\frac{i\Omega_h}{T} M_{23}} \right) = \frac{3 \ell_{\text{AdS}}^2}{16\pi G_N} h_3(x)_{bc} \ .
\end{equation}

\subsection{AdS-Kerr solutions}

The Kerr-AdS black holes form a two-parameter family of solutions, originally discovered by Carter, \cite{Carter:1968ks}, who worked with Boyer-Lindquist-like coordinates $(\mathcal{T},\rho,\theta,\Psi)$. To write the metric, we use instead the coordinates $(t,\rho,\chi,\Psi)$ of Chambers and Moss, \cite{Chambers:1994ap}, with
\begin{equation}
    t = \left(1 - \frac{a^2}{\ads^2}\right) \mathcal{T}, \qquad \chi = a\cos\theta\ .
\end{equation}
In these coordinates, the metric reads
\begin{align}
    ds^2 & = \frac{-\Delta_\rho}{(\rho^2+\chi^2)\Sigma^2}\left(dt - \frac{a^2-\chi^2}{a}d\Psi\right)^2 \\
    & + \frac{\Delta_\chi}{(\rho^2+\chi^2)\Sigma^2}\left(dt - \frac{a^2+\rho^2}{a}d\Psi\right)^2 + \frac{\rho^2+\chi^2}{\Delta_\rho}d\rho^2 + \frac{\rho^2+\chi^2}{\Delta_\chi}d\chi^2\,,
\end{align}
with
\begin{equation} \label{eq:BH-inequalities}
    \Delta_\rho = (a^2+\rho^2)\left(1+\frac{\rho^2}{\ads^2}\right) - 2M\rho,\quad \Delta_\chi = (a^2-\chi^2) \left(1-\frac{\chi^2}{\ads^2}\right),\quad \Sigma = 1 - \frac{a^2}{\ads^2}\ .
\end{equation}
The metric describes a black hole of (ADM) mass $M/\Sigma^2$ and angular momentum $aM/\Sigma^2$. There is an event horizon at $\rho=r_+$, the largest root of $\Delta_\rho$. Parameters $a$ and $r_+$ are not arbitrary, but required to obey the inequalities
\begin{equation}\label{inequalitites}
     a \leq r_+ \sqrt{\frac{\ads^2 + 3r_+^2}{\ads^2 - r_+^2}} \quad \text{for} \quad r_+<\frac{\ads}{\sqrt{3}}\,, \qquad\qquad  a<\ads \quad\text{for} \quad r_+\geq \frac{\ads}{\sqrt{3}} \ .
\end{equation}
Some of the important characteristics of black hole, the angular velocity and the temperature of the horizon, read
\begin{equation}\label{horizon-T-Omega}
     \Omega_H = \frac{a \Sigma}{r_+^2 + a^2}, \quad T_H = \frac{1}{\Sigma} \left(\frac{r_+}{2\pi}\left(1+\frac{r_+^2}{\ads^2}\right)\frac{1}{r_+^2 + a^2} - \frac{1}{4\pi r_+}\left(1-\frac{r_+^2}{\ads^2}\right)\right)\ .
\end{equation}
Inequalities \eqref{inequalitites} can be understood as the requirement that the temperature $T_H$ is non-negative. The angular velocity and temperature as measured at asymptotic infinity are
\begin{equation}\label{asymptotic-T-Omega}
    i\Omega = \Sigma\Omega_H + \frac{a}{\ads^2}, \qquad T = \Sigma T_H\ .
\end{equation}
These are identified with the same quantities in the CFT, i.e. in our notation $\mu = i\beta\Omega$. The Einstein's equations obeyed by this solution in our conventions read
\begin{equation}\label{Einsteins-eqns}
    R_{\mu\nu} + \frac{3}{\ads^2} g_{\mu\nu} = 0\ .
\end{equation}

\paragraph{Partition function} The Euclidean action for the Kerr-AdS black hole is given by (see e.g. \cite{Gibbons:2004ai} for the way the action is regularised and computed)
\begin{equation}
    S_{BH} = \frac{-\pi \left(r_+^2 + a^2\right)^2 \left(\frac{r_+^2}{\ads^2} - 1\right)}{\Sigma \left(3r_+^2 + (\ads^2 + a^2)r_+^2 - a^2 \ads^2\right)}\ .
\end{equation}
At large temperature, the partition function $Z=e^{-\frac{S_{BH}}{G_N}}$ expands as, \cite{Benjamin:2023qsc},
\begin{equation}
    \log Z = \frac{16\pi^3 \ads^2 T^2}{27 G_N (1+\Omega^2)} \left(1 - \frac{9(2+\Omega^2)}{16\pi^2 T^2} + O\left(\frac{1}{T^4}\right)\right)\ .
\end{equation}
From here, one infers the free energy density, as well as coefficients $c_{1,2}$ that determine first subleading correction to the density of states at $\Delta\gg J\gg 1$,
\begin{equation}\label{holographic-free-energy}
    f = \frac{4\pi^2}{27} \frac{\ads^2}{G_N}\,, \qquad c_1 = \frac{1}{12} \frac{\ads^2}{G_N} \,, \qquad c_2 = -\frac{1}{32} \frac{\ads^2}{G_N}\ .    
\end{equation}
We refer the reader to \cite{Benjamin:2023qsc} for details about the asymptotic density of states.

\subsection{Conformal boundary}

The analysis of the boundary is best performed in Fefferman-Graham coordinates $(\mathcal{T},z,\theta,\varphi)$, in which the metric takes the form
\begin{equation}
    ds^2 = \frac{\ads^2}{z^2} \left(dz^2 + h_0 + \frac{z^2}{\ads^2} h_2 + \frac{z^3}{\ads^3} h_3 + O(z^6)\right)\,,
\end{equation}
where
\begin{equation}
    h_0 = ds_\partial^2 = -d\mathcal{T}^2 + \ads^2 \left(d\theta^2 + \sin^2\theta\, d\varphi^2 \right)\ .
\end{equation}
The boundary coordinates $(\mathcal{T},\theta,\varphi)$ are those on the Lorentzian cylinder with the time $\mathcal{T}$ and the usual angles $(\theta,\varphi)$ on the two-sphere. The Fefferman-Graham metric is required to satisfy $g_{zz} = \ads^2/z^2$ and $g_{zb}=0$ for $b\in\{\mathcal{T},\theta,\varphi\}$. Solving these conditions in increasing orders in $z$ gives the coordinate transformation between Chambers-Moss and Fefferman-Graham coordinate systems, with fixed relations
\begin{equation}
    t = \Sigma \mathcal{T}, \qquad \Psi = \varphi - \frac{a}{\ads^2} \mathcal{T}\ .
\end{equation}
The process computes boundary metrics $h_2$, $h_3$ etc. The change of coordinates to appropriate order was found in \cite{Cardoso:2013pza}. This gives the result
\begin{equation}\label{holographic-1-pt-function}
    h_3 = \frac{\left(a^2 + r_+^2\right)(1+r_+^2)}{3r_+(1 - a^2 p^2)} \begin{pmatrix}
        2+a^2 p^2 & 0 & -3 a p^2\\
        0 & \frac{1-a^2 p^2}{p^2} & 0\\
        -3 a p^2 & 0 & p^2 \left(1 + 2 a^2 p^2\right)
    \end{pmatrix}\ .
\end{equation}
We have set the AdS radius to $\ads=1$ in the last expression. One readily verifies that $h_3\equiv h$ is traceless and conserved,
\begin{equation}
    h^b{}_b = 0, \quad \nabla_b h^{bc} = 0\,,
\end{equation}
with respect to the boundary metric $h_0$. To expand $h$ in thermal one-point blocks, we first Wick-rotate the boundary metric to obtain the Euclidean cylinder $\mathbb{R} \times S^2$ and then transport the resulting tensor to the flat $\mathbb{R}^3$. Finally, contracting indices with a polarisation vector, we obtain
\begin{align}
    \langle T_{bc} \rangle_{q,y} = \frac{3}{16\pi G_N} (i r)^{-3}Z^2 \frac{\left(1+r^2_+\right)\left(a^2 + r^2_+\right)}{r_+ (a^2 p^2 -1)^{5/2}} &\Big(1 + 2p \sqrt{1-p^2} \left(z + a\sqrt{1+z^2}\right)\\
    & + p^2 \left(a^2 - 1 + (a^2 + 1) z^2 + 2 a z \sqrt{1+z^2}\right)\ . \nonumber
\end{align}
We observe that the final expression satisfies Ward identities discussed in Section \ref{S:CFT at finite temperature}. To connect to the analysis of the main text, the variables $a$ and $r_+$ are to be expressed in terms of asymptotic temperature and angular momentum, $T$ and $\Omega$, using \eqref{horizon-T-Omega} and \eqref{asymptotic-T-Omega},
\begin{equation}
    \Omega = -i\frac{a\left(1+r^2_+\right)}{a^2 + r_+^2}\,, \qquad T = \frac{r_+\left(1 + r_+^2\right)}{2\pi\left(r_+^2 + a^2\right)} - \frac{1 - r_+^2}{4\pi r_+}\ .
\end{equation}
The stress tensor one-point function becomes particularly simple for zero angular chemical potential
\begin{align}\label{holographic-T-zero-ch-potential}
    \langle T_{bc} \rangle_q & = i (i r)^{-3} Z^2 \left(pz + \sqrt{1-p^2}\right)^2 \frac{16\pi^3 + (8\pi^2 + 3\beta^2 )\sqrt{4\pi^2 - 3\beta^2}}{72\beta^3}\\[2mm]
    & = r^{-3} Z^2 \left(pz + \sqrt{1-p^2}\right)^2\ \left( -\frac{4\pi^2}{9 G_N} \beta^{-3} + \frac{3}{64 \pi^2 G_N} \beta + \frac{3}{256\pi^4 G_N}\beta^3 + \dots \right)\nonumber\ .
\end{align}
The partition function and the stress tensor one-point function \eqref{holographic-T-zero-ch-potential} are readily checked to satisfy the thermodynamic relation \eqref{thermodynamic-relaion}. The one-point function is exponentially small as $q\rightarrow 0$ and therefore does not admit a meaningful conformal block decomposition by working order by order in $q$. This is expected since the inequalities \eqref{eq:BH-inequalities} imply that the temperature of the dual CFT for a non-spinning black hole is bounded to be above a minimal value $T_{\text{min}} = \sqrt{3}/(2\pi)$. On the other hand, the one-point function shows the expected behaviour $\langle T\rangle \sim b_T \beta^{-3}$ at large temperatures $q \rightarrow 1$. In particular, by inspecting the leading coefficient in the high-temperature expansion of \eqref{holographic-T-zero-ch-potential} and using the relation $b_T = -d f$, one reproduces the free energy density given in \eqref{holographic-free-energy}.

\bibliographystyle{JHEP}
\bibliography{bibliography}

\end{document}